\RequirePackage{fix-cm}

\documentclass[smallextended]{svjour3}     

\smartqed  

\usepackage{natbib}
\usepackage{graphicx}
\usepackage{epstopdf}
\usepackage{graphicx}
\usepackage{mathtools,breqn}
\usepackage{tabularx,booktabs}
\usepackage{array}
\usepackage{dcolumn}
\usepackage{subcaption}
\usepackage{mwe}
\usepackage{multirow}
\usepackage{tabu}
\usepackage{algorithm}
\usepackage[noend]{algpseudocode}
\usepackage{amsfonts}
\usepackage{subfig}
\usepackage{epsfig}
\usepackage[section]{placeins}

\def\calc{\mathcal{C}}
\def\ovr{O}
\newcolumntype{F}{>{\bfseries}c}

\def\st{\textit{ s.t. }}

\begin{document}
	\title{Toward Early and Order-of-Magnitude Cascade Prediction in Social Networks\thanks{Some of the authors of this paper are supported by by AFOSR Young Investigator Program (YIP) grant FA9550-15-1-0159, ARO grant W911NF-15-1-0282, and the DoD Minerva program.}
		\footnotetext{\noindent U.S. provisional patent 62/201,517. A non-provisional patent is currently being filed.}
	}
	
	\author{Ruocheng Guo         \and
		Elham Shaabani \and
		Abhinav Bhatnagar \and
		Paulo Shakarian 
	}

	\institute{Ruocheng Guo, Elham Shaabani, Abhinav Bhatnagar, Paulo Shakarian \at
		Arizona State University \\
		\email{\{rguosni, shaabani, abhatn, shak\}@asu.edu}          
	}
	
	\date{Received: date / Accepted: date}

	\maketitle

\begin{abstract}
When a piece of information (microblog, photograph, video, link, etc.) starts to spread in a social network, an important question arises: will it spread to ``viral'' proportions -- where ``viral'' can be defined as an order-of-magnitude increase.  However, several previous studies have established that cascade size and frequency are related through a power-law - which leads to a severe imbalance in this classification problem.  In this paper, we devise a suite of measurements based on ``structural diversity'' -- the variety of social contexts (communities) in which individuals partaking in a given cascade engage.  We demonstrate these measures are able to distinguish viral from non-viral cascades, despite the severe imbalance of the data for this problem.  Further, we leverage these measurements as features in a classification approach, successfully predicting microblogs that grow from 50 to 500 reposts with precision of 0.69 and recall of 0.52 for the viral class - despite this class comprising under 2\% of samples.  This significantly outperforms our baseline approach as well as the current state-of-the-art.  We also show this approach also performs well for identifying if cascades observed for 60 minutes will grow to 500 reposts as well as demonstrate how we can tradeoff between precision and recall.
\end{abstract}

\section{Introduction}

When a piece of information (microblog, photograph, video, link, etc.) starts to spread in a social network, an important question arises: will it spread to ``viral'' proportions -- where ``viral'' is defined as a significant (i.e. order-of-magnitude) increase in the number of individuals re-posting the information. However, several previous studies~\citep{watts,cheng2014can} have established that cascade size and frequency are related through a power-law - which leads to a severe imbalance in this classification problem.  In this paper, we devise a suite of measurements based on ``structural diversity'' that are associated with the growth of a viral cascade in a social network.  Structural diversity refers to the variety of social contexts in which an individual engages and is typically instantiated (for social networks) as the number of distinct communities represented in an individual's local neighborhood~\citep{ugander2012structural,zhang2013social,shakWsdm14,li2015influential}.  Previously, Ugander et al. identified a correlation between structural diversity and influence~\citep{ugander2012structural}.  
We demonstrate these measures are able to distinguish viral from non-viral cascades, despite the severe imbalance of the data for this problem.  Further, we leverage these measurements as features in a classification approach, successfully predicting microblogs that grow to 500 reposts from 50 (size-based experiments) or the first-hour observations (time-based experiments).  The main contributions of the paper are as follows:

\begin{itemize}
	\item We develop a suite of structural diversity-based measurements that are indicative of cascade growth.
	\item We are able to identify cascades of 50 reposts that grow to 500 reposts with a precision of 0.69 and recall of 0.52 for the viral class (200 out of 13,285 samples).
	
	\item We are able to identify cascades that have advanced for 60 minutes that will reach 500 reposts with a precision of 0.65 and recall of 0.53 for the viral class  (200 out of 3,444 samples).
	\item We demonstrate how to trade-off between precision and recall for the above-mentioned problems.  For instance, to predict cascades that reach 500 nodes, we can obtain precision of 0.78 or recall of 0.71 at the expense of the other.
	\item We demonstrate that our approach is stable for alternative definitions of "viral" (i.e. microblogs that grow to sizes above or below 500 reposts).
	
\end{itemize}
We note that our results on the prediction of cascades rely solely upon the use of our structural diversity based measures for features and limited temporal features - hence the prediction is based on network topology alone (no content information was utilized).  We also achieved these results while maintaining the imbalance of the dataset - where we leave the ratio of 'viral' and 'non-viral' samples as it is.  This differs from some previous studies (i.e. \citep{jenders13}) which balance the data before conducting classification experiments.  
Further, we note that we obtained prediction of order-of-magnitude increases in the size of the cascade - which also differs from other work (i.e. \citep{cheng2014can}) which focus on identifying cascades that double in size.  The remainder of the paper is organized as follows.  In Section~\ref{sec:prelim} we introduce our notation and describe the dataset used in this paper.  This is followed by an introduction of our structural diversity measurements for cascades in Section~\ref{sec:measSec}.  Then we describe our experimental results where we examined both the behavior of these measurements and the performance of classifiers built using these measurements in Section ~\ref{sec:clf}. Finally, we discuss related work in Section~\ref{sec:rwSec}.

\section{Technical Preliminaries}
\label{sec:prelim}

Here we introduce necessary notation and describe our social network data.  We represent a social network as $G=(V,E)$ where $V$ is the set of nodes and $E$ as a set of directed edges with sizes $|V|,|E|$ respectively.  
The intuition behind edge $(v,v')$ is that it is possible that $v'$ repost a microblog from $v$ since $v'$ did this previously.  
This intuition stems from how we create the edges in our network: $(v,v')$ is an edge if $v'$ reposted from $v$ once or more during a specified time period (for our experiments, May 1 to July 31, 2011).
We also assume a partition over nodes that specifies a community structure.  We assume that such a partition is \textbf{static} (based on the same time period from which the edges were derived) and that the \textit{partition} $\calc$ consists of $k$ communities: $\{C_1, C_2, ...,C_k \}$, each is a set of nodes. 
There are many possible methods to derive the communities (if user-reported communities are not available) - for instance: the Louvain algorithm \citep{blondel2008fast}, Infomap \citep{rosvall2008maps}, Smart Local Move (SLM) \citep{waltman2013smart} and Label Propagation \citep{raghavan2007near}. Previous work such as \citep{weng2014predicting,grabowicz2012social} showed the effectiveness of communities detected by these algorithms for different applications. In this paper, We utilize the Louvain algorithm, Infomap algorithm and SLM algorithm to identify communities in the social network $G$ due to their scalability for large social network. For these algorithms, the number of communities is not an argument as input but rather produced as part of the output of these algorithms.
Note that we require $\calc$ to be a partition over nodes - hence we disallow for overlapping communities.  This is consistent with the community structure derivations from previous, related work ~\citep{ugander2012structural,zhang2013social,shakWsdm14,li2015influential} which also required a partition over nodes such as strongly connected components.  As such, we leave the study of structural diversity in the case of overlapping communities to future efforts.

\paragraph{Cascades.}
A cascade $\tau = (U,R)$ consists of all nodes ($U$) who posted or reposted a certain original microblog and the reposting relationships between them, treated as edges ($R$). 
Naturally, any cascade is a subgraph of the social network $G$.
In order to predict the final size, snapshots of a cascade can be taken by different time since adoption of the seed adopter (denoted by $t$).
Then a snapshot of cascade $\tau$ introduces a subset $\tau_{t} = (U_{t},R_{t})$ of $\tau$. We refer to $U_{t}$ as \textit{adopters}. Moreover, we also call the out-neighbors of adopters in $G$ but \textbf{not among the adopters} as \textit{exposed users} and denote them as $N_{G}(U_{t})$.
For each node $v \in N_{G}(U_{t})$, we define the adopters who exposes the cascade to $v$ as its \textit{exposers}.
For convenience, we also define function $u_{ea} : v \rightarrow u$ to return the earliest adopter $u$ among exposers of $v \in N_{G}(U_{t})$.

For size-based experiments, the time $t$ for taking snapshot of a cascade is decided by a given cascade size $m$. We use $t(m)$ to denote the smallest $t$ such that $|U_t|=m$ is true for a certain cascade.
Accordingly, to get the corresponding order number $n$ of an adopter $u \in U_{t}$, we define function $Index:u \rightarrow n$ where $n \in [1,|U_t|]$.
To maintain a unique order of reposts, a very small random number is added to each $t(n)$ for all integers $n \in  [1,|U_t(m)|] $. We have not found this to be a significant issue in this dataset. For convenience and simplicity, we use $t$ to stand for both $t(m)$ in size-based and $t$ in time-based experiments later.

For a given snapshot $\tau_t = (U_t,R_t)$, then we want to divide the set $N_{G}(U_t)$ into two sets, namely \textit{recently exposed users} ($\mathcal{F}_t$) and \textit{past exposed users} ($\mathcal{N}_t$).
Intuitively, this division is done based on how long it is since $v \in N_{G}(U_t)$ is true (when it is possible for $v$ to make a repost) till the snapshot $\tau_t$ is taken.
Formally, given a node $v \in N_{G}(U_t)$, we decide whether it is a recently or past exposed user:
\begin{equation}
	t_{expose}(v) = t - t(Index(u_{ea}(v)))
\end{equation}

As defined before, $t(n)$ denotes the earliest time $t$ when $|U_t|=n$ is true.
Then the value of $t_{expose}(v)$ is the number of time periods since the earliest adoption among its exposers till when the snapshot of cascade is taken.

A positive constant $\lambda$ is set as a threshold on $t_{expose}(v)$ (we will discuss how this constant is set in the last paragraph of~\ref{sec:prelim}), the \textit{recently exposed users} and \textit{past exposed users} are defined as follows:
\begin{equation}
	\mathcal{F}_t = \left\{v \in N_{G}(U_t) \st t_{expose}(v) \le \lambda\right\}
\end{equation}
\begin{equation}
	\mathcal{N}_t = N_{G}(U_t) \setminus \mathcal{F}_t
\end{equation}

\begin{figure*}
	\begin{subfigure}[t]{0.99\textwidth}
		\centering
		\includegraphics[width=79mm]{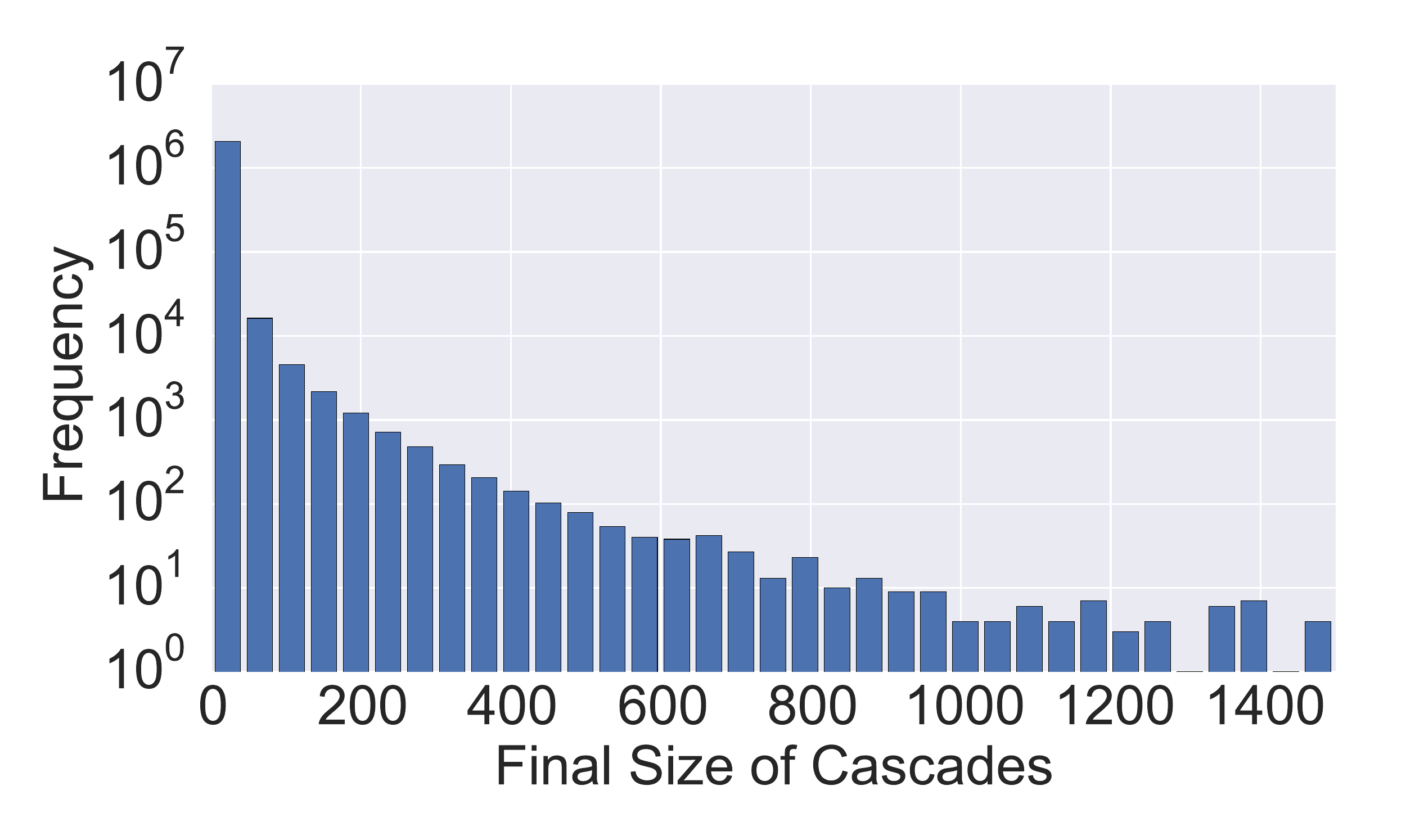}
		\caption{The histogram of cascade size for August, 2011}
		\label{fig:hist_cas_size}
	\end{subfigure}
	\vfill
	\begin{subfigure}[t]{0.99\textwidth}
		\centering
		\includegraphics[width=79mm]{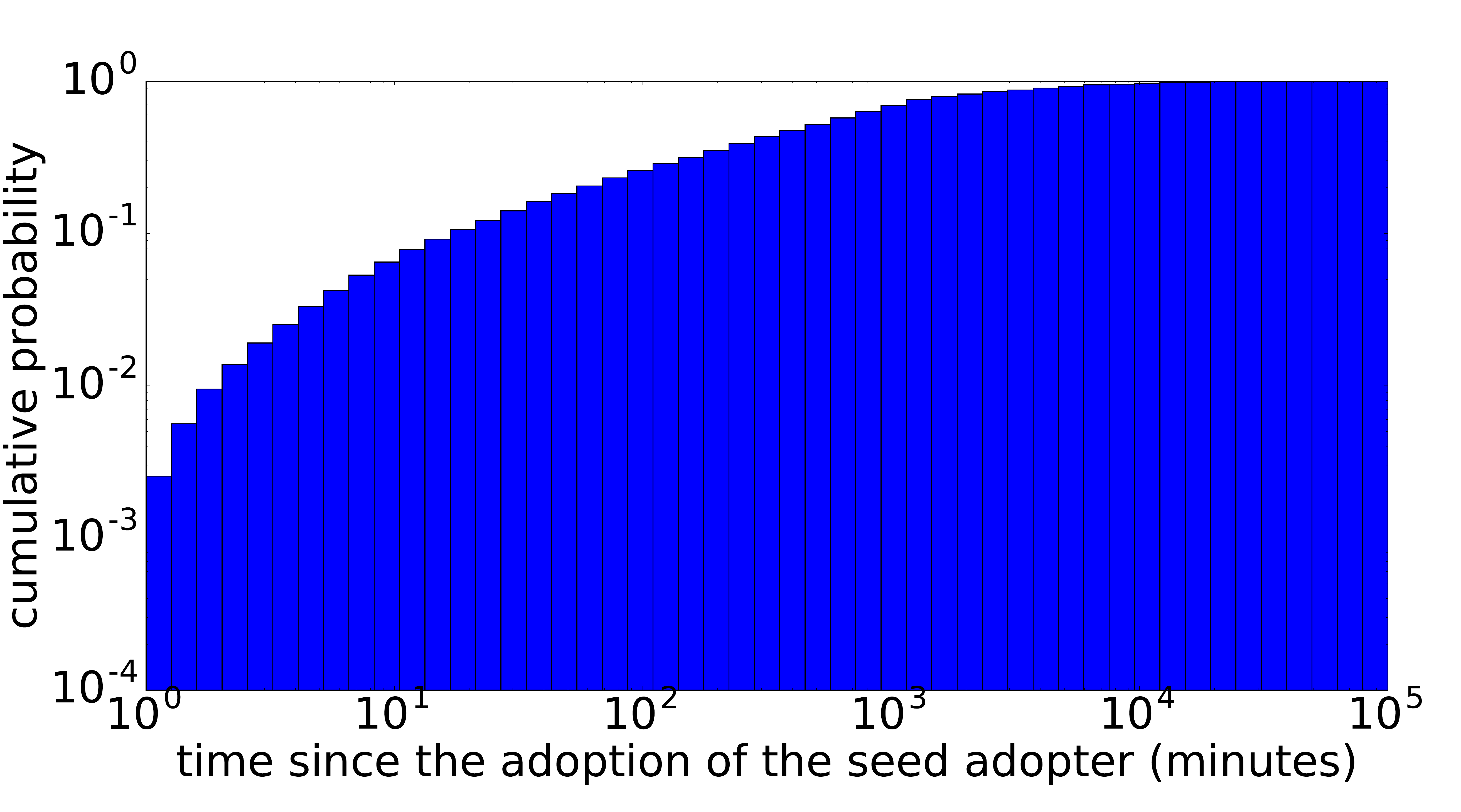}
		\caption{{CDF of adoption time since adoption of the seed adopter (minutes) for May-July, 2011}}
		\label{fig:time_since_root}
	\end{subfigure}
	\hfill
	\begin{subfigure}[t]{0.99\textwidth}
		\centering
		\includegraphics[width=79mm]{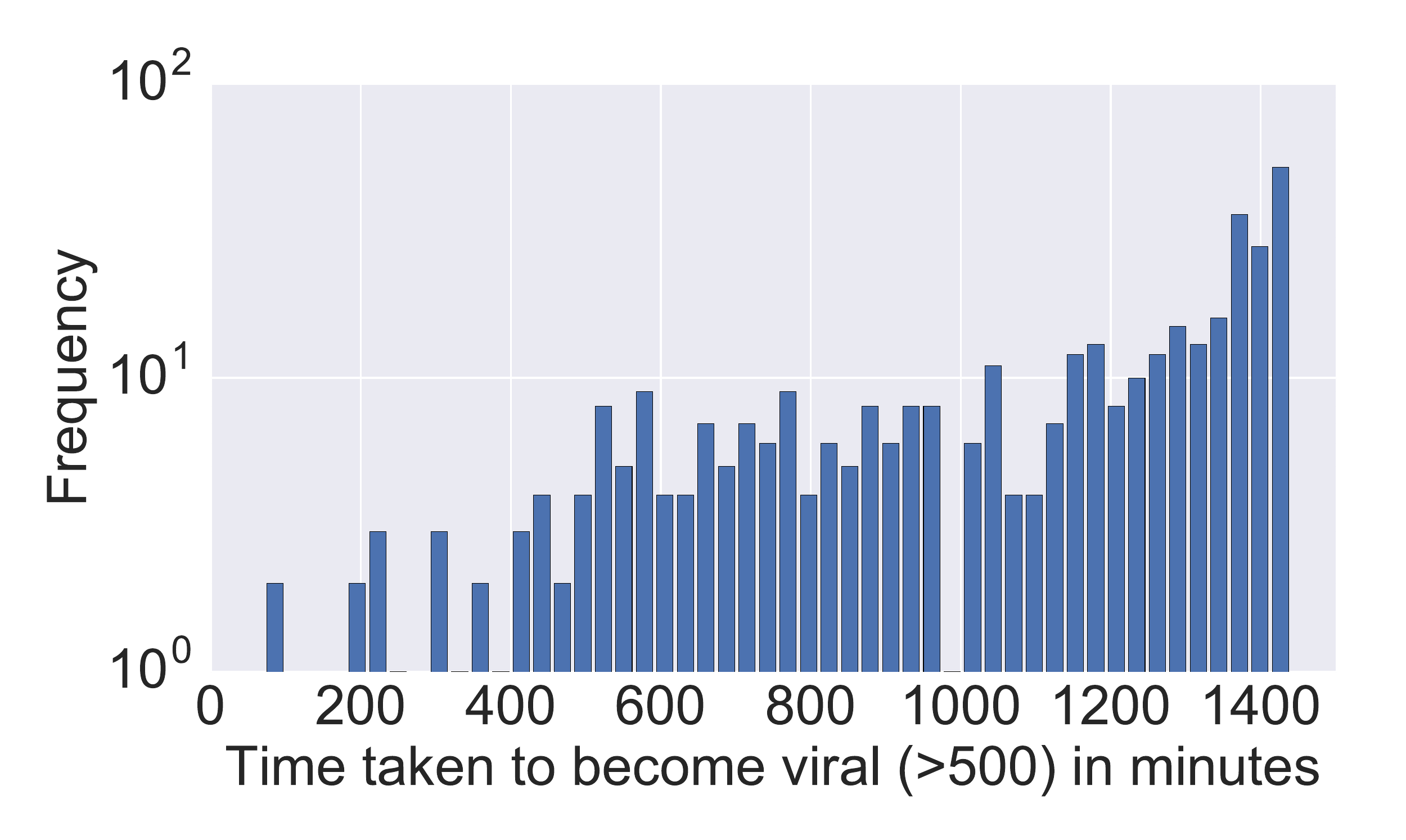}
		\caption{{Histogram of time (minutes) 'viral' cascades took to reach size of 500.}}
		\label{fig:time_to_viral}
	\end{subfigure}
	
	\caption{Network Dataset Statistics }
	
\end{figure*}

\paragraph{Sina Weibo Dataset.}  The dataset we used was provided by WISE 2012 Challenge\footnote{http://www.wise2012.cs.ucy.ac.cy/challenge.html}. It included a sample of microblogs posted on Sina Weibo\footnote{http://weibo.com} from 2009 to 2012. In this dataset, we are provided with time and user information for each microblog and subsequent repost which enabled us to derive a corpus of cascades. For every repost in this dataset, the reposting relationship is provided as \textit{uid: $v'$ tab $v$} which indicates this message is a reposted from user $v$ by $v'$. From this data, we derived our social network $G=(V,E)$ that was created from microblogs (including original posts and reposts) published during May 1, 2011 to July 31, 2011 (the 3-month period).  
For this network, the number of active nodes in August (the time period we studied for cascade prediction) is 5,910,608, while 5,664,625 of them at least have one out-neighbor.
During the month of August, there were 22,182,704 microblogs.
Of these, 9,323,294 are reposts. 2,252,368 different of original posts succeeded to make at least one user repost, while 1,920,763 ($86.6\%$) of them were written by authors who at least published one microblog during the 3-month period mentioned before.  
For this dataset, although different from a power-law noted previoulsly in \citep{watts,cheng2014can}, the histogram of final cascade size (see Figure~\ref{fig:hist_cas_size}) still shows that only quite few cascades went 'viral'. 
Therefore, we could demonstrate that this dataset is more representative of cascade behavior observed in real world than work like \citep{jenders13} which conducted biased sampling to artificially provide balanced classes.

We select the threshold constant $\lambda$ as 30 minutes since vast majority of all the reposts in May-July, 2011 occurred within 30 minutes since adoption of the seed adopter (see Figure~\ref{fig:time_since_root}).
To justify this selection,
knowing that $\lambda$ is a threshold on $t_{expose}(v)$ which is upper bounded by $t$, the proportion of exposed users became adopter with $t_{expose}(v) \le 30 (min)$ should be more than that of those did the repost with $t \le 30 (min)$. 
This implies why it is necessary to distinguish recently and past exposed users due to the significant difference in probability to adopt. 
In Figure~\ref{fig:time_to_viral}, we show distribution of how long it takes for viral cascades to reach 500 nodes - note that the average value here is approximately 18 hours (which is significantly greater than what we study in our time-based classification problem).

\smallskip

\begin{table}[!tbp]%
	\renewcommand{\arraystretch}{1}
	\caption{\textmd{Properties of the Social Network and Cascades}}
	\label{tab:net_prop}
	\centering
	\begin{tabular}{| p{5.5cm}| c|}
		\hline
		\textbf{Network Properties} & \textbf{Value} \\ \hline \hline
		Vertices (Nodes) & 17,996,803 \\ \hline
		Edges & 52,472,547 \\ \hline
		Average degree & 5.83 \\ \hline
		Average clustering coefficient &  0.107  \\ \hline
		Connected components & 4974 \\ \hline \hline
		
		Number of communities (Louvain) & 379,416 \\ \hline
		Average size of communities (Louvain) & 47.5\\ \hline
		Number of communities (Infomap) & 39,922 \\ \hline
		Average size of communities (Infomap) & 450.799 \\ \hline Number of communities (SLM) & 380,854 \\ \hline
		Average size of communities (SLM) &  47.3 \\ \hline \hline
		
		\textbf{Cascade Properties} & \textbf{Value} \\ \hline \hline
		Number of cascades & 2,113,405 \\ \hline
		Number of viral cascades & 208 \\ \hline 
		Number of active nodes in cascades & 5,910,608 \\ \hline
		Average time to become viral & 18 (h) \\ \hline
	\end{tabular}

\end{table}

\section{Structural Diversity Measurements in Real Information Cascades}
\label{sec:measSec}
\begin{figure*}
	\begin{subfigure}[t]{0.99\textwidth}
		\centering
		\includegraphics[width=79mm]{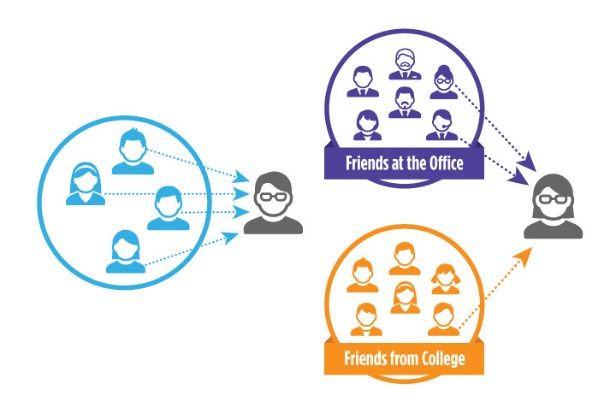}
		
	\end{subfigure}
	
	\caption{An example for structural diversity: Although the man on the left has more infected in-neighbors, the woman on the right is more likely to be infected by the social contagion. As in-neighbors of her are showing higher structural diversity (from two communities).}
	\label{fig:sd}
	
\end{figure*}

Found by \citep{ugander2012structural}, an individual is more likely to be infected by a `social contagion' if his/her `infected' in-neighbors are distributed over more connected components of social network users. For example, as shown in Figure~\ref{fig:sd}, although the man on the left has more infected in-neighbors, the woman on the right is more likely to be infected by the social contagion. As in-neighbors of her are showing higher structural diversity (from two communities).
Translated into the terminologies introduced in this paper, they showed that an exposed user is more likely to become an adopter with exposers of high structural diversity.
If this effect is aggregated over all the exposed users of a cascade, the significance to measure the relationship between structural diversity of adopters would be revealed.
Moreover, we also extend our experiments to measure that of exposed users.
Instead of connected components, we consider structural diversity described by communities.
In this section we introduce a suite of various structural diversity measurements.
We study these measurements as cascades progress in Section~\ref{measSec} and then leverage them as features for our classification problem in Section~\ref{sec:clf}.
We introduce these measurements as follows. 

\smallskip

\paragraph{Number of communities.} For a given set of node $S \in \left\{U_t,F_t,N_t\right\}$ we can retrieve the associated communities $\calc(S)$ by the partition of the social network $\calc(G)$.
Formally:  
\[\calc(S)=\{C_i \in \calc(V) \st S \cap C_i \neq \emptyset\}\] where $\calc(V)$ is the partition of the social network $G$, introduced in Section~\ref{sec:prelim}.  
We measure the number of communities represented by $|\calc(S)|$ for $S \in \left\{U_t,F_t,N_t\right\}$.

\paragraph{Gini impurity.}  For $S \in \left\{U_t,F_t,N_t\right\}$ in a cascade $\tau_t$, the gini impurity $I_G(S)$ proposed by \citep{breiman1984classification} for splitting samples in decision tree, intuitively, is a scalar describing how much the distribution of nodes in $S$ over communities in $\calc(S)$ differs from \textbf{the uniform distribution}. Here the uniform distribution stands for the situation where $|C_i| = \frac{|S|}{|\calc(S)|}$ for all $C_i \in \calc(S) $. To show the extreme values, $ I_G(S) \approx 1$ means the nodes are uniformly distributed over a large quantity of communities while $I_G(S) \approx 0$ implies most of the nodes in $S$ are from the few 'dominant' communities.
Formally, we define gini impurity as follows:
\begin{equation}
	I_G(S) =  1 -\sum_{C_i \in \calc(S)}(\frac{ |C_i|}{|S|})^2
\end{equation}
We study the gini impurity $I_G(S)$ for $S \in \left\{U_t,F_t,N_t\right\}$ for each cascade.  
We note that the impurity of the adopter set $I_G(U_t)$ behaves similar to the entropy of this set (a measurement introduced in \citep{weng2014predicting}).  However, as we will see in the next two sections, we found that the impurity of the recently exposed users is a more discriminating feature.

\paragraph{Overlap.} For $\left\{S_a,S_b\right\} \subset \left\{U_t,F_t,N_t\right\}$, the overlap ($\ovr(S_a,S_b)$) is simply the number of shared communities between the sets of nodes $S_a$ and $S_b$.  
Formally: 
\begin{equation}
	\ovr(S_a,S_b)=|\calc(S_a)\cap \calc(S_b)|
\end{equation}
The intuition behind overlap stems directly from the original structural diversity results of the related work \citep{ugander2012structural} - for instance a large overlap value $O(U_t,F_t)$ is likely to indicate that the local neighborhoods of many of the recently exposed users will exhibit high structural diversity - hence increasing the probability to become adopters in the future.

\paragraph{Baseline measures.} In addition to the aforementioned structural diversity measurements, we also examine two baseline measurements dealing with time and size.

\paragraph{Average time to adoption.}  The average time to adoption for adopters in the cascade snapshot of size $m$: $ \frac{1}{m}\sum_{i=1}^m t(m) $.

\paragraph{Number of nodes.} The cardinality of adopters, recently and past exposed users $|U_t|$,$|F_t|$,$|N_t|$.
\smallskip

\section{Structural Diversity Measurement Study}
\label{measSec}
Here we examine the behavior of the various structural diversity measurements as viral and non-viral cascades progress.  In this section, we define a cascade as viral if the number of reposts eventually reaches a threshold (denoted $TH$) of $500$ (in the next section we will explore various values for $TH$). Only the distributions of feature values computed based on Louvain algorithm are exhibited in this section as it provides best results in both size-based and time-based classification tasks (See Section~\ref{sec:clf}).
All the measurements are computed by cascade snapshots with five populations of nodes with $m = \left\{10, 30, 50, 100, 200\right\}$ (or $t(m)$ accordingly) and five values of time since adoption of the seed adopter with $t = \left\{40, 60, 100, 150, 300\right\}$.
Table~\ref{tab:samples} shows the number of samples our analysis covers in both classes for each value of $m$ and $t$.
For each time $t$ we perform analysis on measurements for those cascade snapshots with no less than 5 adopters at the time so that the enough information can be provided from $U_t$,$\mathcal{F}_t$ and $\mathcal{N}_t$ for the prediction task.  
For each size $m$, we consider the cascades with $ |U_t| =  m $ adopters at the corresponding time $ t(m) $, $ t(m) $ can vary for different cascades. 
Hence, cascades with final size less than $ m $ are ignored in our analysis. This leads to that the number of non-viral cascades decreases as $ m $ increases.  
We examined a total of 24 measurements discussed in the previous section (12 for size-based and 12 for time-based analysis, listed as $A_m$ and $A_t$ respectively in Table~\ref{tab:features}).  
For each measurement, for each $m$ and $t$ describing the diffusion process, we attempted to identify statistically significant difference between viral and non-viral classes.  
For this, we performed KS tests for each pair of measurements.  In every test, $p \leq 10^{-13} $, so the null hypothesis is rejected for all cases, which means each pair of the distributions are significantly different.  We choose KS test over T test and Chi-square test as it is sensitive to both the location and shape of the distribution as well as it does not require each distribution to cover all possible values of the other.
As notations of the box plots in the following subsections, A and M denotes mean and median for each box plot respectively.  

\smallskip

\begin{table}[!t]%
	\renewcommand{\arraystretch}{1}
	\caption{\textmd{Number of samples analyzed in different stages}}
	\label{tab:samples}
	\centering
	\begin{tabular}{| p{1.5cm}| c| c|}
		\hline
		
		\textbf{$m$} & \textbf{Samples} & \textbf{Viral Samples (\%)} \\ \hline
		10 & 98,832 & $0.2 \%$ \\ \hline
		30 & 26,733 & $0.7 \%$ \\  \hline
		50 & 13,285 & $1.5 \%$ \\  \hline
		100 & 4,722 & $4.2 \%$ \\  \hline
		200 & 1,324 & $15 \%$ \\  \hline \hline
		
		\textbf{$ t $ (min)} & \textbf{Samples} & \textbf{Viral Samples (\%)} \\ \hline
		40 & 2,234 & $7 \%$ \\  \hline
		60 & 3,444 & $5 \%$ \\ \hline
		100 & 5,767 & $3 \%$ \\ \hline
		150 &  8,349 & $2 \%$  \\ \hline
		300 & 15,350 & $1 \%$\\ \hline
		
	\end{tabular}
\end{table}

\subsection{Size Progression}

\begin{figure}[!tbp]
	\begin{subfigure}[b]{0.49\textwidth}
		\centering
		\includegraphics[width=39mm]{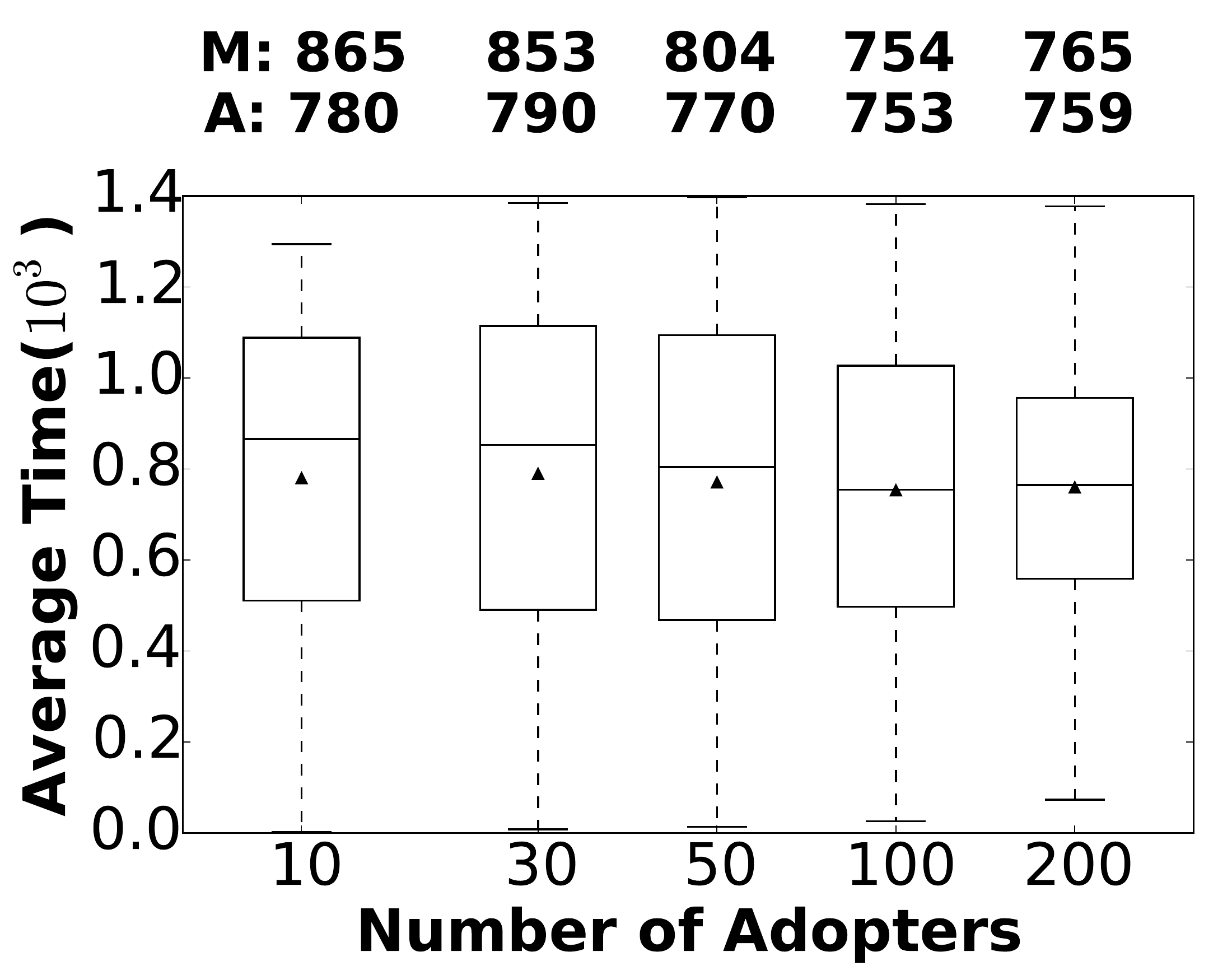}
		\caption{Non-viral cascades}
		\label{fig:nv_avg_time}
	\end{subfigure}
	\hfill
	\begin{subfigure}[b]{0.49\textwidth}
		\centering
		\includegraphics[width=39mm]{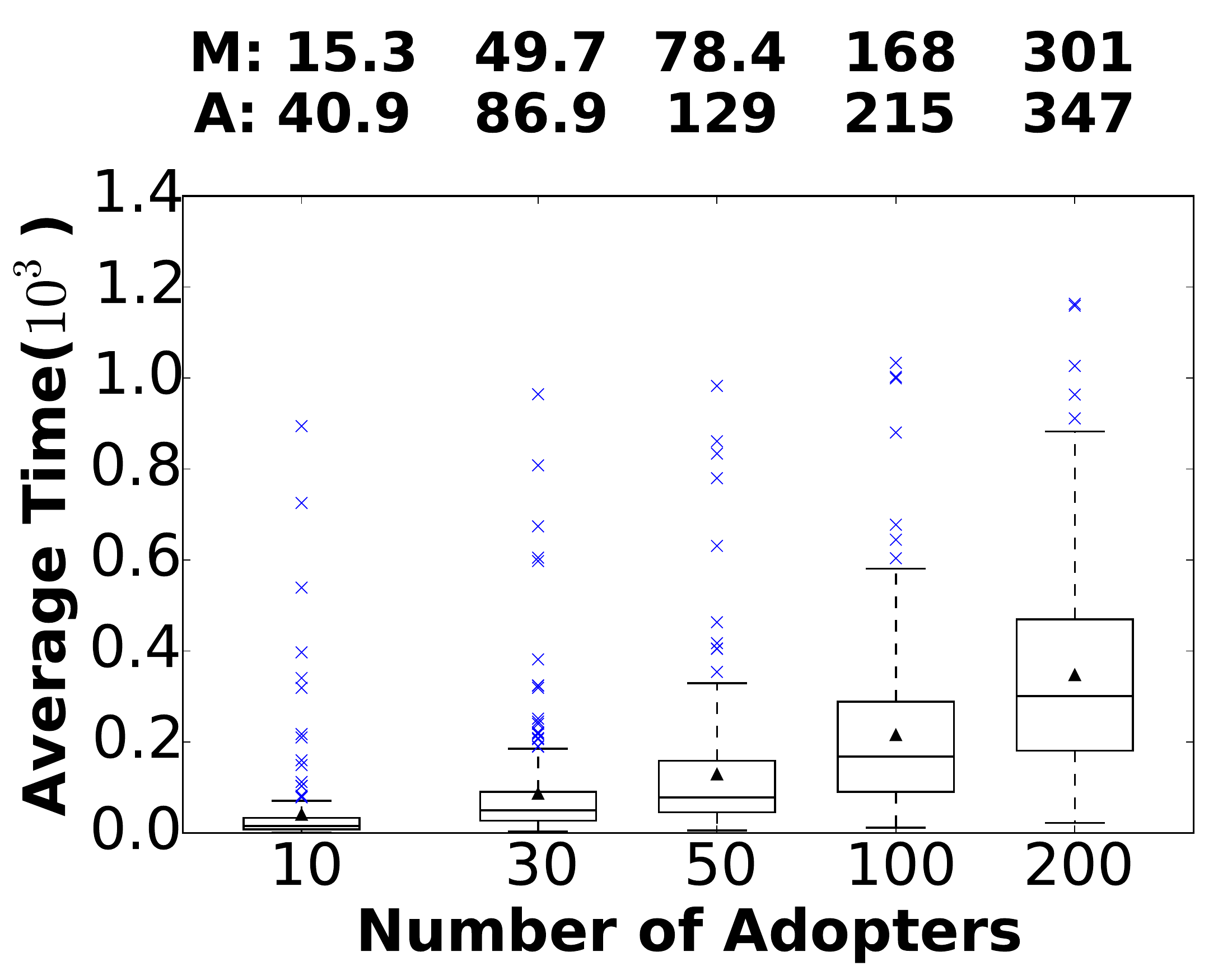}
		\caption{Viral cascades}
		\label{fig:v_avg_time}
	\end{subfigure}
	\caption{Average time (minutes in $10^3$) since adoption of the seed adopter to each later adoption}
	\label{fig:avg_time}
\end{figure}

\paragraph{Average time to adoption.}  As a baseline measurement, we study the average time to adoption for each $m$ of the cascade process (Figure~\ref{fig:avg_time}).  
As expected, viral cascades exhibit shorter average time since adoption of the seed adopter till each later adoption.  While we note that significant differences are present - especially in the early stages of the cascade, the whiskers of the non-viral class indicate a significant proportion of non-viral cascades that exhibit rapid adoption.  We believe this is likely due to the fact that certain cascades may have very high appeal to specialized communities.

\begin{figure}[!tbp]
	
	\begin{subfigure}[b]{0.49\textwidth}
		\centering
		\includegraphics[width=39mm]{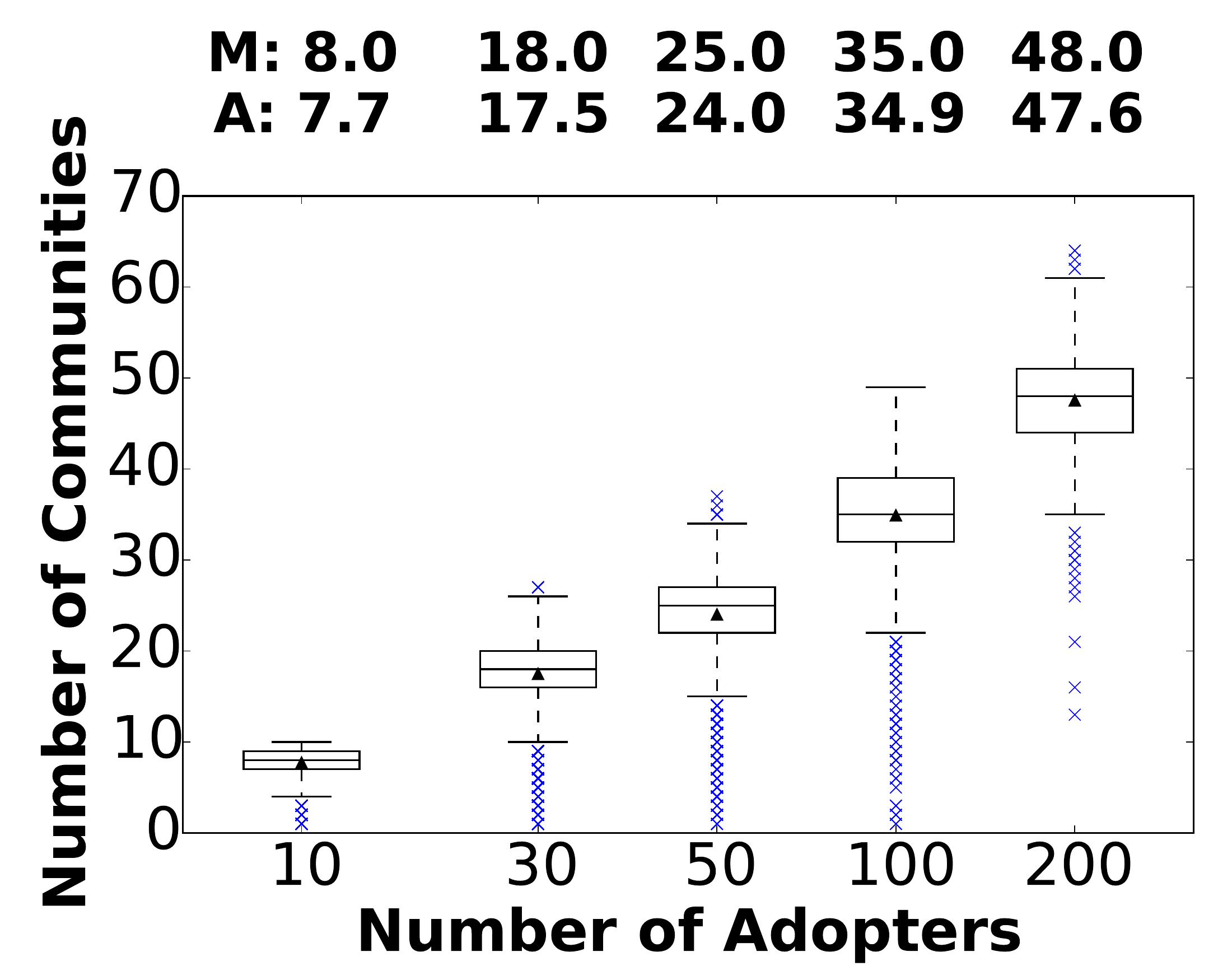}
		\caption{Number of communities amongst adopters ($|\calc(U_t)|$) for non-viral cascades}
		\label{fig:S_nv_comm_a}
	\end{subfigure}
	\hfill
	\begin{subfigure}[b]{0.49\textwidth}
		\centering
		\includegraphics[width=39mm]{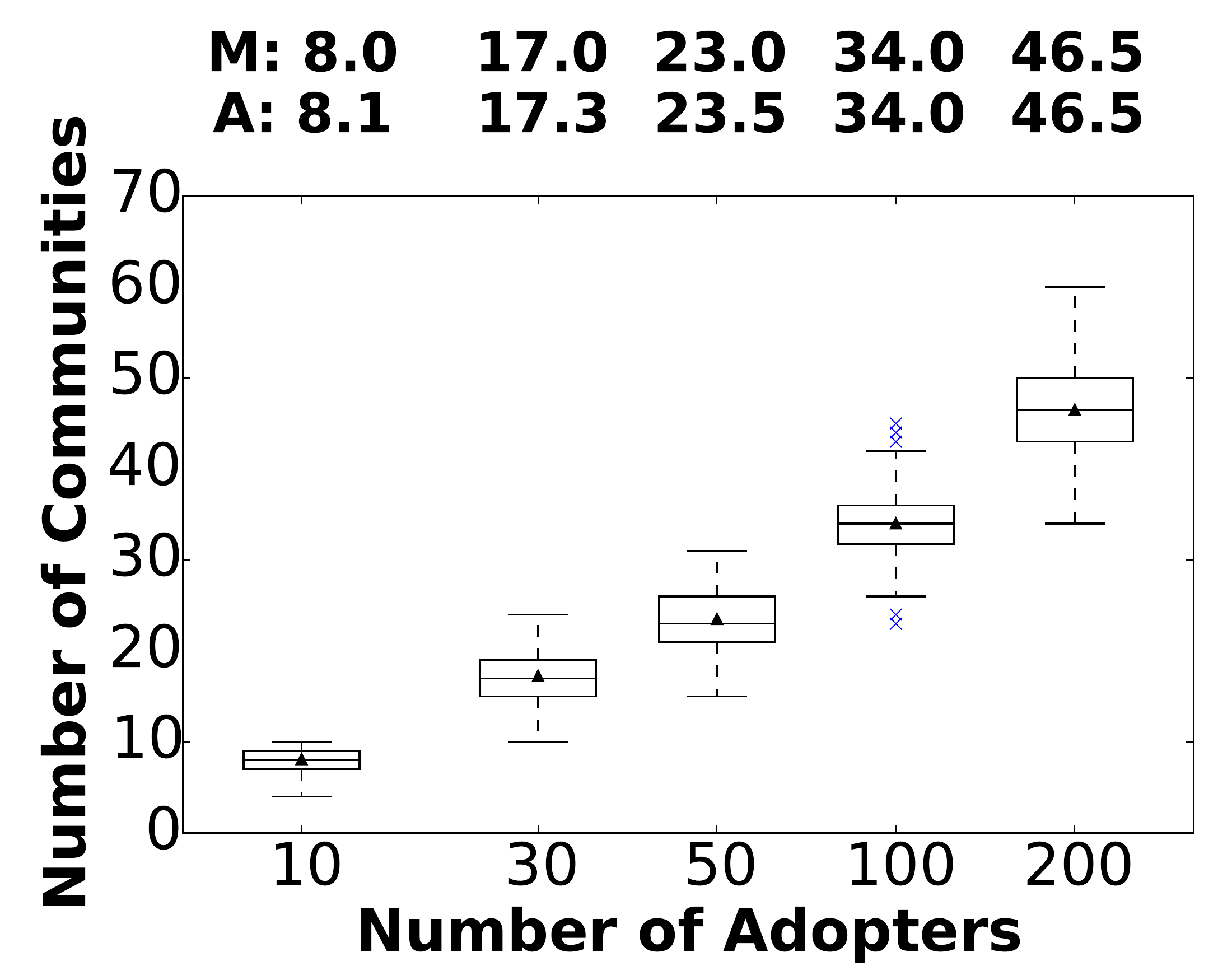}
		\caption{Number of communities amongst adopters ($|\calc(U_t)|$) for viral cascades}
		\label{fig:S_v_comm_a}
	\end{subfigure}
	\hfill
	\begin{subfigure}[b]{0.49\textwidth}
		\centering
		\includegraphics[width=39mm]{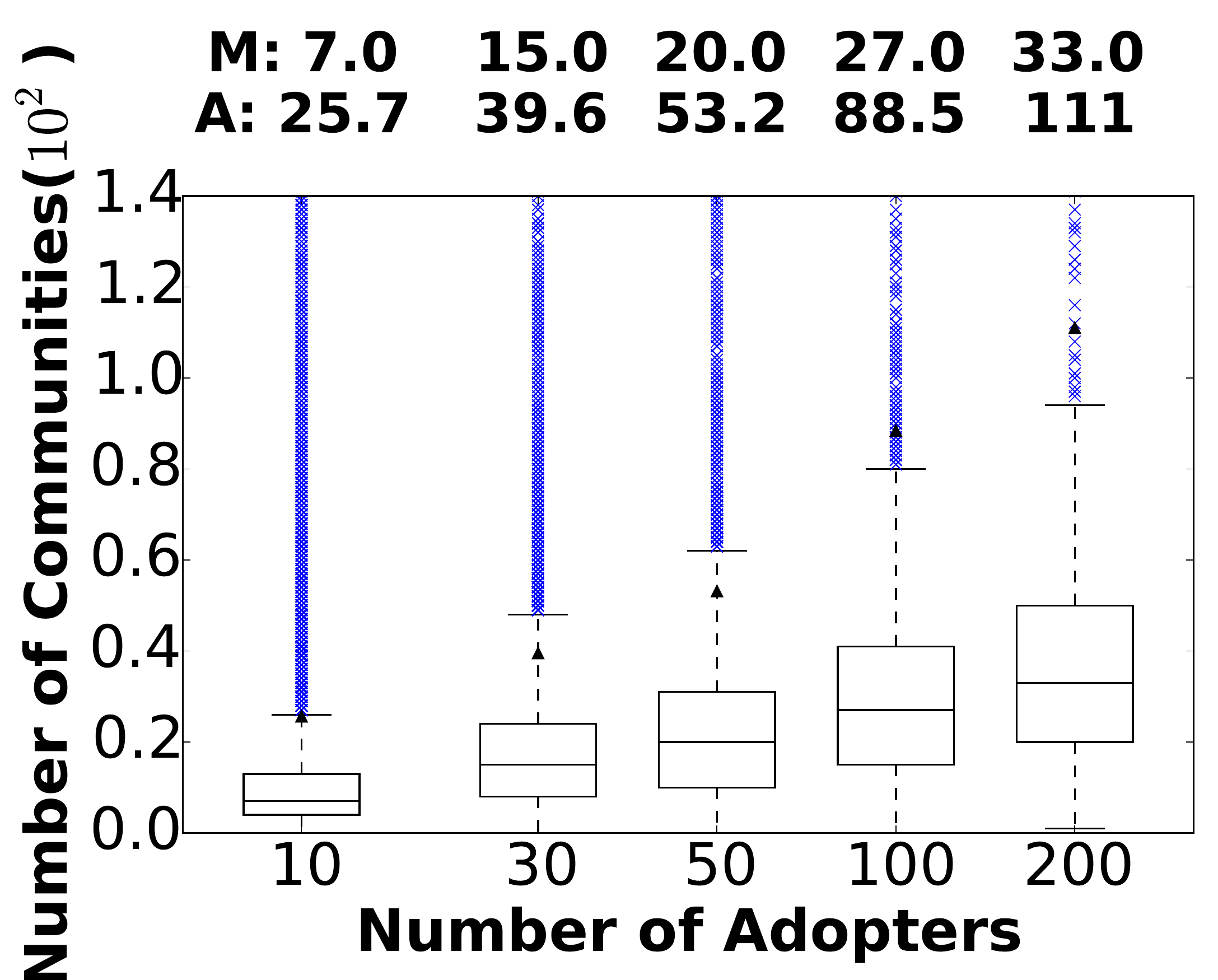}
		\caption{Number of communities amongst recently exposed users ($|\calc(\mathcal{F}_t)|$) for non-viral cascades}
		\label{fig:S_nv_comm_f}
	\end{subfigure}
	\hfill
	\begin{subfigure}[b]{0.49\textwidth}
		\centering
		\includegraphics[width=39mm]{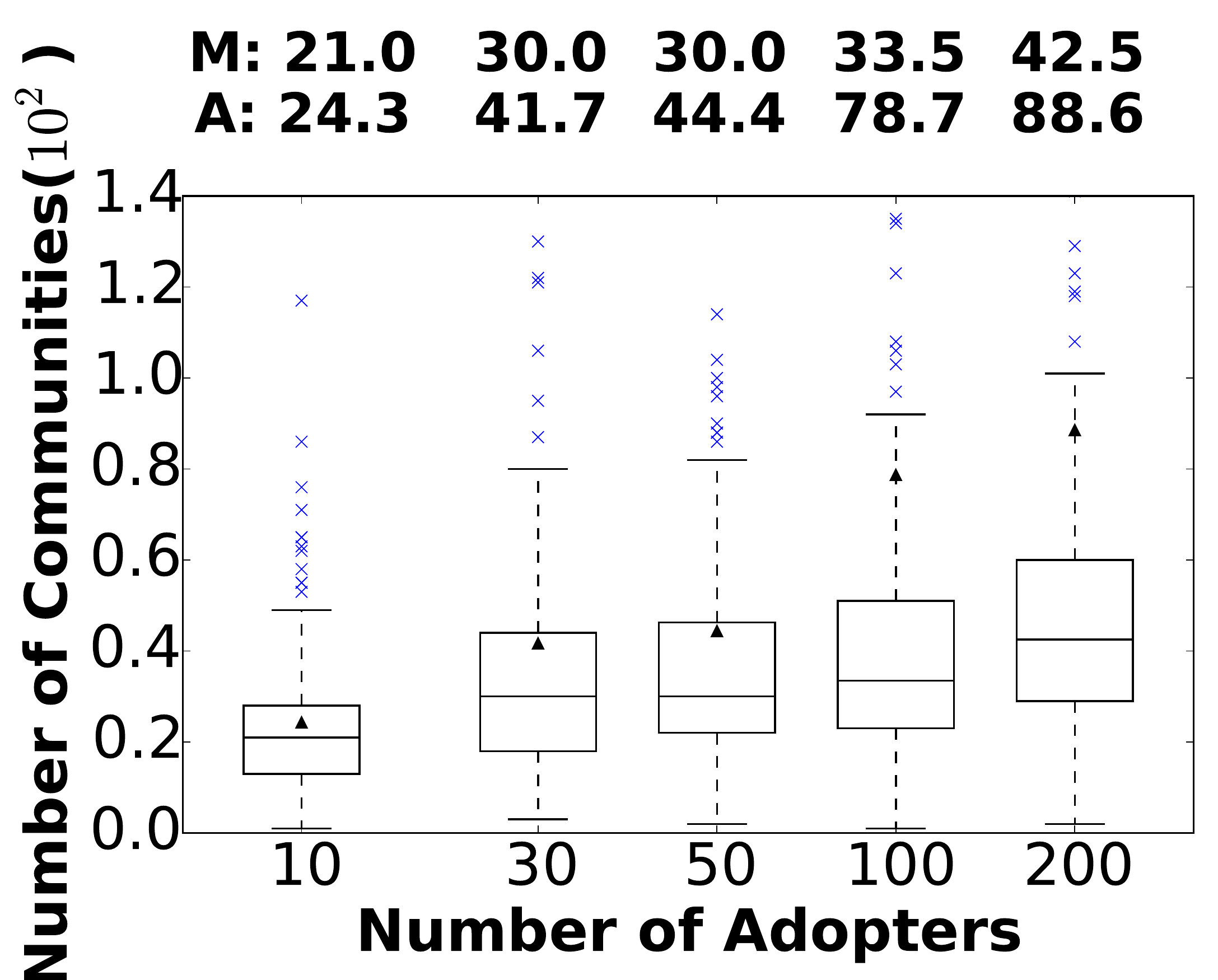}
		\caption{Number of communities amongst recently exposed users ($|\calc(\mathcal{F}_t)|$) for viral cascades
		}
		\label{fig:S_v_comm_f}
	\end{subfigure}
	\ifx
	\begin{subfigure}[b]{0.49\textwidth}
		\centering
		\includegraphics[width=39mm]{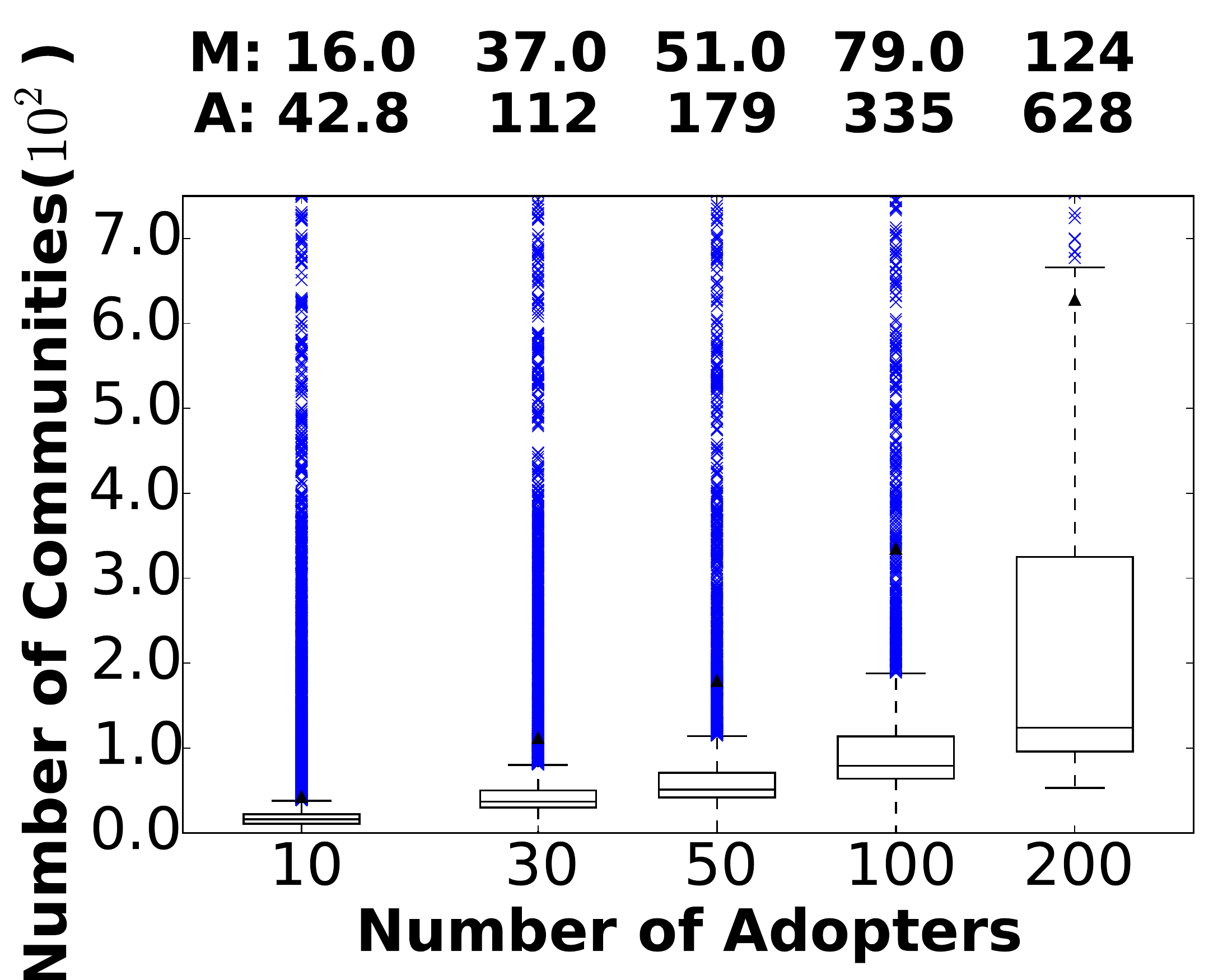}
		\caption{Number of communities amongst past exposed users ($|\calc(\mathcal{N}_t)|$) for non-viral cascades}
		\label{fig:S_nv_comm_na}
	\end{subfigure}
	\hfill
	\begin{subfigure}[b]{0.49\textwidth}
		\centering
		\includegraphics[width=39mm]{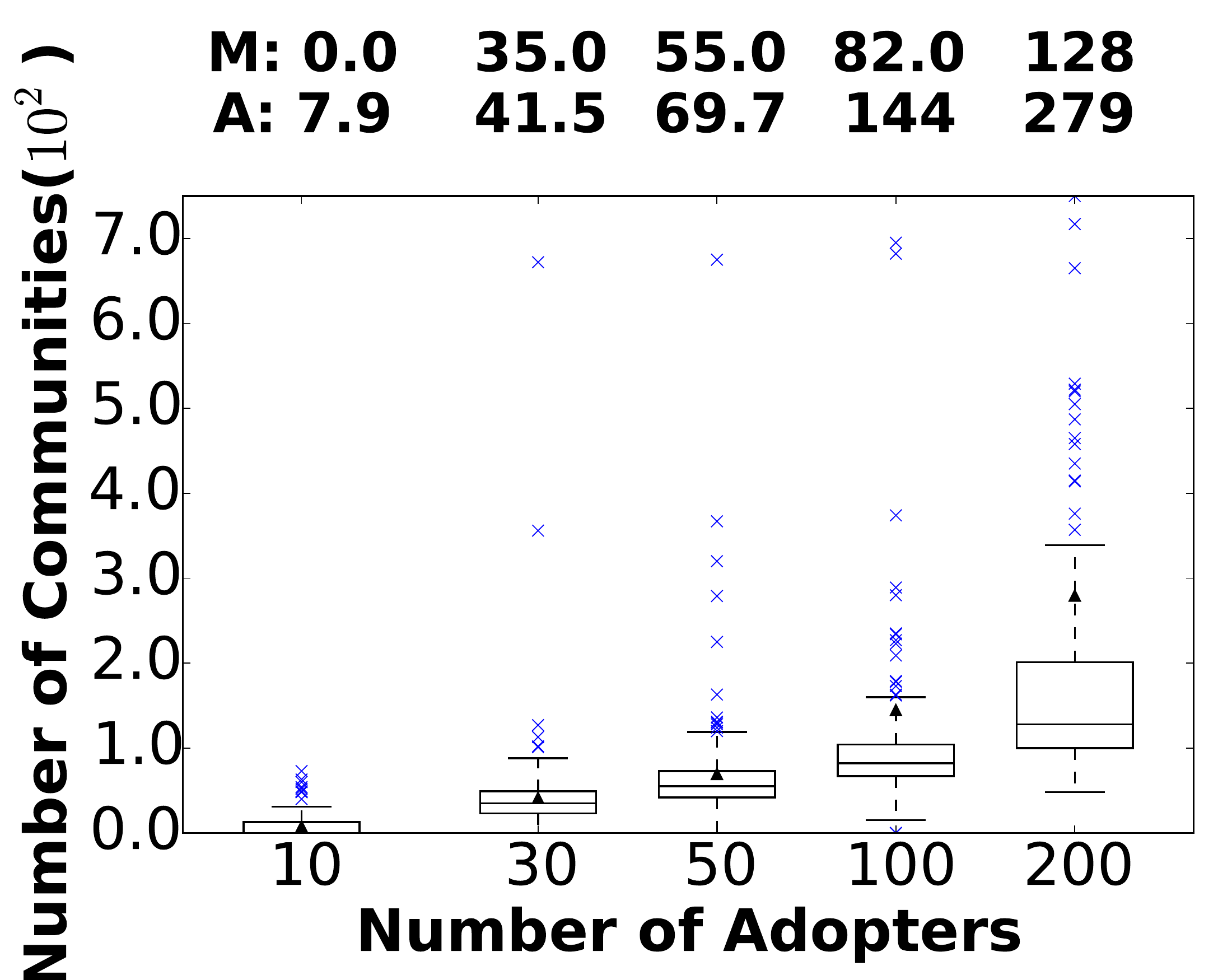}
		\caption{Number of communities amongst past exposed users ($|\calc(\mathcal{N}_t)|$) for viral cascades
		}
		\label{fig:S_v_comm_na}
	\end{subfigure}
	\fi
	\caption{Number of communities for $ m = \left\{10,30,50,100,200\right\} $}
	\label{fig:S_comm}
\end{figure}

\paragraph{Number of communities.}  Figure~\ref{fig:S_comm} displays how the number of communities $ |\calc(S)| $ increases over $ m = \left\{10,30,50,100,200\right\} $ for the sets $ S = \left\{U_t,\mathcal{F}_t\right\}$.  We note that $|\calc(U_t)|$ (the communities represented in the set of adopters) was shown to be a useful feature in \citep{weng2014predicting} for tasks where the target class had fewer reposts than in this study.  
Here, we note that while statistically significant differences exist, the average and median values at each of the examined stages are generally similar.  On the other hand, the communities represented by the set of rencently exposed users ($\mathcal{F}_t$) shows viral cascades have stronger capability to keep set of rencently exposed users with many communities than non-viral ones.  
We also noted that the median of $|\calc(\mathcal{N}_t)|$ shows viral cascades start with smaller $|\calc(\mathcal{N}_t)|$. However, it increases faster in viral cascades as nodes in rencently exposed users become past exposed users (not pictured) as $m$ increases.

\begin{figure}[!tbp]
	\begin{subfigure}[b]{0.49\textwidth}
		\centering
		\includegraphics[width=39mm]{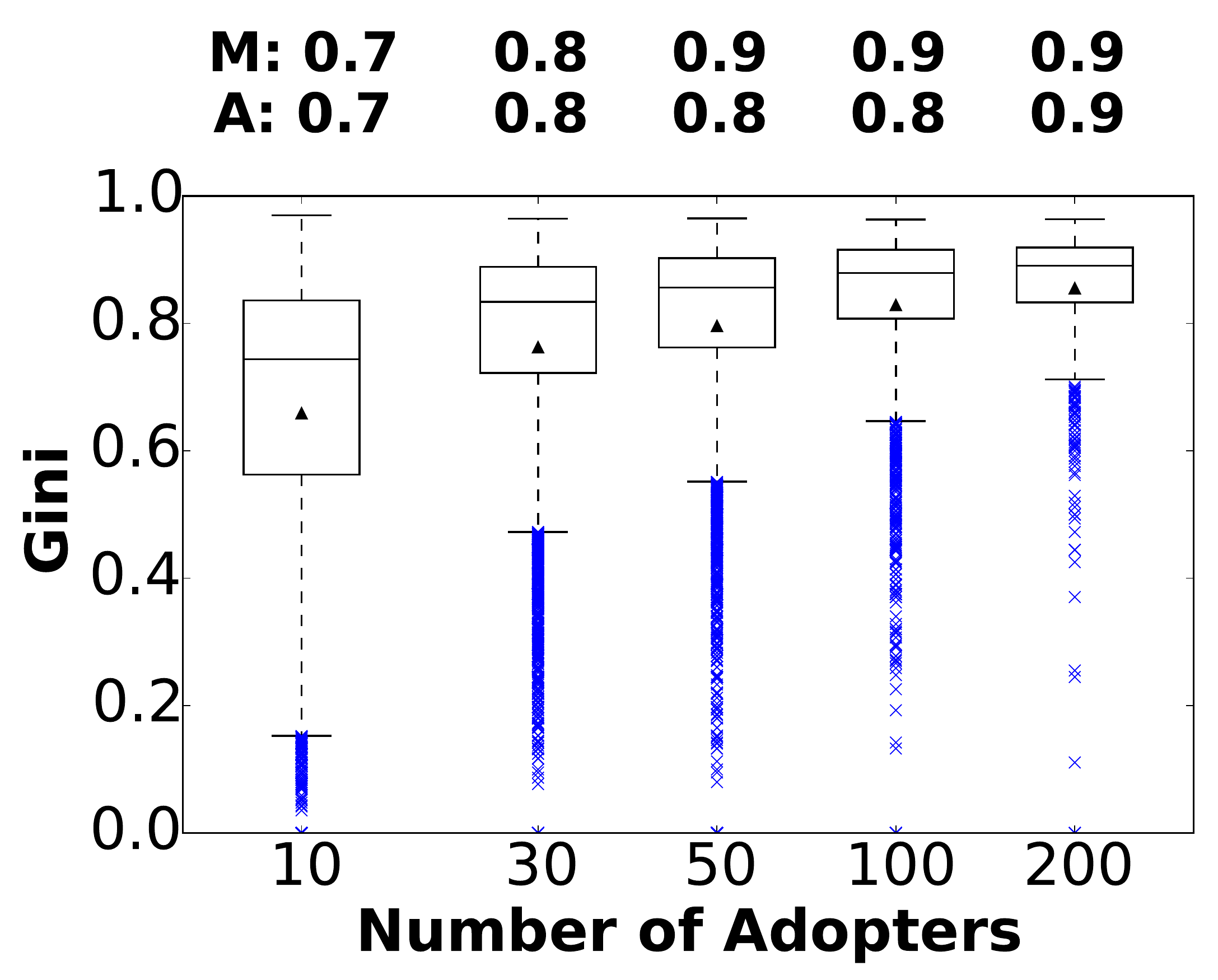}
		\caption{Gini impurity of recently exposed users ($I_G(\mathcal{F}_t)$) for non-viral cascades}
		\label{fig:S_nv_ent_f}
	\end{subfigure}
	\hfill
	\begin{subfigure}[b]{0.49\textwidth}
		\centering
		\includegraphics[width=39mm]{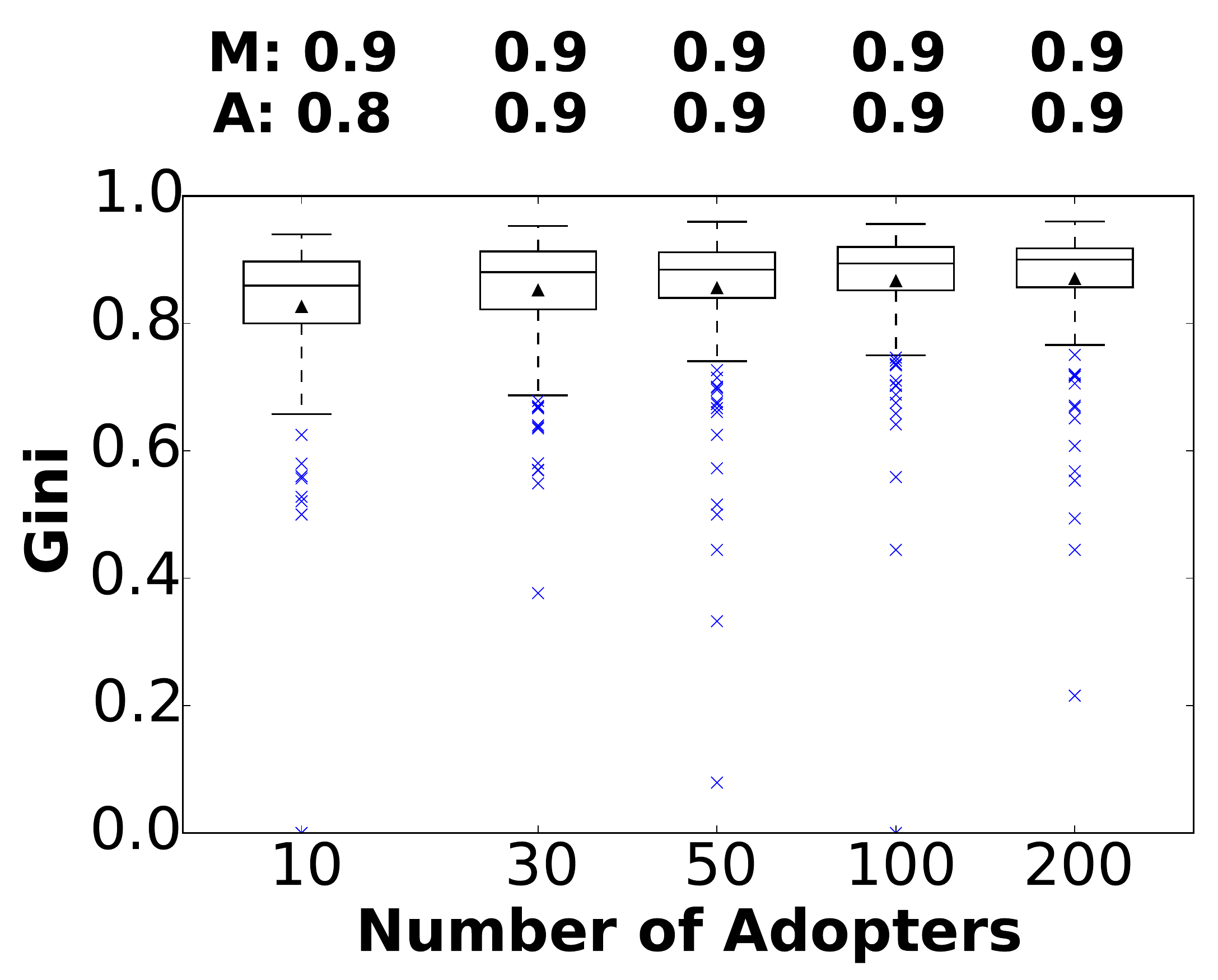}
		\caption{Gini impurity of recently exposed users ($I_G(\mathcal{F}_t)$) for viral cascades}
		\label{fig:S_v_ent_f}
	\end{subfigure}
	\begin{subfigure}[b]{0.49\textwidth}
		\centering
		\includegraphics[width=39mm]{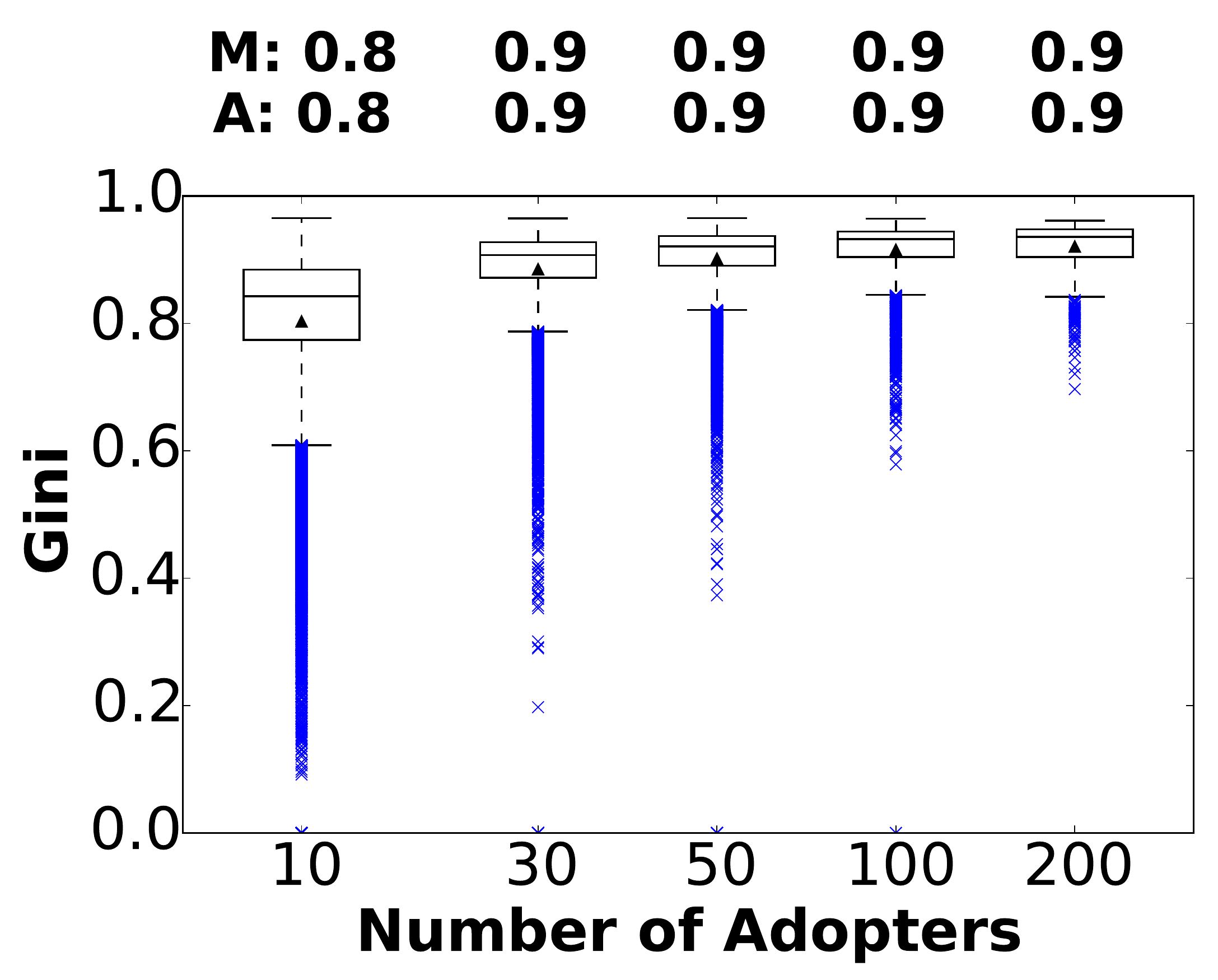}
		\caption{Gini impurity of past exposed users ($I_G(\mathcal{N}_t)$) for non-viral cascades}
		\label{fig:S_nv_ent_na}
	\end{subfigure}
	\hfill
	\begin{subfigure}[b]{0.49\textwidth}
		\centering
		\includegraphics[width=39mm]{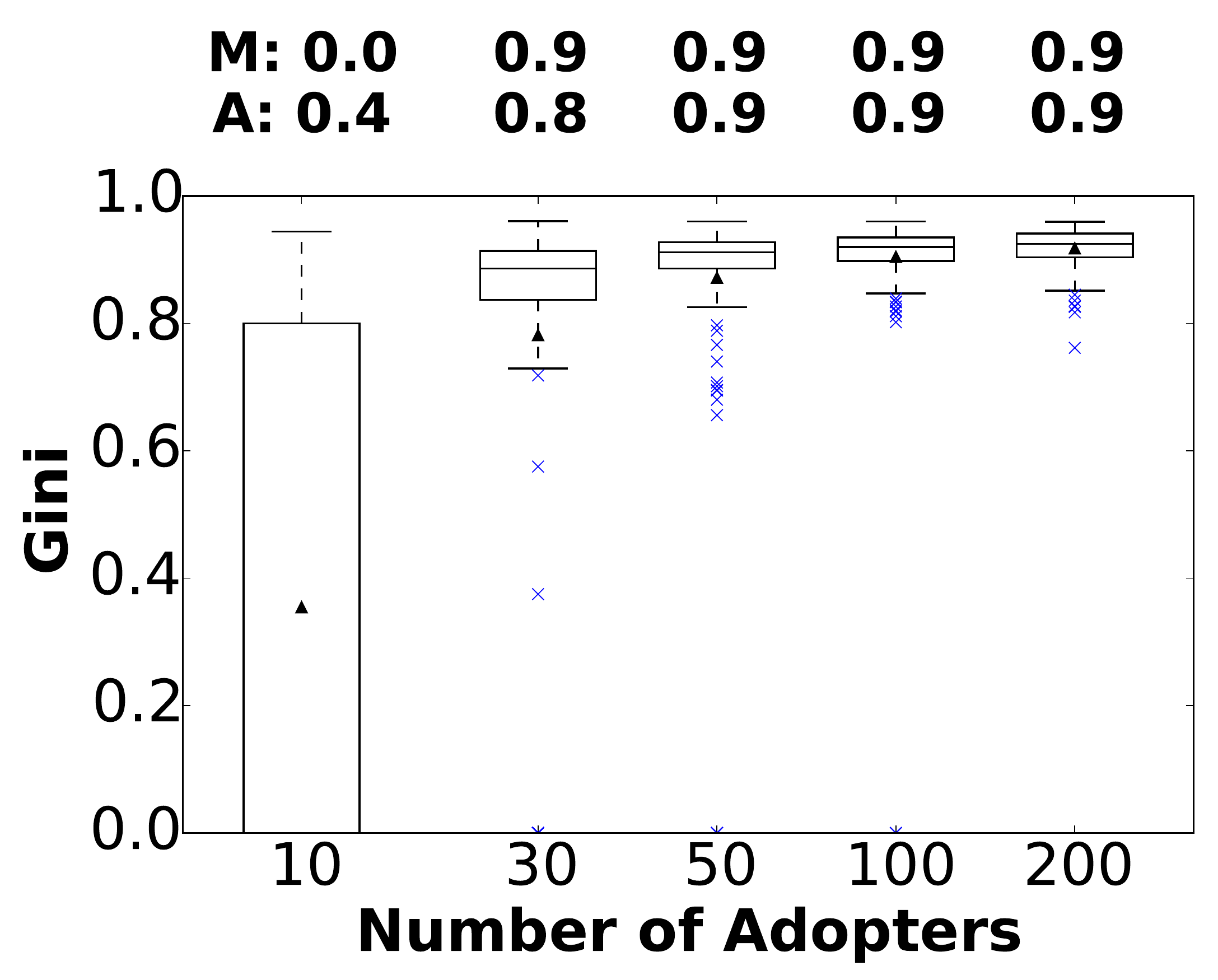}
		\caption{Gini impurity of past exposed users ($I_G(\mathcal{N}_t)$) for viral cascades}
		\label{fig:S_v_ent_na}
	\end{subfigure}
	\caption{Gini impurity for $ m = \left\{10,30,50,100,200\right\} $}
	\label{fig:S_gini}
\end{figure}

\paragraph{Gini impurity.}  Cascades in both classes tend to accumulate diversity in the process of collecting more adopters - and we have also noted that a related entropy measure (studied in \citep{weng2014predicting}) performed similarly.  We also observed that viral cascades can show larger gini impurity in recently exposed users measured by $ I_G(\mathcal{F}_t) $ in early stages ($ m = \left\{10,30,50\right\} $).
However, perhaps most striking, non-viral cascades gain more uniformly distributed nodes over communities in non-adopters, shown by $ I_G(\mathcal{N}_t) $ (Figure~\ref{fig:S_gini}).  We believe that this is due to non-viral cascades likely have an appeal limited to a relatively small number of communities - hence those \textbf{not} adopting the trend may represent a distribution of nodes over communities which is more different from a uniform distribution.

\begin{figure}[!tbp]
	\begin{subfigure}[b]{0.49\textwidth}
		\centering
		\includegraphics[width=39mm]{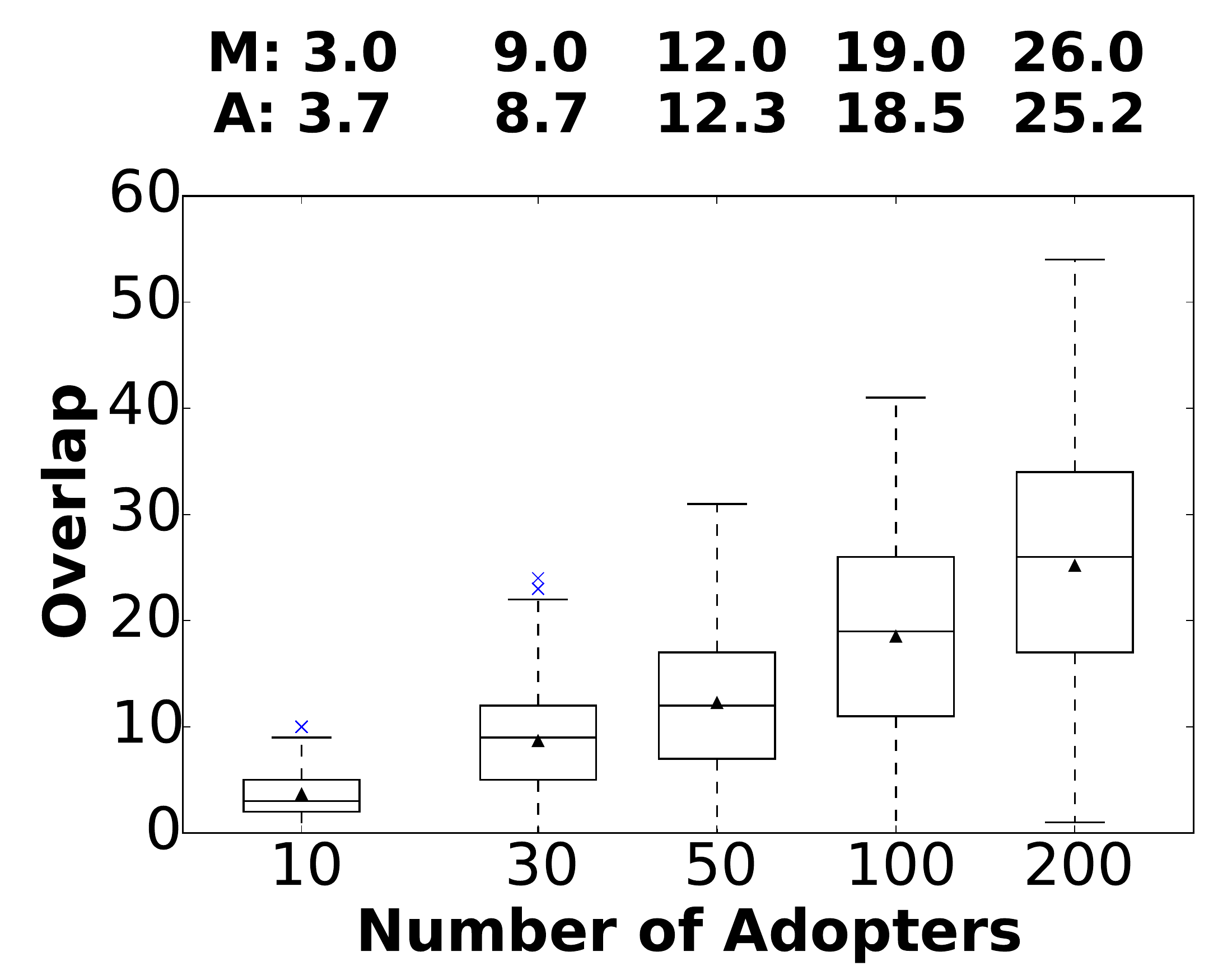}
		\caption{Overlap of adopters and recently exposed users ($O(U_t,\mathcal{F}_t)$) for non-viral cascades}
		\label{fig:S_nv_ol_af}
	\end{subfigure}
	\hfill
	\begin{subfigure}[b]{0.49\textwidth}
		\centering
		\includegraphics[width=39mm]{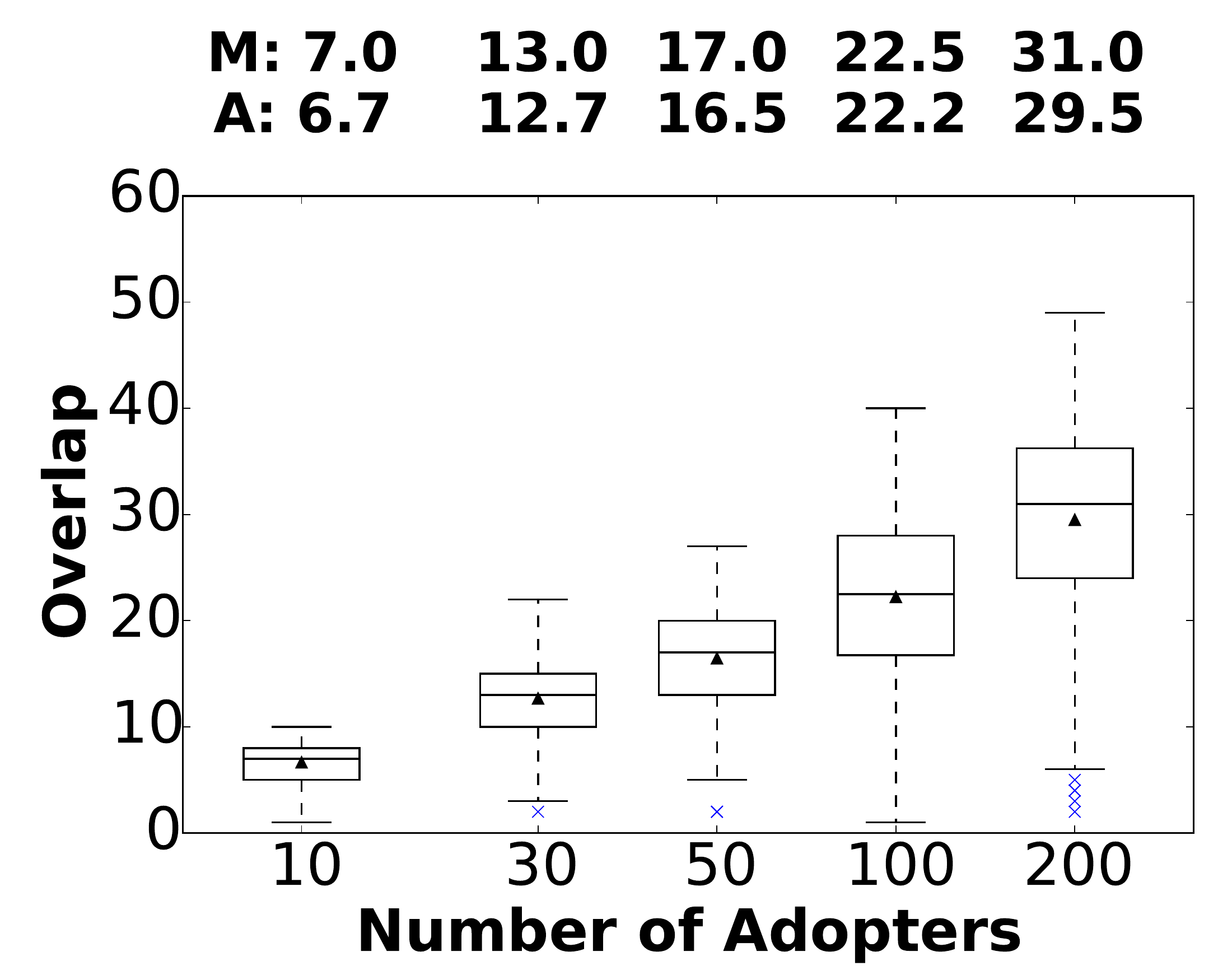}
		\caption{Overlap of adopters and recently exposed users ($O(U_t,\mathcal{F}_t)$) for viral cascades}
		\label{fig:S_v_ol_af}
	\end{subfigure}
	\begin{subfigure}[b]{0.49\textwidth}
		\centering
		\includegraphics[width=39mm]{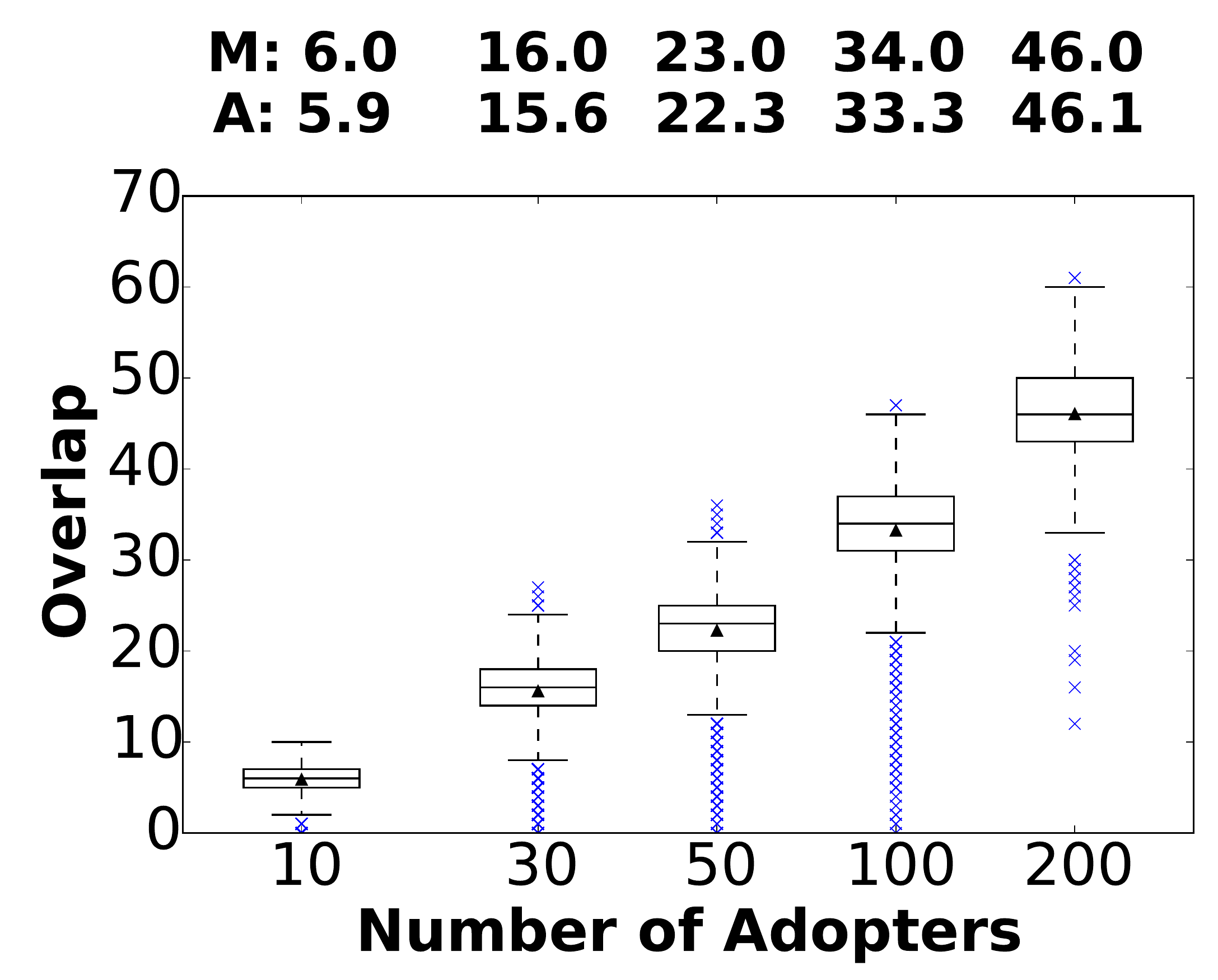}
		\caption{Overlap of adopters and past exposed users ($O(U_t,\mathcal{N}_t)$) for non-viral cascades}
		\label{fig:S_nv_ol_ana}
	\end{subfigure}
	\hfill
	\begin{subfigure}[b]{0.49\textwidth}
		\centering
		\includegraphics[width=39mm]{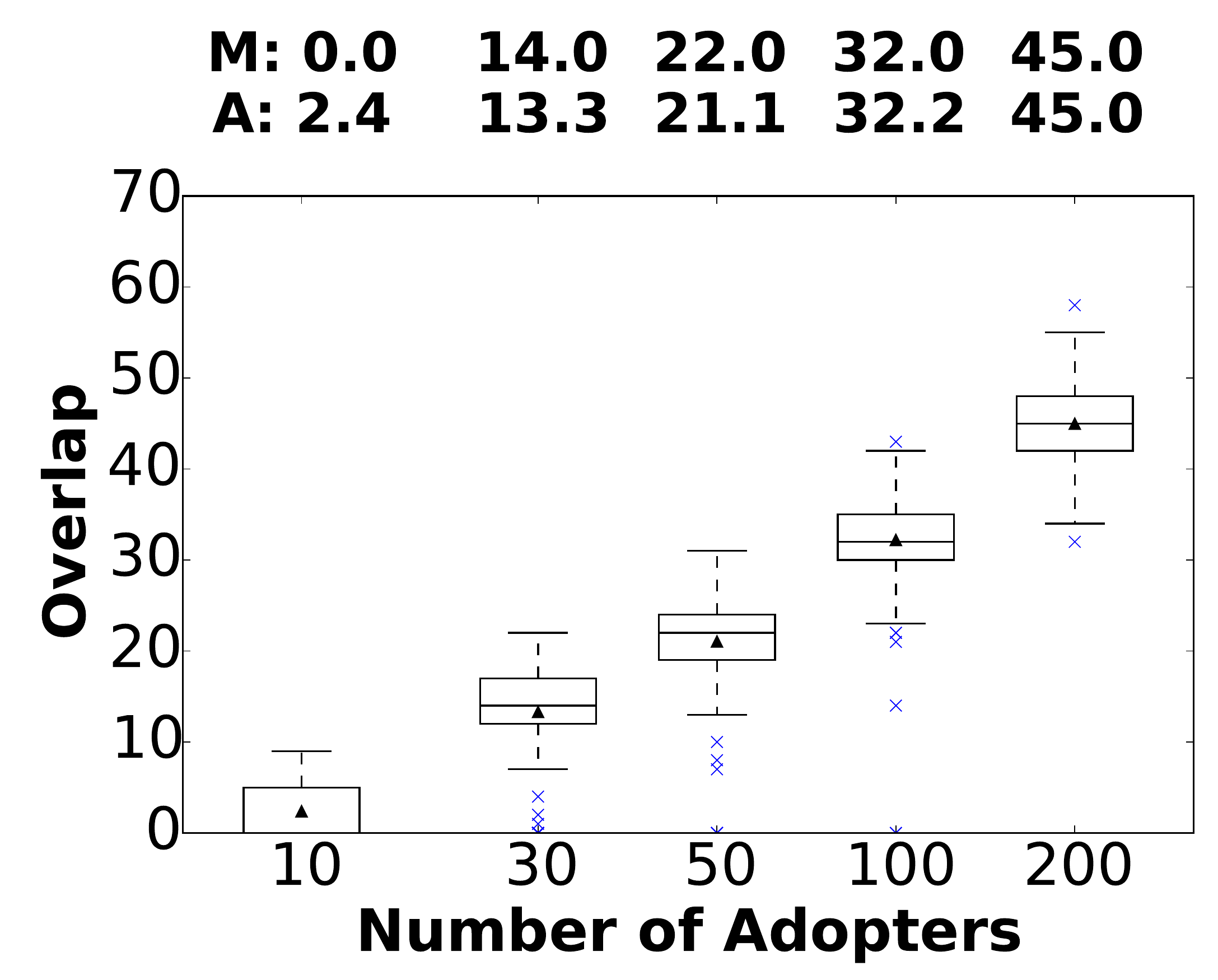}
		\caption{Overlap of adopters and past exposed users ($O(U_t,\mathcal{N}_t)$) for viral cascades}
		\label{fig:S_v_ol_ana}
	\end{subfigure}
	\begin{subfigure}[b]{0.49\textwidth}
		\centering
		\includegraphics[width=39mm]{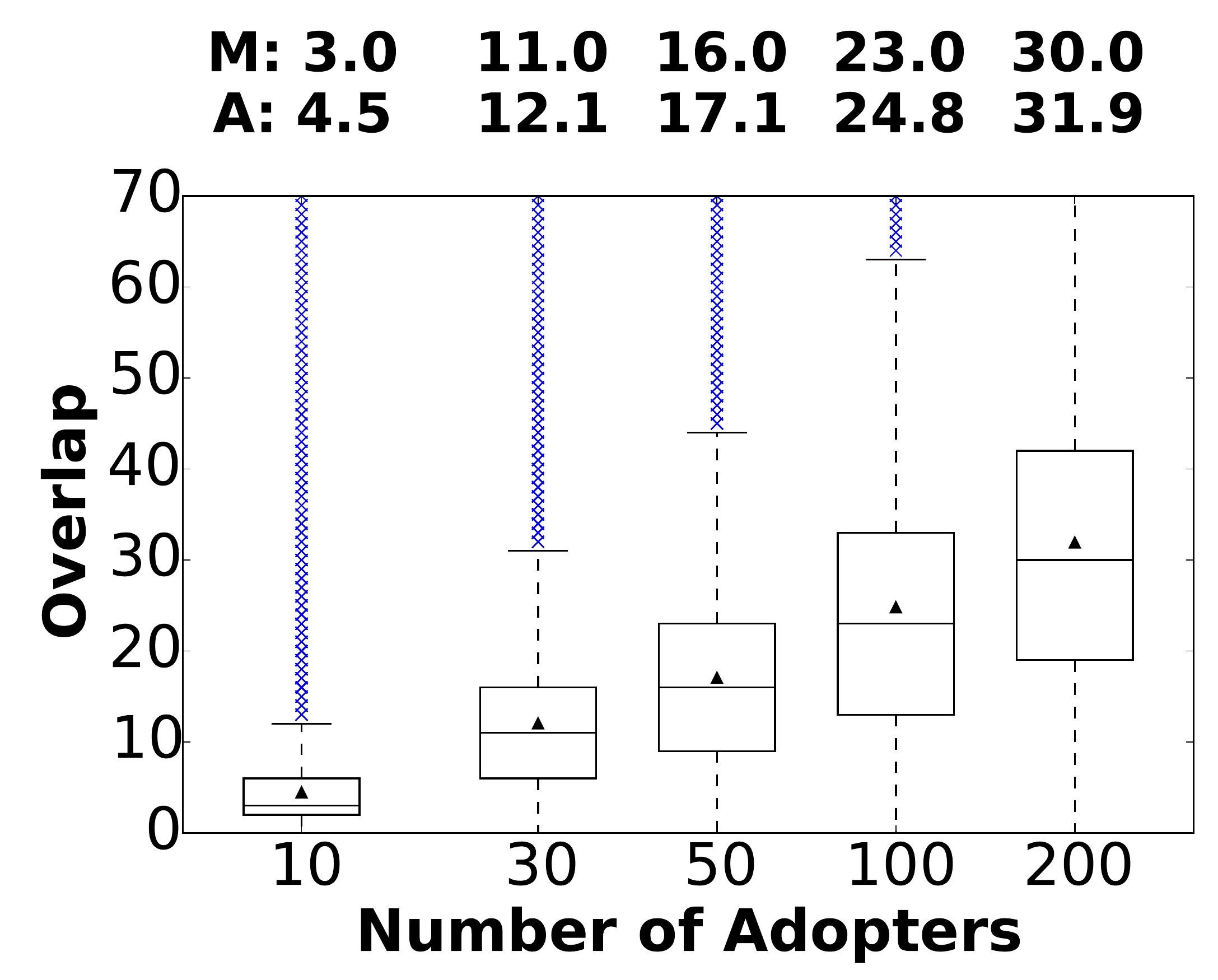}
		\caption{Overlap of recently and past exposed users ($O(\mathcal{F}_t,\mathcal{N}_t)$) for non-viral cascades}
		\label{fig:S_nv_ol_fna}
	\end{subfigure}
	\hfill
	\begin{subfigure}[b]{0.49\textwidth}
		\centering
		\includegraphics[width=39mm]{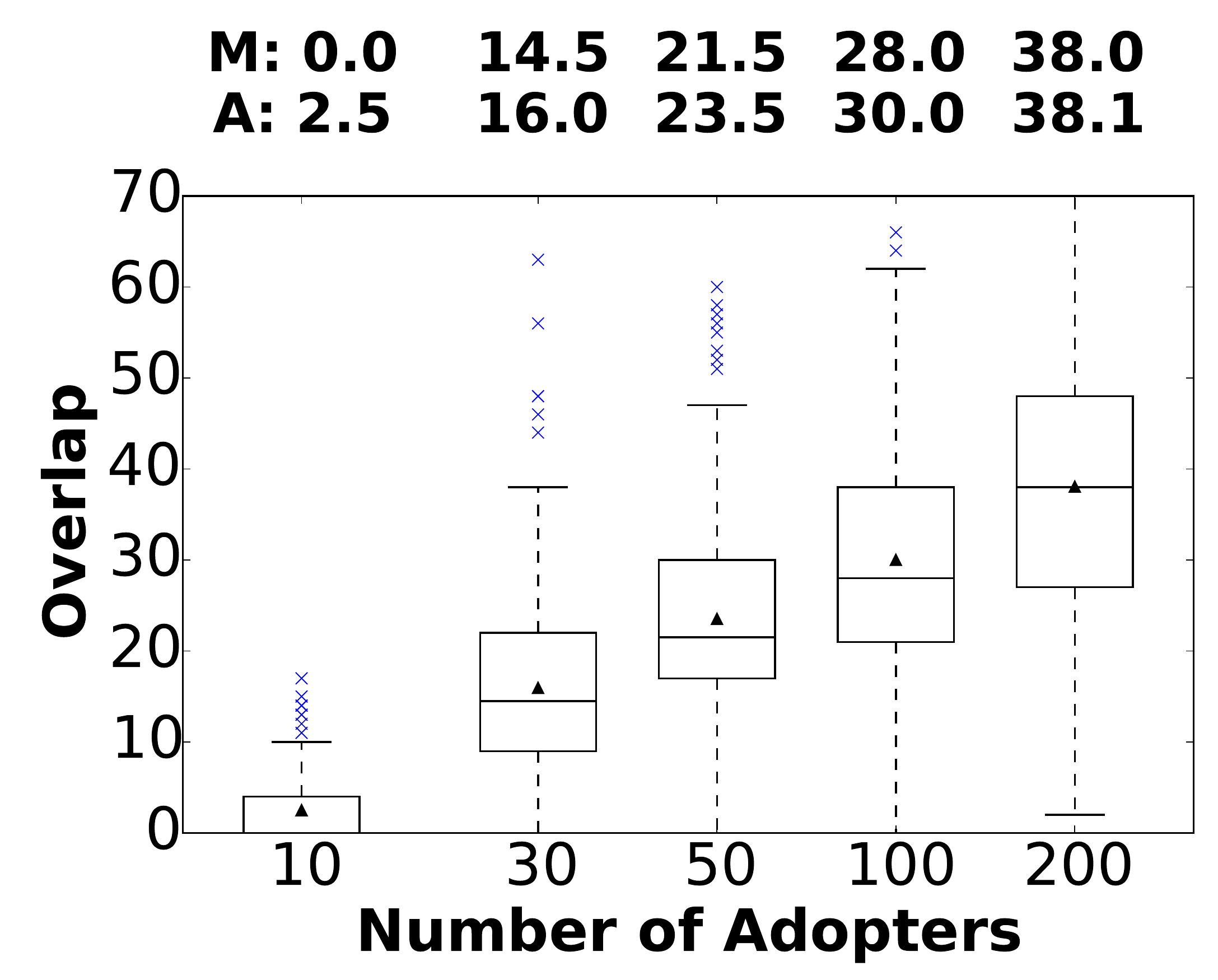}
		\caption{Overlap of recently and past exposed users ($O(\mathcal{F}_t,\mathcal{N}_t)$) for viral cascades}
		\label{fig:S_v_ol_fna}
	\end{subfigure}
	\caption{Overlap for $ m = \left\{10,30,50,100,200\right\} $}
	\label{fig:S_ol}
	
\end{figure}

\paragraph{Overlap.}  We found that overlap grows with the number of adopters in the three types of overlap considered.  For $O(U_t,\mathcal{F}_t)$, viral cascades start with a larger initial value and keep leading non-viral ones in the diffusion process of first 200 nodes (Figure~\ref{fig:S_ol}).  We consider that viral cascades also take advantage of the densely linked communities to help them become viral.
However, in the case of $O(U_t,\mathcal{N}_t)$ and $O(\mathcal{F}_t,\mathcal{N}_t)$, viral cascades begin with lower value but grow much faster than non-viral cascades.

\subsection{Time Progression}

\begin{figure}[!tbp]
	\begin{subfigure}[b]{0.49\textwidth}
		\centering
		\includegraphics[width=39mm]{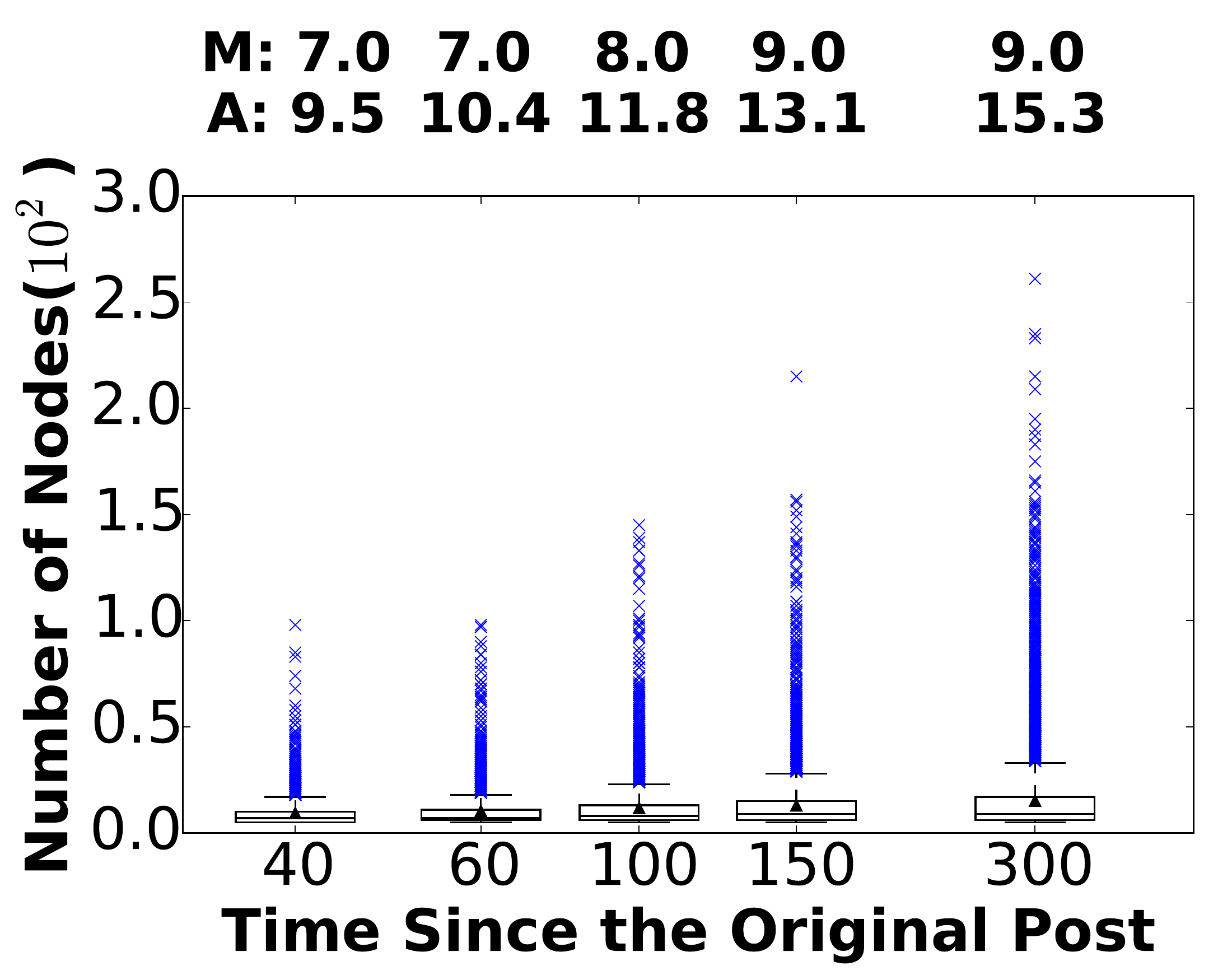}
		\caption{Number of adopters ($ |U_t| $) for non-viral cascades}
		\label{fig:cs_nonviral}
	\end{subfigure}
	\hfill
	\begin{subfigure}[b]{0.49\textwidth}
		\centering
		\includegraphics[width=39mm]{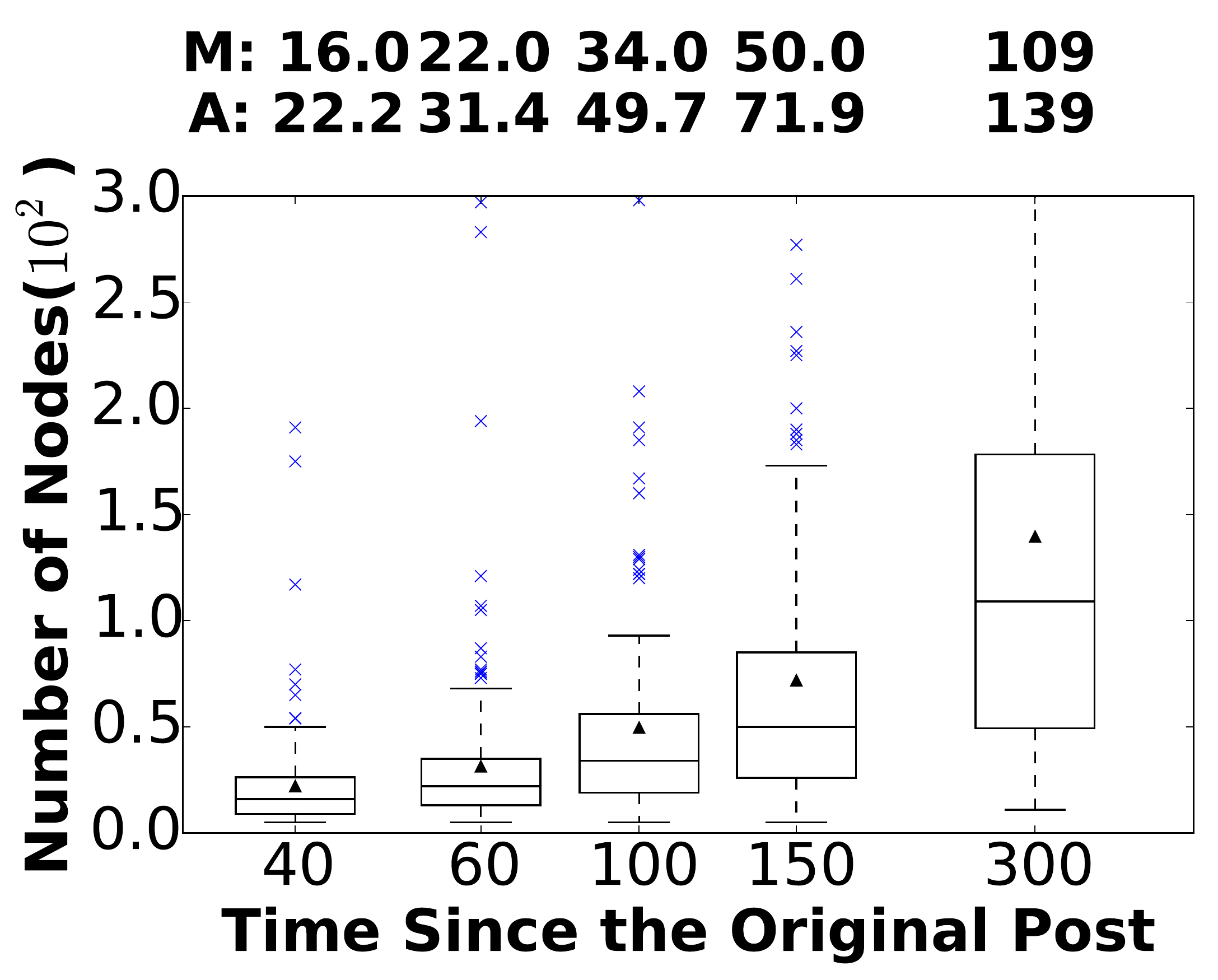}
		\caption{Number of adopters ($ |U_t| $) for viral cascades}
		\label{fig:cs_viral}
	\end{subfigure}
	\caption{Number of adopters for $ t \in \left\{40,60,100,150,300\right\} $ (min)}
	\label{fig:T_size}
\end{figure}

\paragraph{Number of adopters.}  As a baseline measurement, we study the number of adopters at regular time intervals and, as expected, found a clear difference between the two classes.  Figure~\ref{fig:T_size} shows how $ |U_t| $ changes over 40, 60, 100, 150 and 300 minutes.  Although there is an obvious difference in early stages (40-60 minutes) between the two distributions, we will see in the next section that this alone does not provide adequate performance for our prediction task (see Section~\ref{sec:clf}).

\begin{figure}[!tbp]
	
	\begin{subfigure}[b]{0.49\textwidth}
		\centering
		\includegraphics[width=39mm]{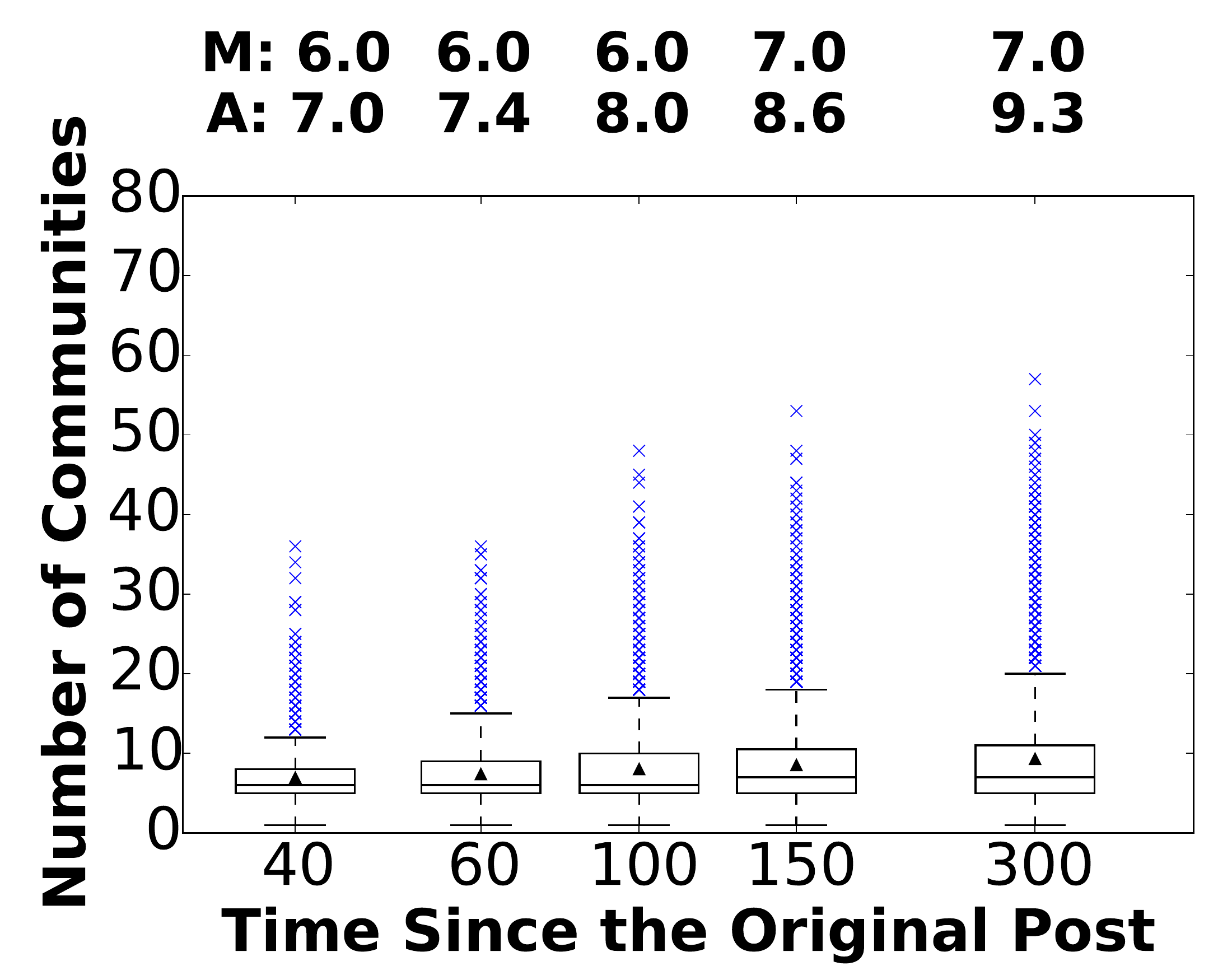}
		\caption{Number of communities amongst adopters ($|\calc(U_t)|$) for non-viral cascades}
		\label{fig:T_nv_comm_a}
	\end{subfigure}
	\hfill
	\begin{subfigure}[b]{0.49\textwidth}
		\centering
		\includegraphics[width=39mm]{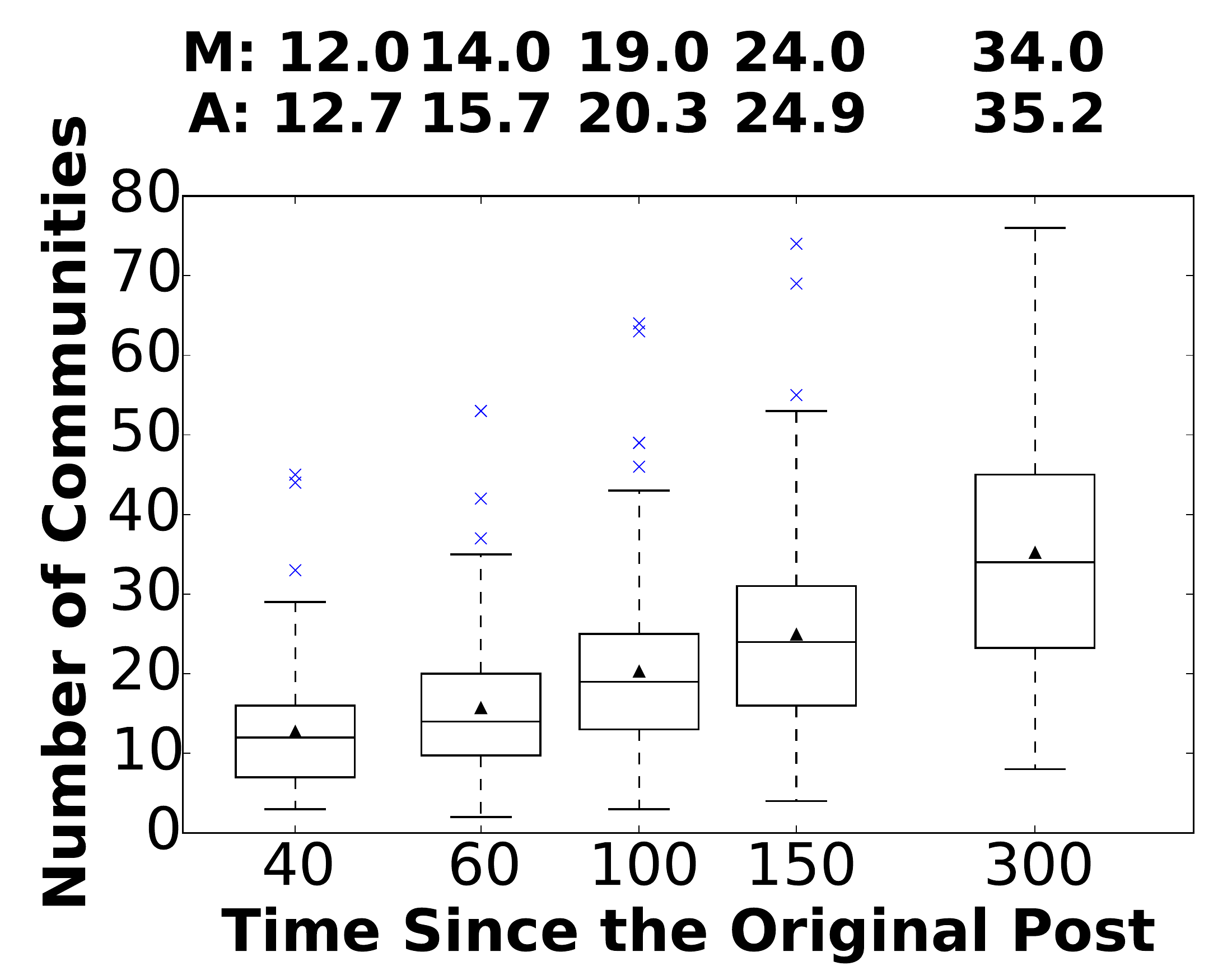}
		\caption{Number of communities amongst adopters ($|\calc(U_t)|$) for viral cascades}
		\label{fig:T_v_comm_a}
	\end{subfigure}
	\hfill
	\begin{subfigure}[b]{0.49\textwidth}
		\centering
		\includegraphics[width=39mm]{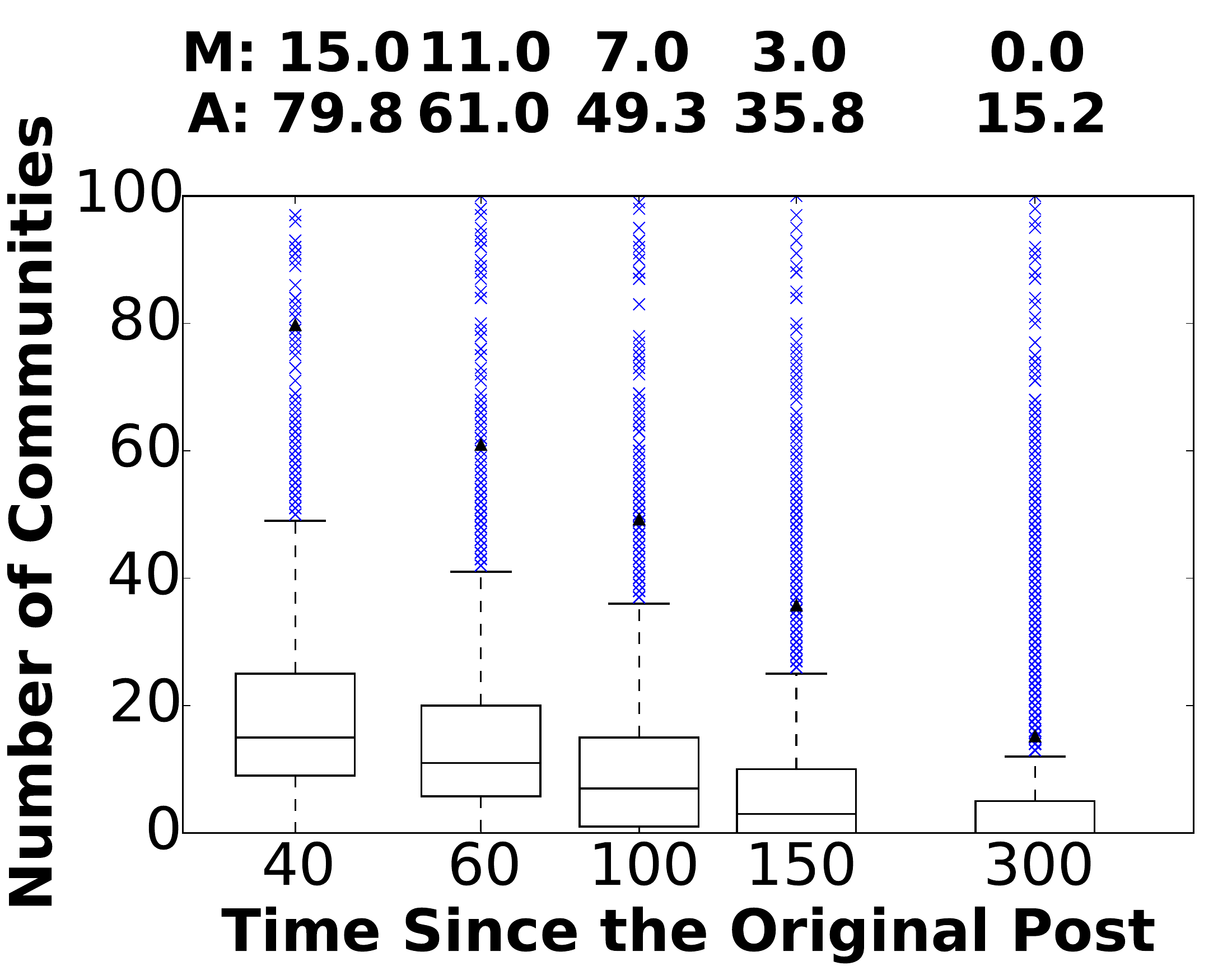}
		\caption{Number of communities amongst recently exposed users ($|\calc(\mathcal{F}_t)|$), non-viral cascades}
		\label{fig:T_nv_comm_f}
	\end{subfigure}
	\hfill
	\begin{subfigure}[b]{0.49\textwidth}
		\centering
		\includegraphics[width=39mm]{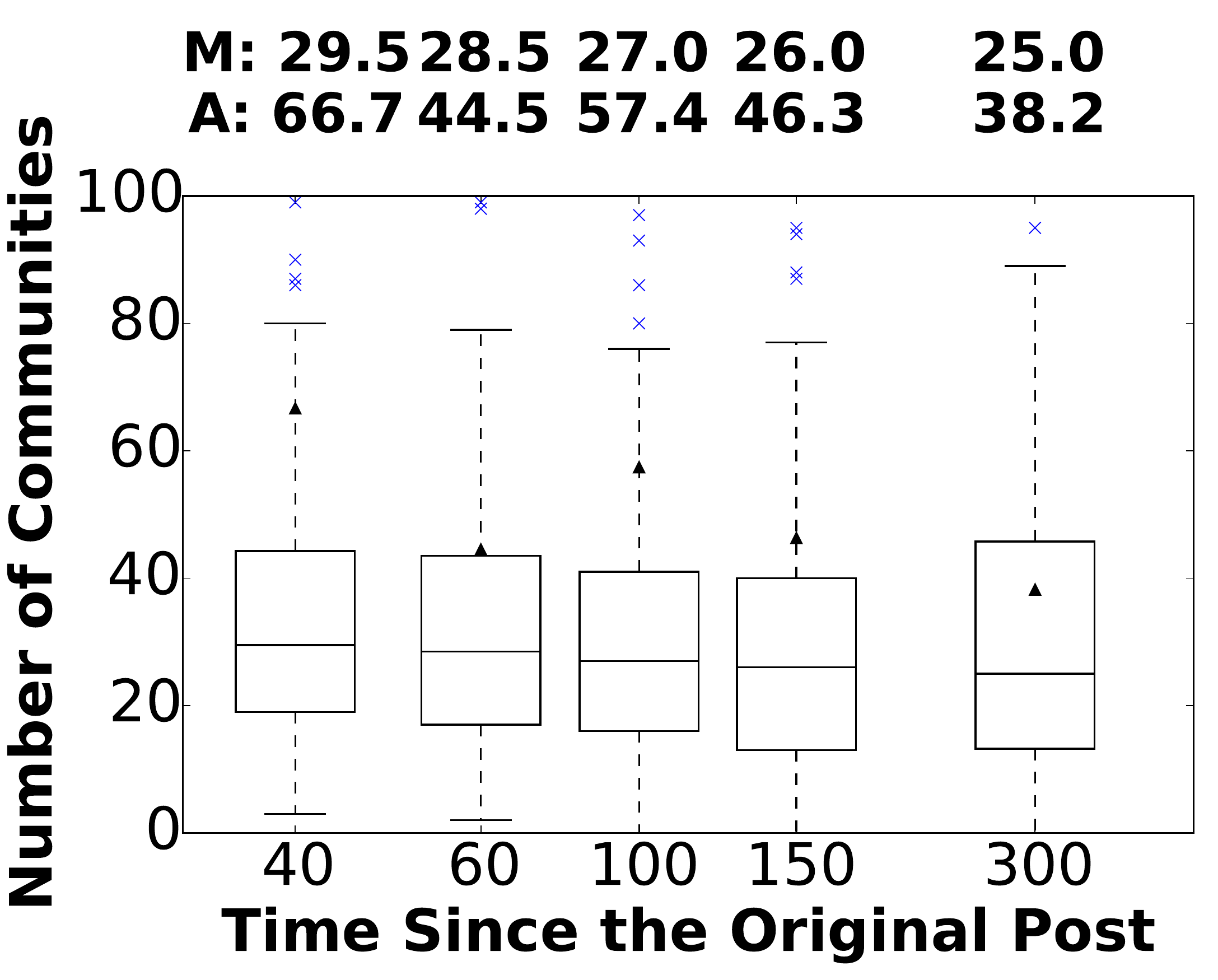}
		\caption{Number of communities amongst recently exposed users ($|\calc(\mathcal{F}_t)|$) for viral cascades}
		\label{fig:T_v_comm_f}
	\end{subfigure}
	\hfill
	\begin{subfigure}[b]{0.49\textwidth}
		\centering
		\includegraphics[width=39mm]{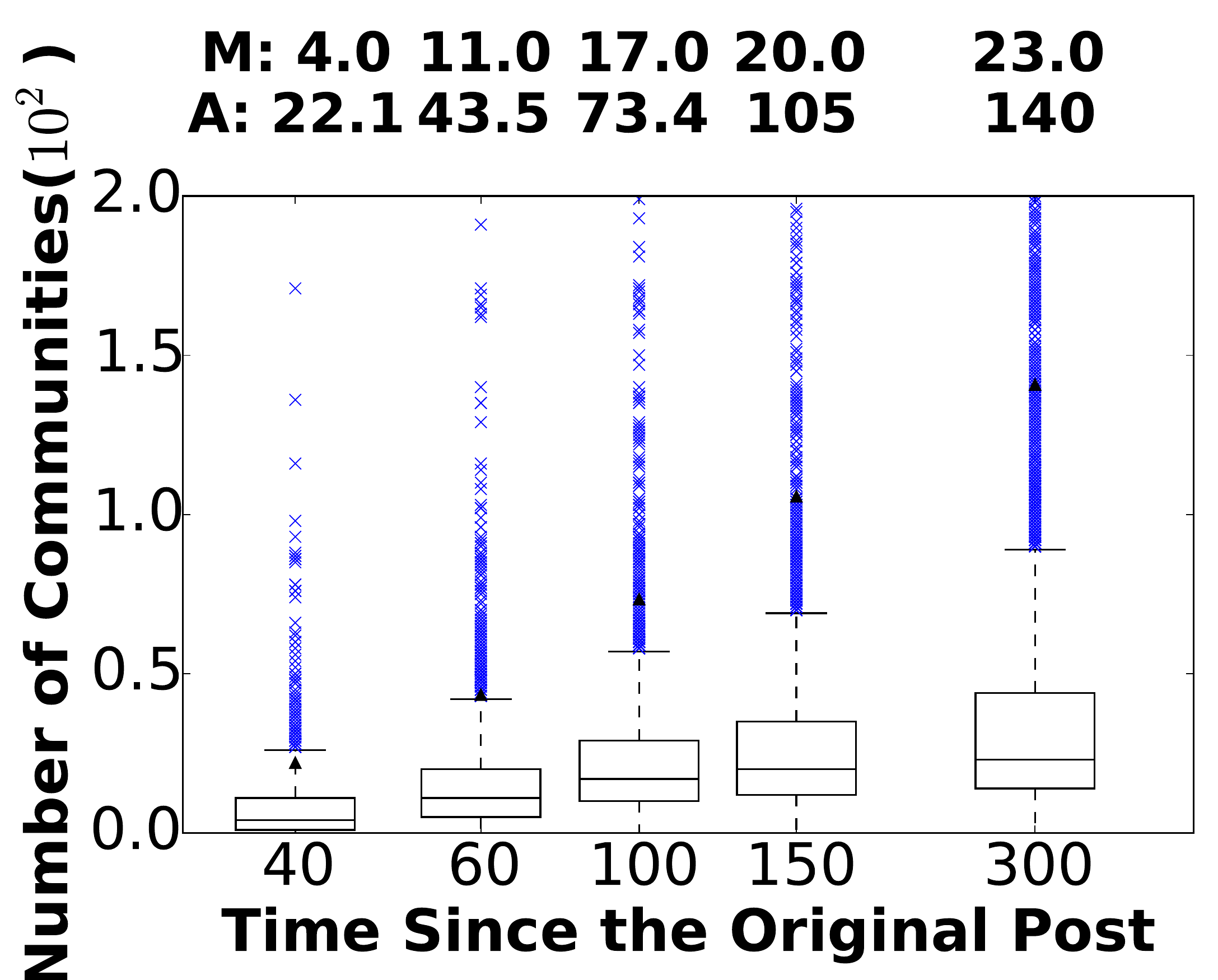}
		\caption{Number of communities amongst past exposed users ($|\calc(\mathcal{N}_t)|$) for non-viral cascades}
		\label{fig:T_nv_comm_na}
	\end{subfigure}
	\hfill
	\begin{subfigure}[b]{0.49\textwidth}
		\centering
		\includegraphics[width=39mm]{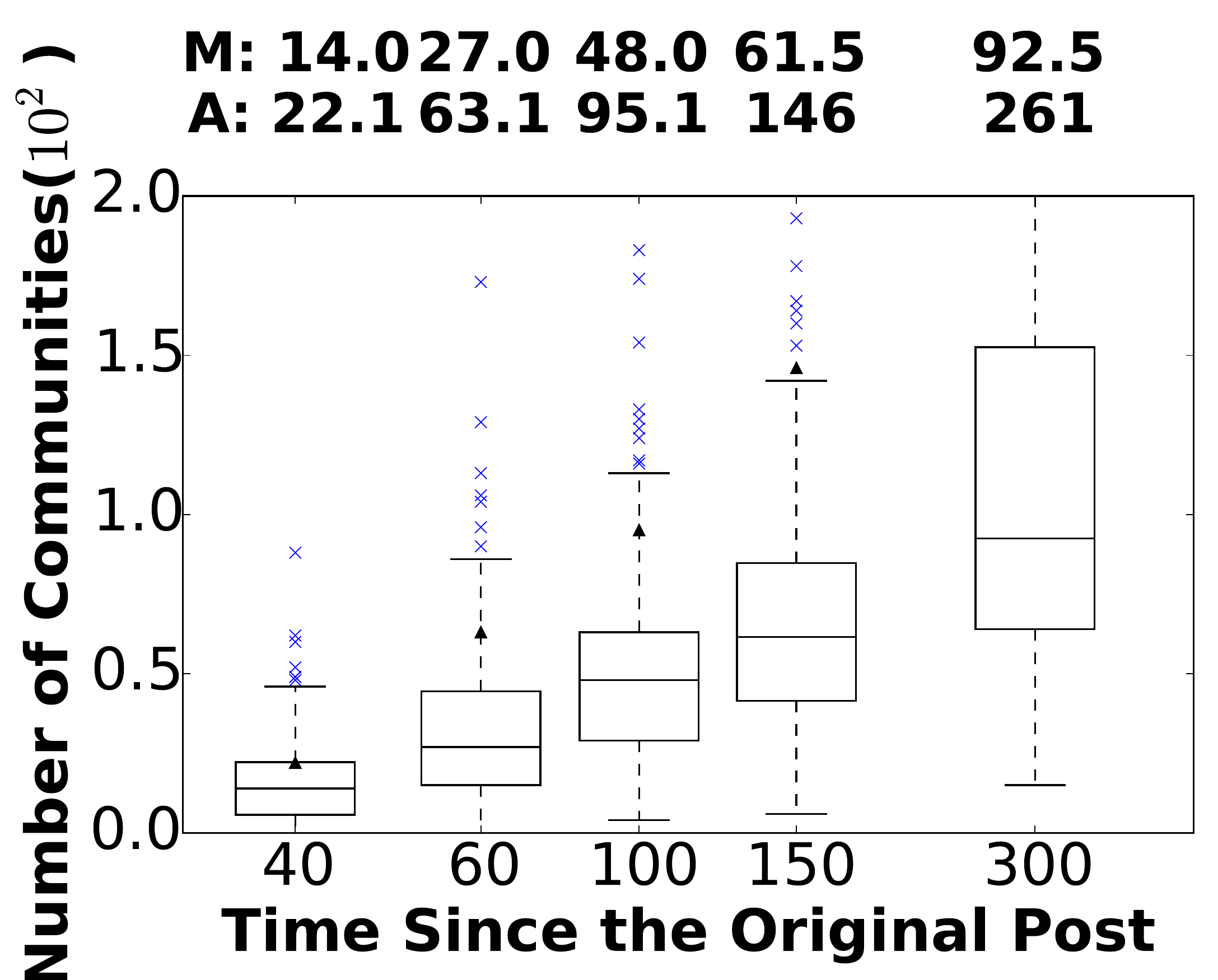}
		\caption{Number of communities amongst past exposed users ($|\calc(\mathcal{N}_t)|$) for viral cascades}
		\label{fig:T_v_comm_na}
	\end{subfigure}
	\caption{Number of communities for $ t = \left\{40,60,100,150,300\right\} $ (min)}
	\label{fig:T_comm}
\end{figure}

\paragraph{Number of communities.}  Figure~\ref{fig:T_comm} shows how $|\calc(S)|$ for $S \in \left\{U_t,\mathcal{F}_t,\mathcal{N}_t\right\}$ changes over time.  The value of $|\calc(S)|$ increases over time for $ U_t $ and $ \mathcal{N}_t $ but decreases for $\mathcal{F}_t$.  Here, the differences are somewhat more pronounced than for the size-progression measurements (compare with Figure~\ref{fig:S_comm}).   Viral cascades are more likely to have more communities in any one of $U_t$, $\mathcal{F}_t$, $\mathcal{N}_t$ than non-viral ones.  
For adopters and non-adopters, $|\calc(U_t)|$ and $ |\calc(\mathcal{N}_t)| $ value of viral cascades increases faster than that of non-viral ones over time. 
While for recently exposed users, $ |\calc(\mathcal{F}_t)| $ of non-viral cascades decreases more than viral ones in the same amount of time.

\begin{figure}[!tbp]
	\begin{subfigure}[b]{0.49\textwidth}
		\centering
		\includegraphics[width=39mm]{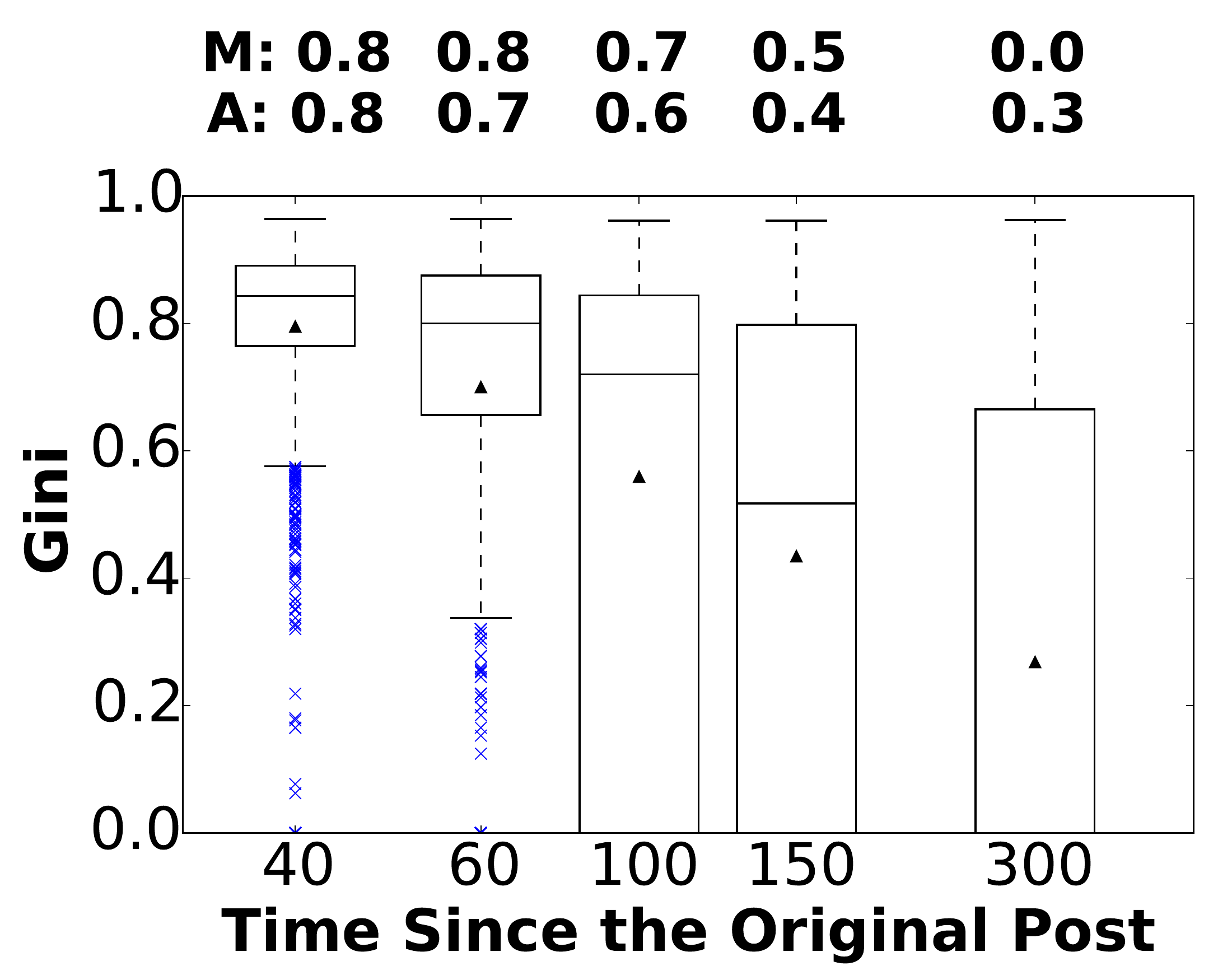}
		\caption{Gini impurity of recently exposed users ($I_G(\mathcal{F}_t)$) for non-viral cascades}
		\label{fig:T_nv_ent_f}
	\end{subfigure}
	\hfill
	\begin{subfigure}[b]{0.49\textwidth}
		\centering
		\includegraphics[width=39mm]{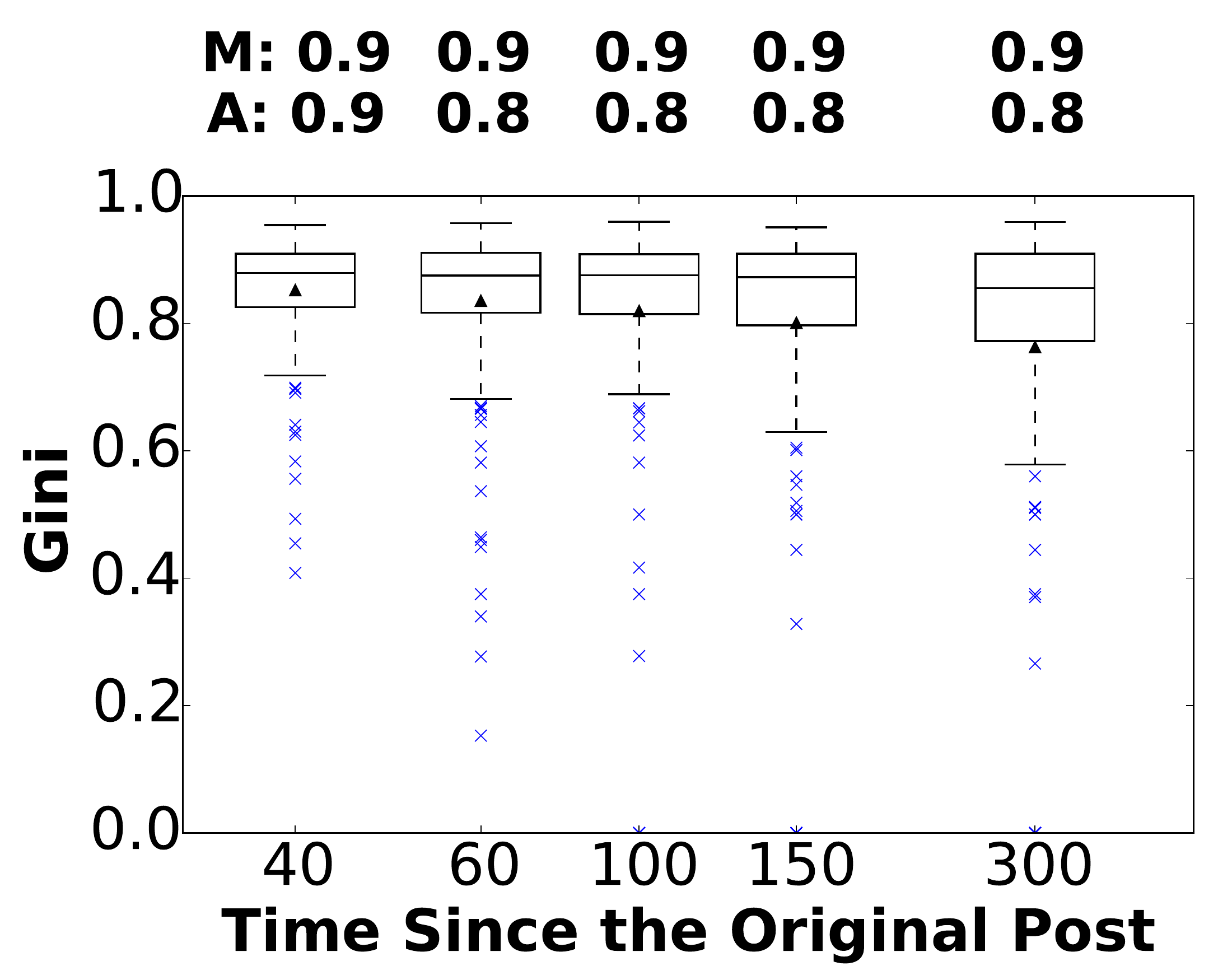}
		\caption{Gini impurity of recently exposed users ($I_G(\mathcal{F}_t)$) for viral cascades}
		\label{fig:T_v_ent_f}
	\end{subfigure}
	\hfill
	\begin{subfigure}[b]{0.49\textwidth}
		\centering
		\includegraphics[width=39mm]{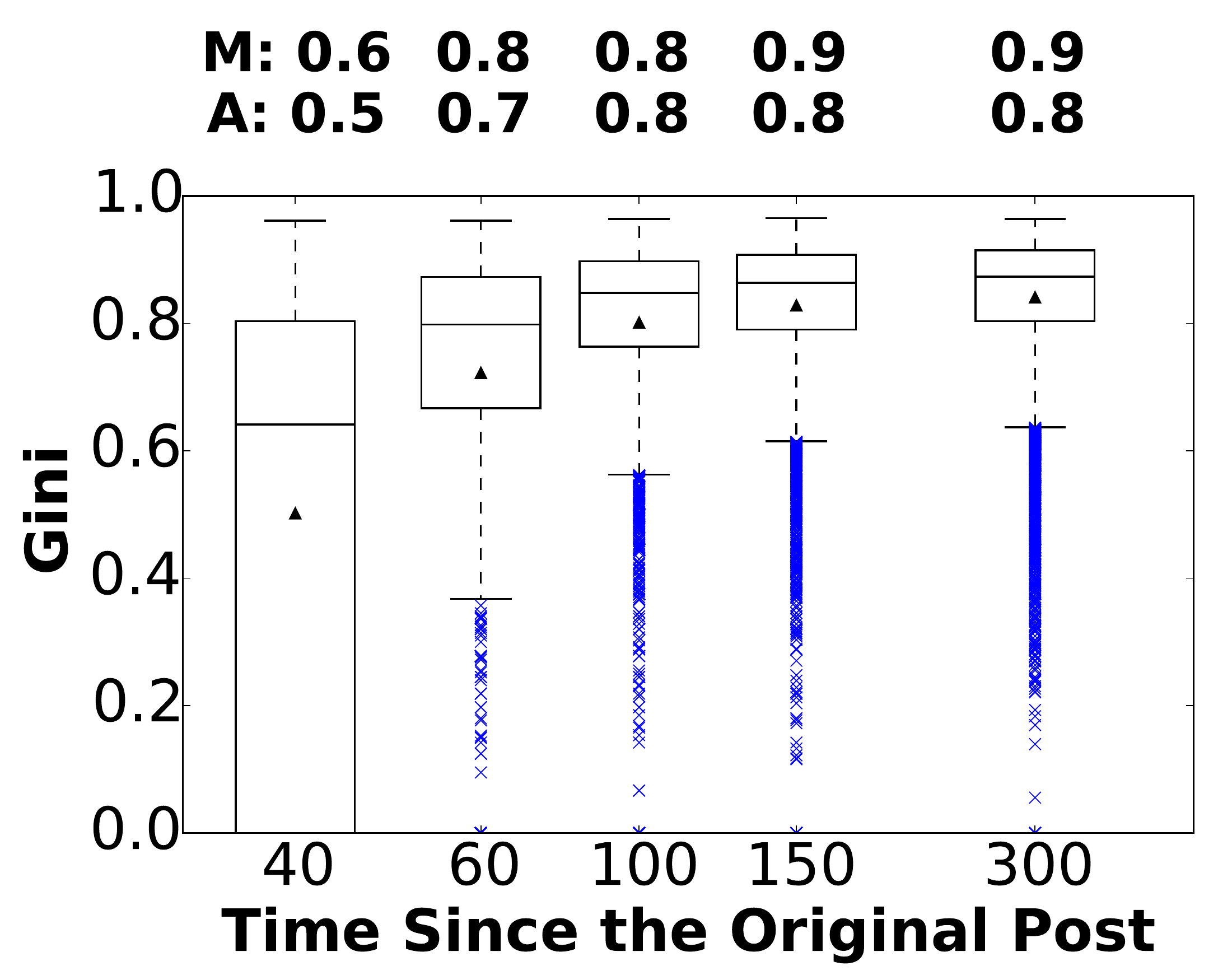}
		\caption{Gini impurity of past exposed users ($I_G(\mathcal{N}_t)$) for non-viral cascades}
		\label{fig:T_nv_ent_na}
	\end{subfigure}
	\hfill
	\begin{subfigure}[b]{0.49\textwidth}
		\centering
		\includegraphics[width=39mm]{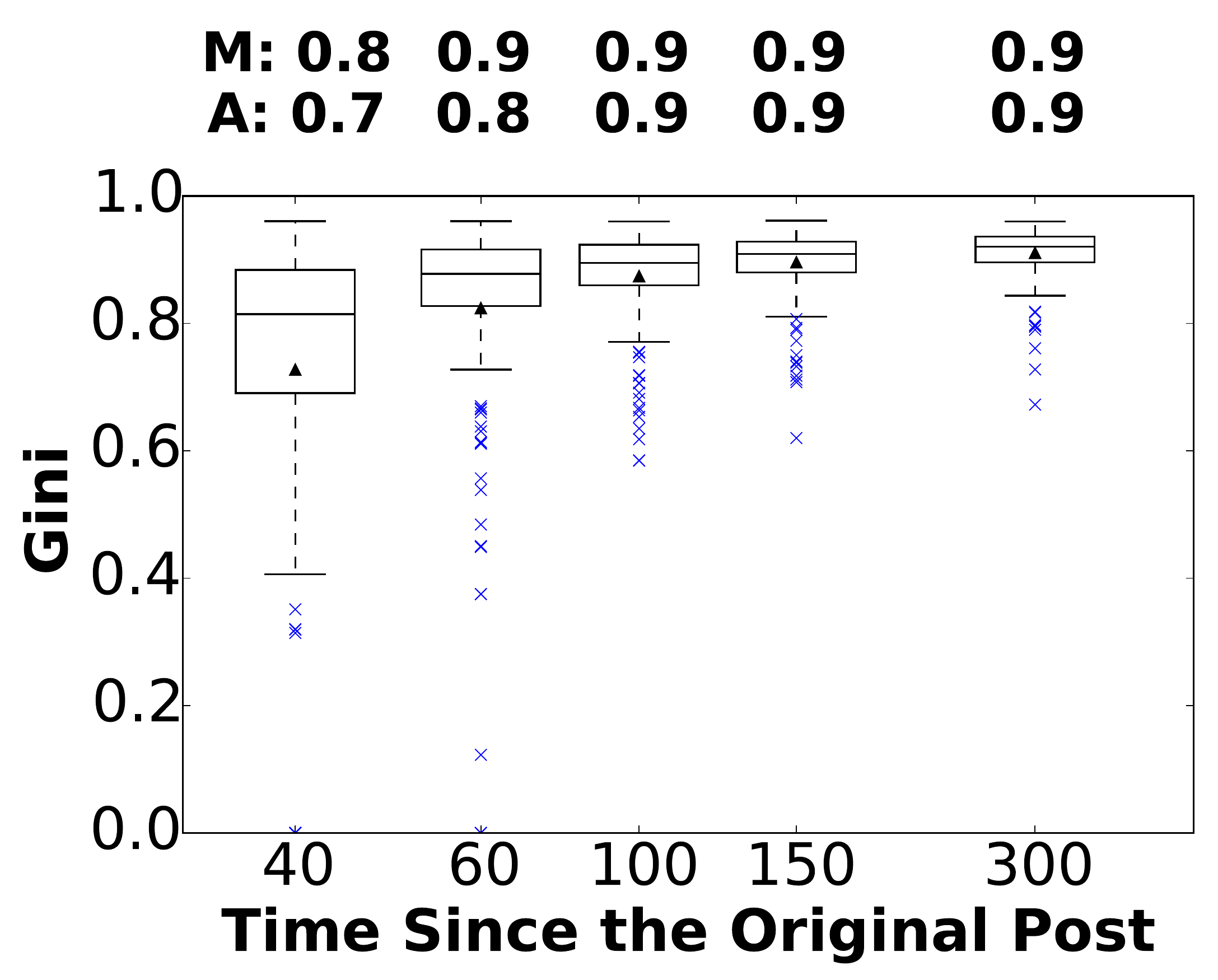}
		\caption{Gini impurity of past exposed users ($I_G(\mathcal{N}_t)$) for viral cascades}
		\label{fig:T_v_ent_na}
	\end{subfigure}
	\caption{Gini impurity for $ t = \left\{40,60,100,150,300\right\} $ (min)}
	\label{fig:T_gini}
\end{figure}

\paragraph{Gini impurity.}  
It takes less than $ \lambda =30 $ minutes for a considerable portion of viral cascades to reach size $ m = 30 $. 
This explains the difference between size-based and time-based gini impurity values in initial-stage cascades (compare Figure~\ref{fig:S_gini} and Figure~\ref{fig:T_gini}). 
In terms of size-based gini impurity of the non-adopters ($I_G(\mathcal{N}_t) $), the values of viral cascades are smaller than those of non-viral cascades when $ m $ is small. However, when $t$ is small, larger gini impurity ($ I_G(\mathcal{N}_t)$) amongst non-adopters are shown in viral cascades. 
Furthermore, as $ m $ increases, although no significant difference is shown by the median and average of $ I_G(\mathcal{N}_t) $, Figure~\ref{fig:S_gini} shows non-viral cascades are more likely to have a value smaller than the lower whisker to become outliers.

\begin{figure}[!tbp]
	\begin{subfigure}[b]{0.49\textwidth}
		\centering
		\includegraphics[width=39mm]{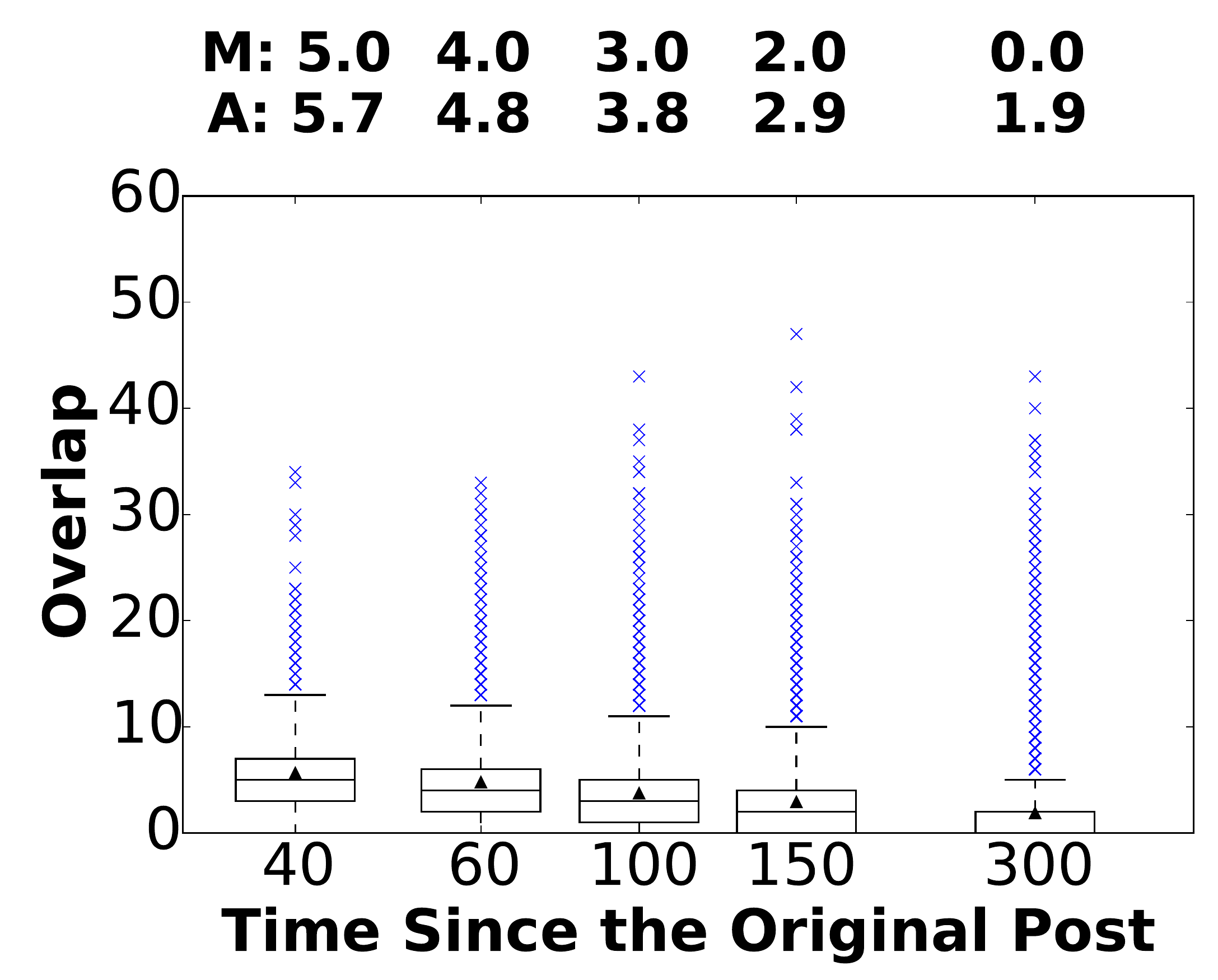}
		\caption{Overlap of adopters and recently exposed users ($O(U_t,\mathcal{F}_t)$) for non-viral cascades}
		\label{fig:T_nv_ol_af}
	\end{subfigure}
	\hfill
	\begin{subfigure}[b]{0.49\textwidth}
		\centering
		\includegraphics[width=39mm]{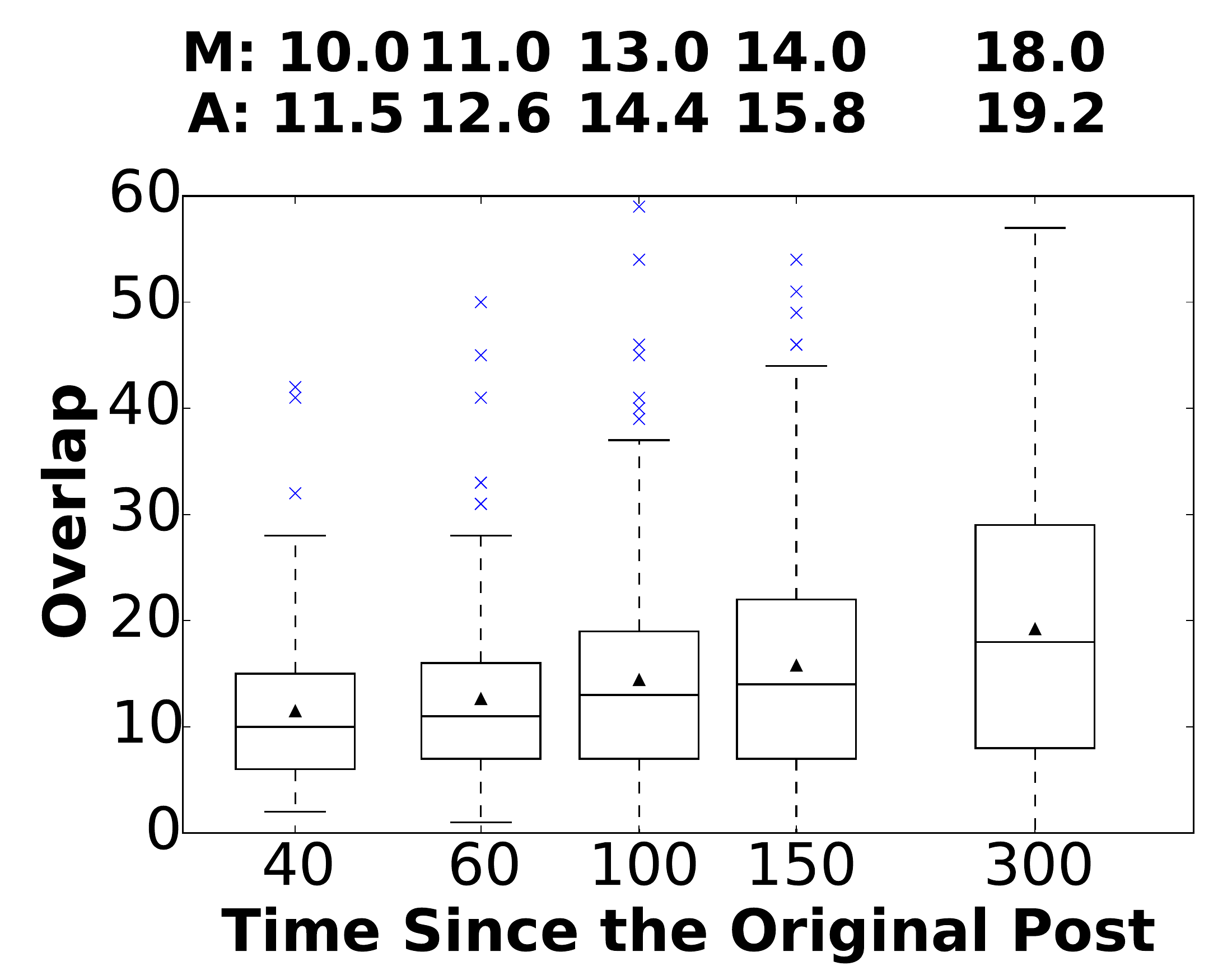}
		\caption{Overlap of adopters and recently exposed users ($O(U_t,\mathcal{F}_t)$) for viral cascades}
		\label{fig:T_v_ol_af}
	\end{subfigure}
	\caption{Overlap for $ t = \left\{40,60,100,150,300\right\} $ (min)}
	\label{fig:T_ol}
\end{figure}

\paragraph{Overlap.}  By definition, overlap is the number of shared communities between two sets of nodes. We found that overlap $ O(U_t,\mathcal{F}_t) $, $O(U_t,\mathcal{N}_t)$ and $O(U_t,\mathcal{N}_t)$ manifest obvious difference between viral and non-viral cascades by values and trend over time.  For instance, in Figure~\ref{fig:T_ol}, we see growth of $ O(U_t,\mathcal{F}_t) $ for the viral cascades compared to the non-viral class.  In fact, over time, this value decreases for non-viral cascades as the set of recently exposed users fades away for non-viral cascades with time.

\section{Classification Experiments}
\label{sec:clf}
Here we examine our experiments for predicting whether a cascade becomes viral - when the number of adopters exceeds a size threshold ($TH=500$) given that either the cascade has 50 adopters ($ m = 50 $) or has progressed for an hour ($ t = 60 $).  We shall refer to these as \textit{size-based} and \textit{time-based} prediction problems.  Based on the distribution of final size of cascades in this dataset (see Figure~\ref{fig:hist_cas_size}), as shown in Table~\ref{tab:samples}, this binary classification task deals with two heavily imbalanced classes.  
Hence, we report performance measurements (precision, recall and \textit{F1} score) for only the minority (viral) class.  
Throughout the course of our experiments, we found that varying  threshold (slightly modifying the definition of ``viral'') for \textit{only} the training set allows for a trade-off between precision and recall.  
We study the trend of performance metrics in two cases:
\begin{itemize}
	\item The threshold for test set is maintained as $ TH_{ts} = 500 $ while the training threshold is varied $ TH_{tr} \in \left\{300,400,500,600,700\right\} $.
	\item The two thresholds are kept as the same $ TH $ while we modify this value $ TH \in \left\{300,400,500,600,700\right\} $.
\end{itemize}

Table~\ref{tab:features} shows the groups of features used in our prediction tasks.  The features introduced in this paper are groups $A_m$ (size-based) and $A_t$ (time-based).
We compare our features (Group $ A_m $, $A_t$) with the community features extracted in \citep{weng2014predicting} (Group $ B_m $,$B_t$) and nodal features of the seed adopter (Group $C_m$ and $C_t$).
Here nodal measures of the seed adopter refer to k-shell number, out-degree, in-degree, pagerank and eigenvector, which are computed based on the social network $G$. 
In previous work \citep{pei2014searching}, k-shell number of the seed adopter node is shown to be correlated to the average size of cascades. However, cascades from the seed adopter nodes with the same k-shell number can end up with quite different size \citep{shakBk}.
As baseline methods, average time to adoption (group $D_m$) is applied to the size-based experiment while cascade size at time $t$ (group $D_t$) is evaluated for time-based prediction. 
We extracted each group of community-based features ($ A_m $, $A_t$, $ B_m $, $B_t$) with all the three community detection algorithms mentioned in Section~\ref{sec:prelim}: Louvain, Infomap and SLM.
Therefore, for both size-based and time-based prediction, there are 8 groups of features.
Among them, $B_m$ and $B_t$ were the best performing feature set in the paper \citep{weng2014predicting} for a comparable task.\footnote{This was their highest-performing set of features for predicting cascades that grew from $50$ to $367$ and $100$ to $417$ reposts.  We also included the baseline feature in this set as we found it improved the effectiveness of this approach.}

Additionally, we study the average size of correctly classified viral cascades and the other viral samples using features in groups $A_m$ and $A_t$.
We also investigate the significance and performance of individual and certain combinations of features introduced in this paper.

\begin{table}[!t]%
	\setlength\extrarowheight{5pt}
	\renewcommand{\arraystretch}{1}
	\caption{\textmd{Features for prediction tasks (size-based and time-based): $A_m$, $B_m$, $A_t$ and $B_t$ are computed based on three community detection algorithms (Louvain, Infomap and SLM)}}
	\label{tab:features}
	\centering
	\begin{minipage}[t]{0.005\textwidth}
		\begin{tabular}[t]{| c| p{4cm}|}
			\hline
			\textbf{Name} & \textbf{Feature(s) over size } \\ \hline 
			
			$A_m$ & 
			$|\calc(\mathcal{F}_t)|$,$|\calc(\mathcal{N}_t)|$,
			$I_G(U_t)$,$I_G(\mathcal{F}_t)$,$I_G(\mathcal{N}_t)$,
			$O(U_t,\mathcal{F}_t)$,$O(U_t,\mathcal{N}_t)$,$O(\mathcal{F}_t,\mathcal{N}_t)$,
			$|\mathcal{F}_t|$,$|\mathcal{N}_t|$, $\frac{1}{m}\sum_{i=1}^m t(i)$
			
			where $t$ stands for $t(m)$
			with $m \in \left\{30,50\right\}$
			\\ \hline
			$B_m$ & Community Features Mentioned in \citep{weng2014predicting} and $\frac{1}{m}\sum_{i=1}^m t(i)$, $m=50$ \\ \hline 
			$C_m$ & Nodal Features and $\frac{1}{m}\sum_{i=1}^m t(i)$, $m=50$ \\ \hline
			$D_m$ & $\frac{1}{m}\sum_{i=1}^m t(i)$, $m=50$ \\ \hline
		\end{tabular}
	\end{minipage}%
	\hfill
	\begin{minipage}[t]{0.49\textwidth}
		\begin{tabular}[t]{| c| p{4cm}|}
			\hline
			\textbf{Name} & \textbf{Feature(s) over  time} \\ \hline
			$A_t$ & 
			
			$|\calc(\mathcal{F}_t)|$,$|\calc(\mathcal{N}_t)|$,
			$I_G(U_t)$,$I_G(\mathcal{F}_t)$,$I_G(\mathcal{N}_t)$,
			$O(U_t,\mathcal{F}_t)$,$O(U_t,\mathcal{N}_t)$,$O(\mathcal{F}_t,\mathcal{N}_t)$,
			$|U_t|$,$|\mathcal{F}_t|$,$|\mathcal{N}_t|$
			
			for time $ t \in \left\{40,60\right\} $(min)
			
			\\ \hline
			$B_t$ & Community Features Mentioned in \citep{weng2014predicting} and $|U_t|$, $t = 60$(min)\\ \hline
			$C_t$ & Nodal Features and $|U_t|$, $t = 60$(min) \\ \hline
			$D_t$ & $|U_t|$, $t=60$(min) \\ \hline

		\end{tabular}

	\end{minipage}
	
\end{table}

\subsection{Cascade Prediction Results}
We split cascades into training set and testing set using ten-fold cross-validation. All classification experiments are repeated for 10 times to ensure the results do not take any advantage of randomness in picking training and testing sets. 
First we carried out the prediction tasks with fixed thresholds for both training and testing $ TH_{tr} = 500$, $TH_{ts} = 500 $.
Then we modify the training threshold $ TH_{tr} \in  \left\{300,400,500,600,700\right\}$ to show how this achieves a tradeoff between precision and recall. The difference in average final size between correctly classified viral cascades and incorrectly classified ones is also monitored over $ TH_{tr} \in \left\{300,400,500,600,700\right\}$ to show the potential to predict exact number of adopters by features in $ A_m $ and $ A_t $.
Furthermore, we modify threshold of both training and testing sets $ TH \in \left\{300,400,500,600,700\right\} $ to show the robustness of our features on related classification problems.
We used the oversampling method SMOTE \citep{chawla2002smote} with random forest classifier to generate synthetic samples for the viral class. 
Other, lesser-performing classifiers were also examined (including SVM, MLP, and other ensemble methods) and are not reported here.  All results shown in this section is a sample mean produced by repeated experiments (10 times) under each combination of variables. Error bars represent one standard deviation.

\begin{figure}[!t]
	\centering
	\includegraphics[width=84mm]{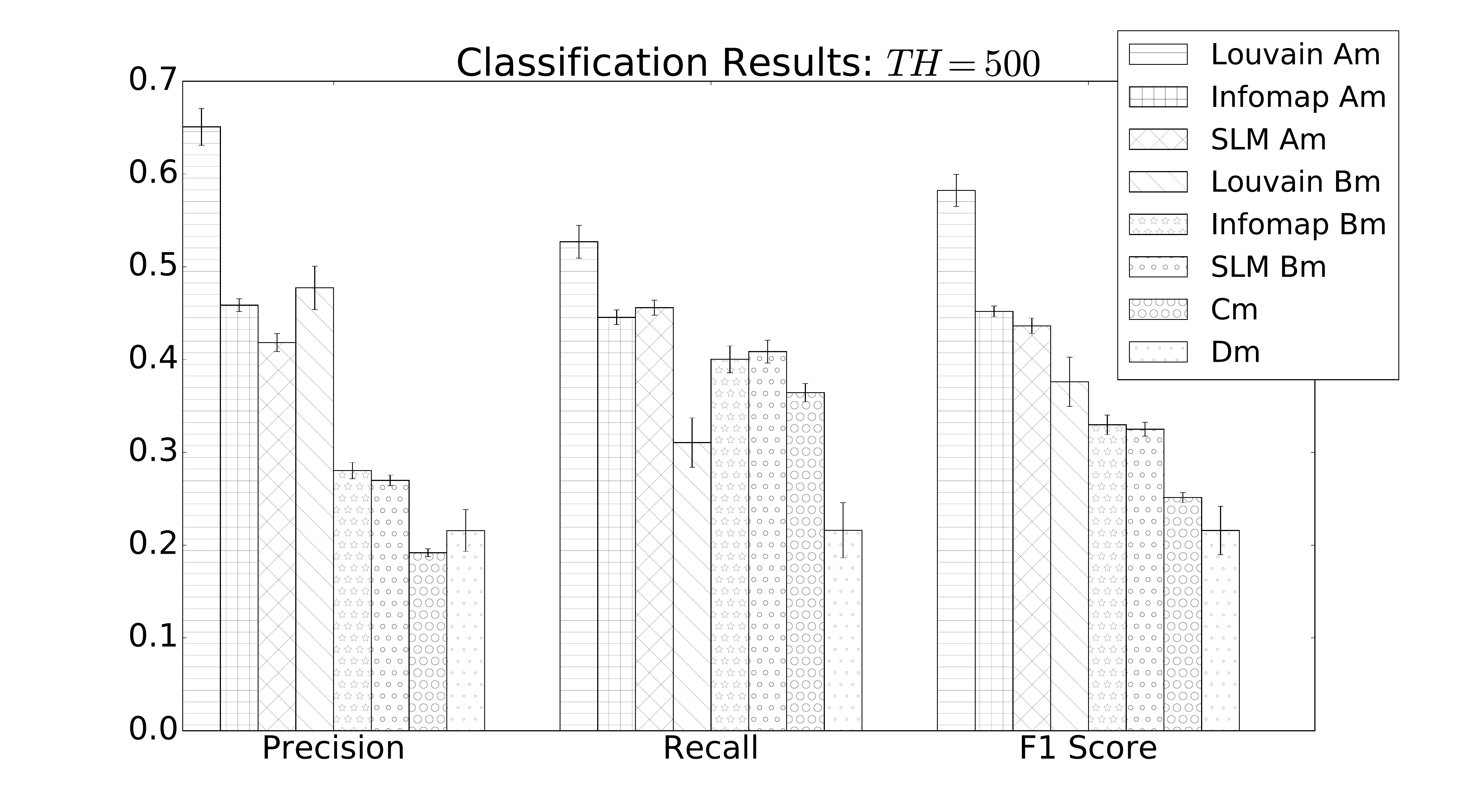}
	\caption{Classification results based on groups of features ($A_m$,$B_m$,$C_m$,$D_m$) extracted with three community detection algorithms (Louvain, Infomap and SLM) when $ m = 50 $ for fixed $ TH_{tr} = 500$, $TH_{ts} = 500 $. Error bars represent one standard deviation.}
	\label{fig:pred_Size}
\end{figure}

\begin{figure}[!tbp]
	\begin{subfigure}[b]{0.49\textwidth}
		\centering
		\includegraphics[width=39mm]{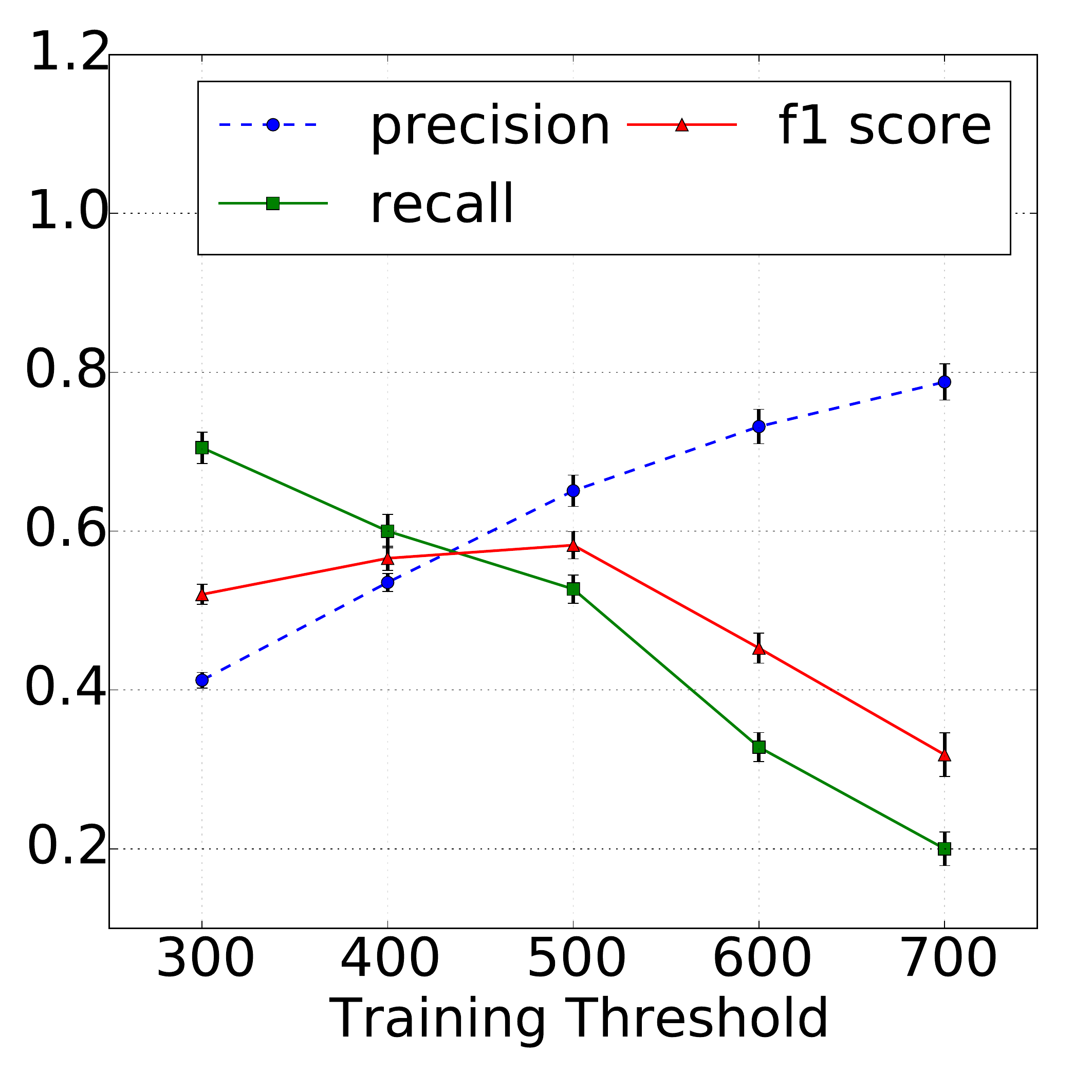}
		\caption{Precision, recall, and F1 score for different training thresholds, using Louvain algorithm.}
		\label{fig:to_Size}
	\end{subfigure}
	\hfill
	\begin{subfigure}[b]{0.49\textwidth}
		\centering
		\includegraphics[width=39mm]{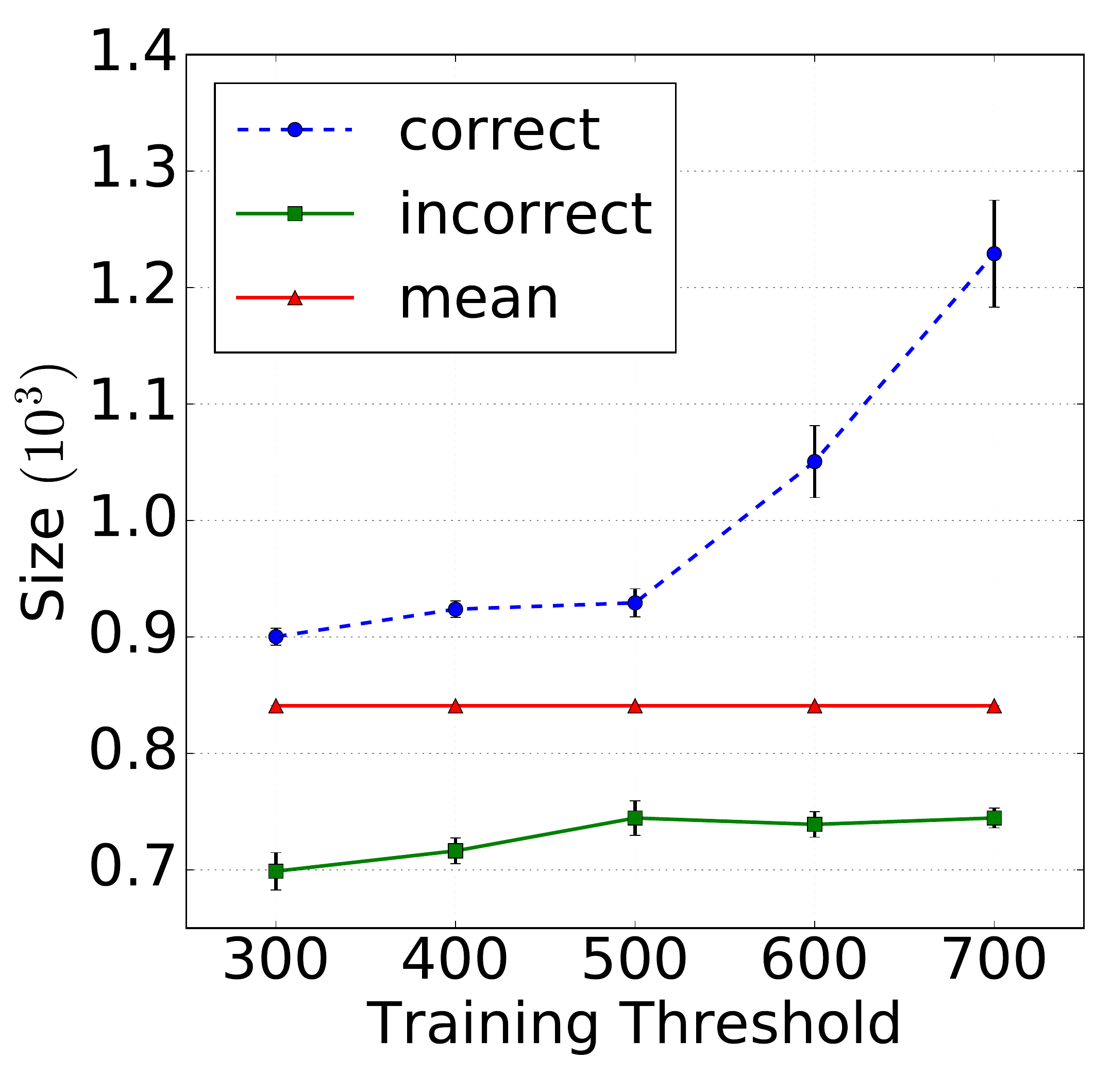}
		\caption{Average final size of viral cascades (correctly classified, mean and incorrectly classified), using Louvain algorithm.}
		\label{fig:tr_th_Size}
	\end{subfigure}
	
	\begin{subfigure}[b]{0.49\textwidth}
		\centering
		\includegraphics[width=39mm]{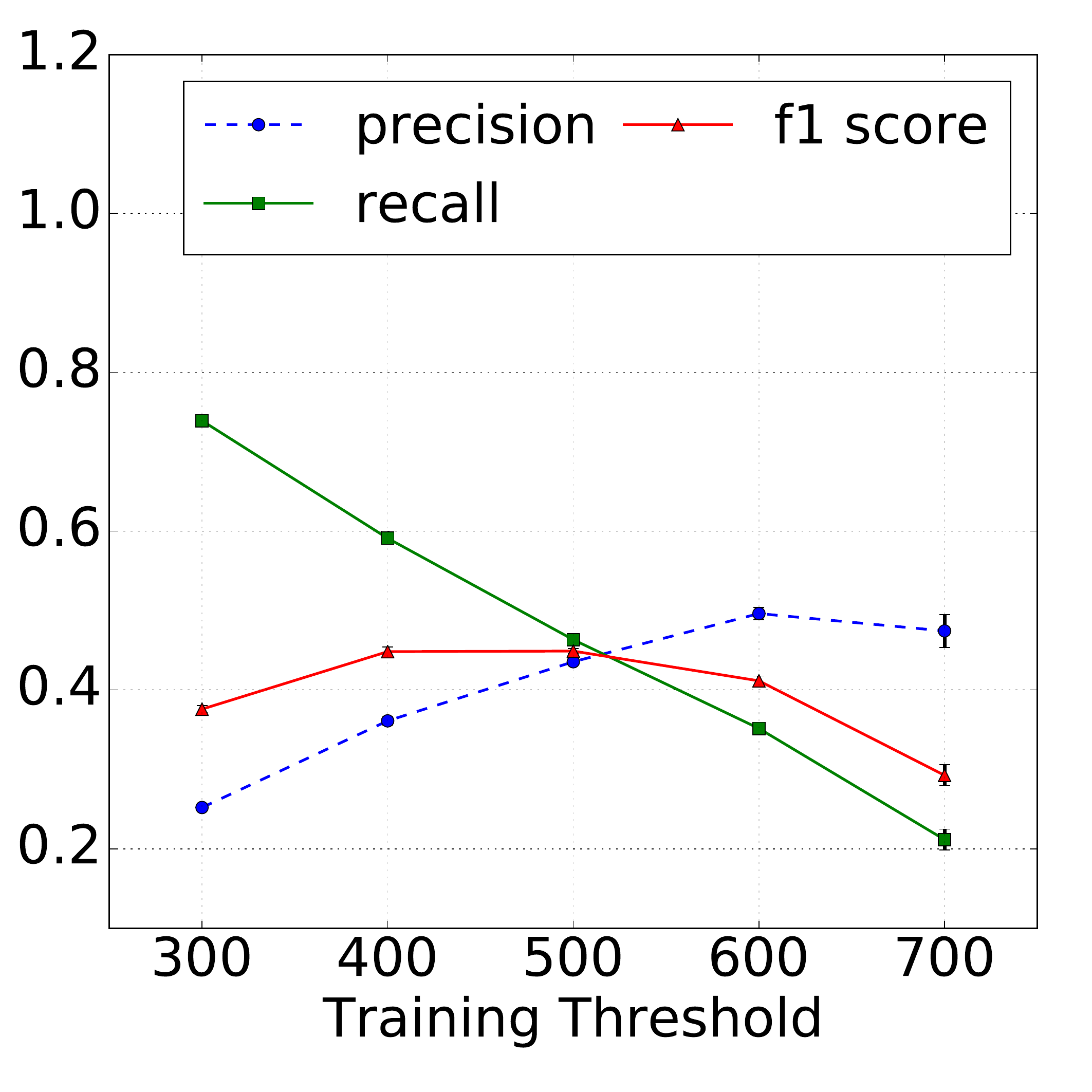}
		\caption{Precision, recall, and F1 score for different training thresholds, using Infomap algorithm.}
		\label{fig:to_Size_infomap}
	\end{subfigure}
	\hfill
	\begin{subfigure}[b]{0.49\textwidth}
		\centering
		\includegraphics[width=39mm]{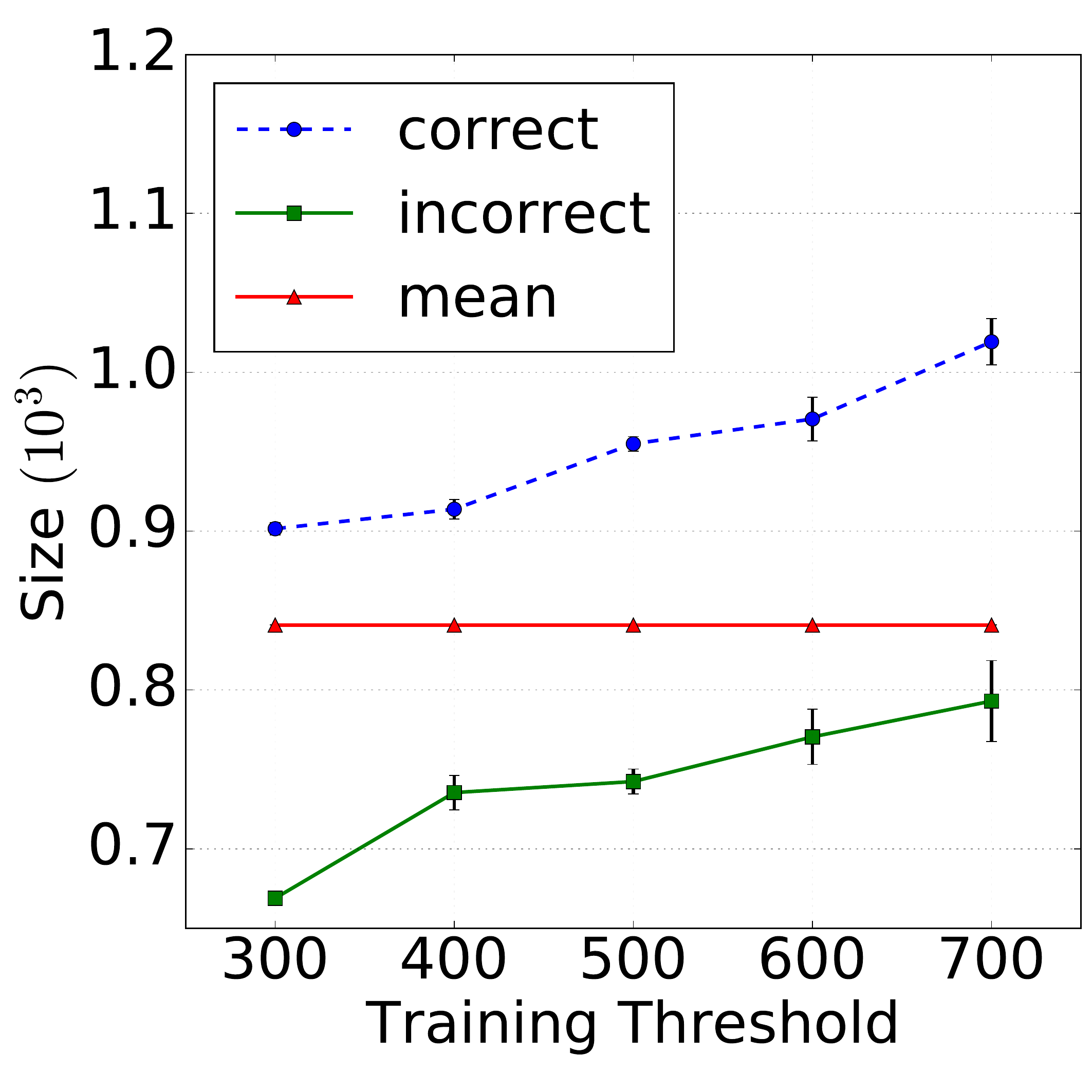}
		\caption{Average final size of viral cascades (correctly classified, mean and incorrectly classified), using Infomap algorithm.}
		\label{fig:tr_th_Size_infomap}
	\end{subfigure}
	
	\begin{subfigure}[b]{0.49\textwidth}
		\centering
		\includegraphics[width=39mm]{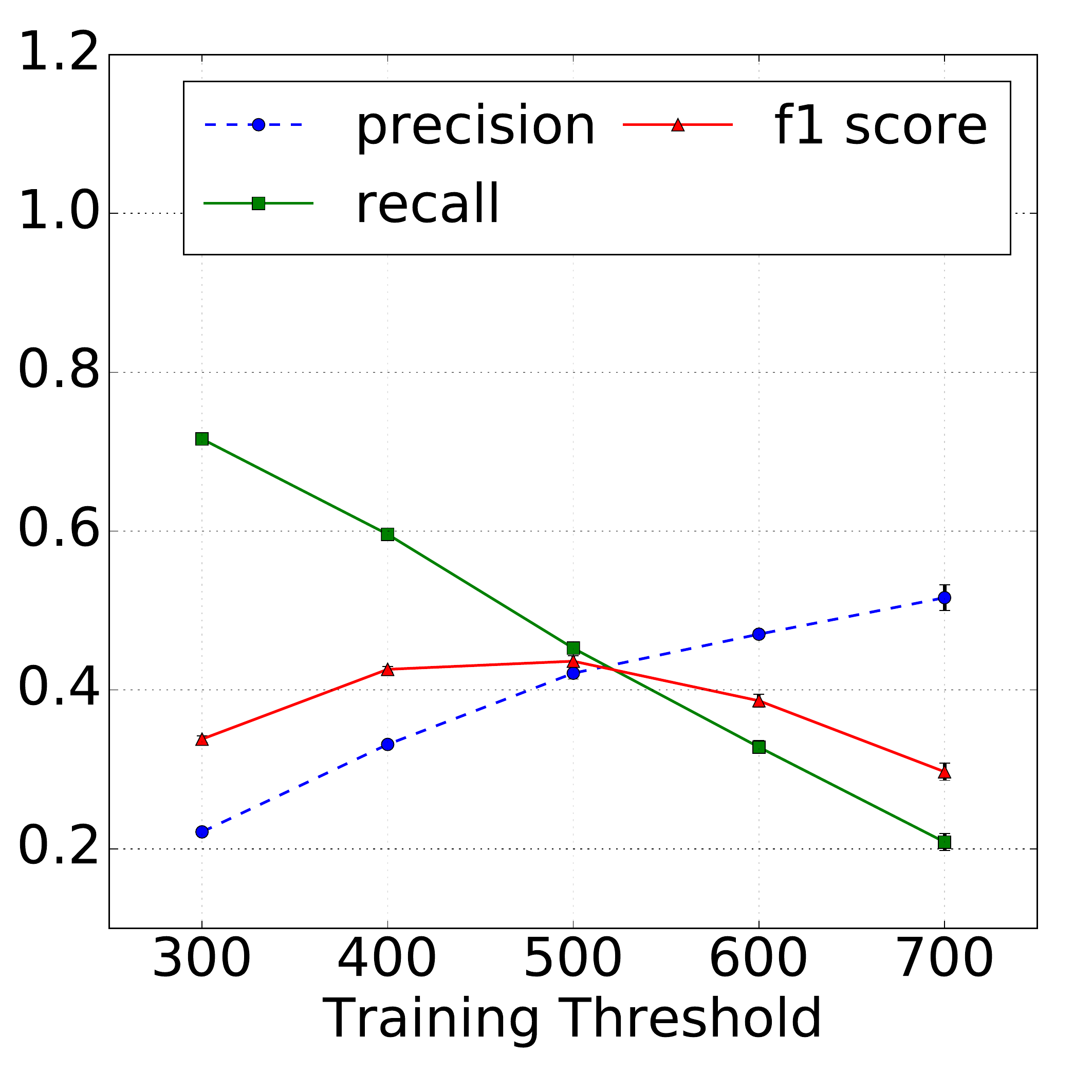}
		\caption{Precision, recall, and F1 score for different training thresholds, using SLM algorithm.}
		\label{fig:to_Size_slm}
	\end{subfigure}
	\hfill
	\begin{subfigure}[b]{0.49\textwidth}
		\centering
		\includegraphics[width=39mm]{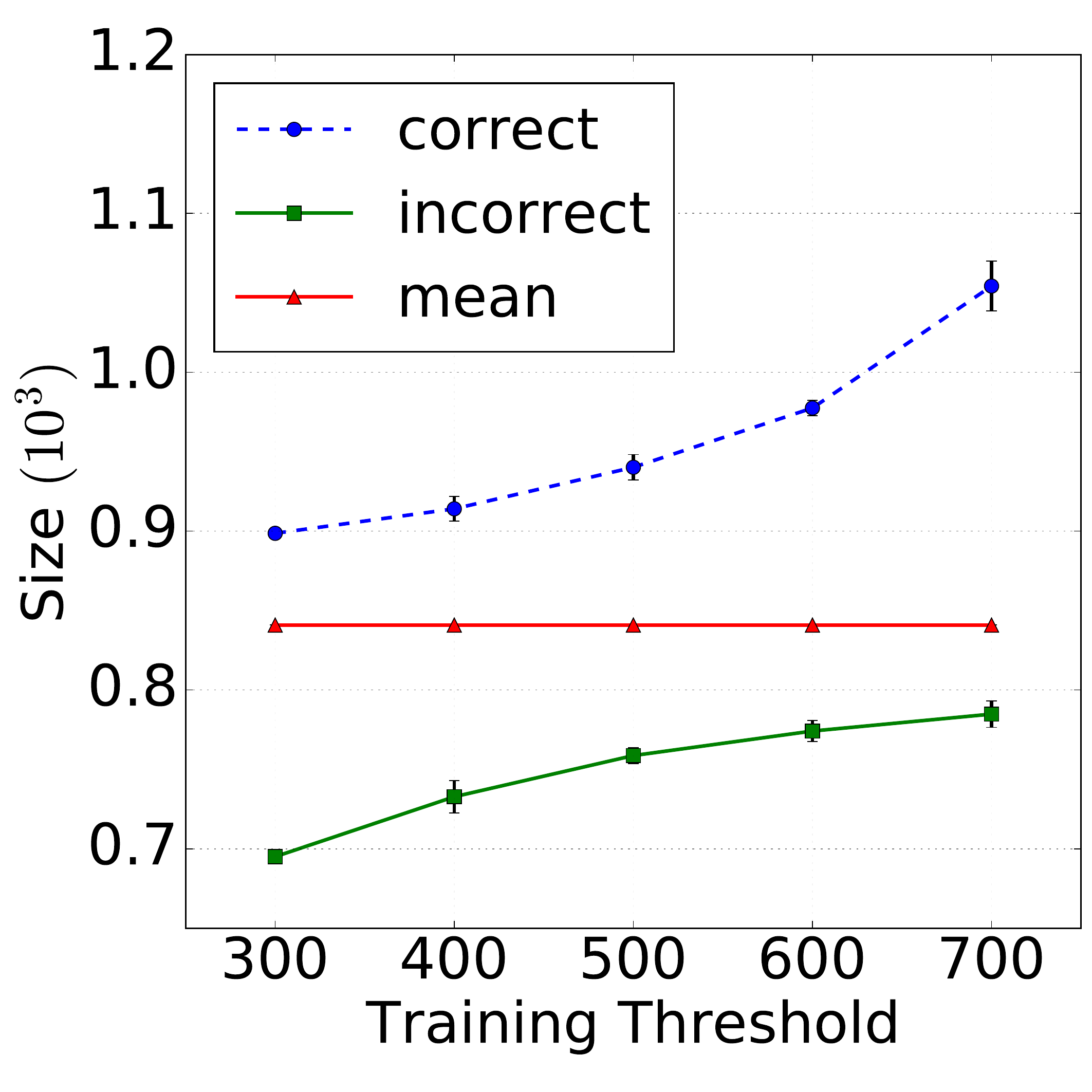}
		\caption{Average final size of viral cascades (correctly classified, mean and incorrectly classified), using SLM algorithm.}
		\label{fig:tr_th_Size_slm}
	\end{subfigure}

	\caption{Prediction results when $TH_{tr} = \left\{300,400,500,600,700\right\}$ for $ A_m $(Louvain, Infomap and SLM). Error bars represent one standard deviation.}
	\label{tngSizeRes}
\end{figure}

\begin{figure}[!tbp]
	\begin{subfigure}[b]{0.49\textwidth}
		\centering
		\includegraphics[width=39mm]{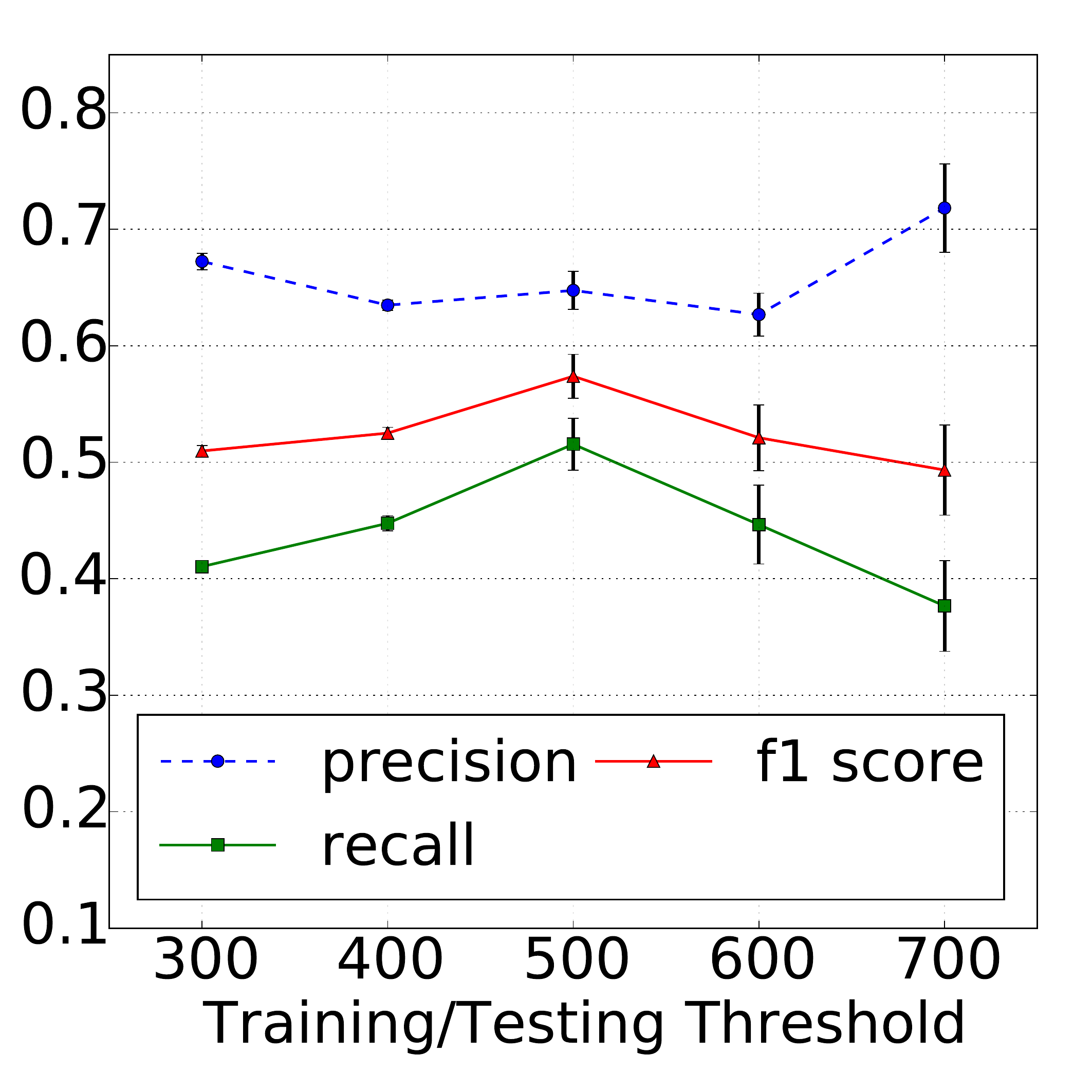}
		\caption{Classification results for features in group $ A_m  $(Louvain)}
		\label{fig:SIZE_A}
	\end{subfigure}
	\hfill
	\begin{subfigure}[b]{0.49\textwidth}
		\centering
		\includegraphics[width=39mm]{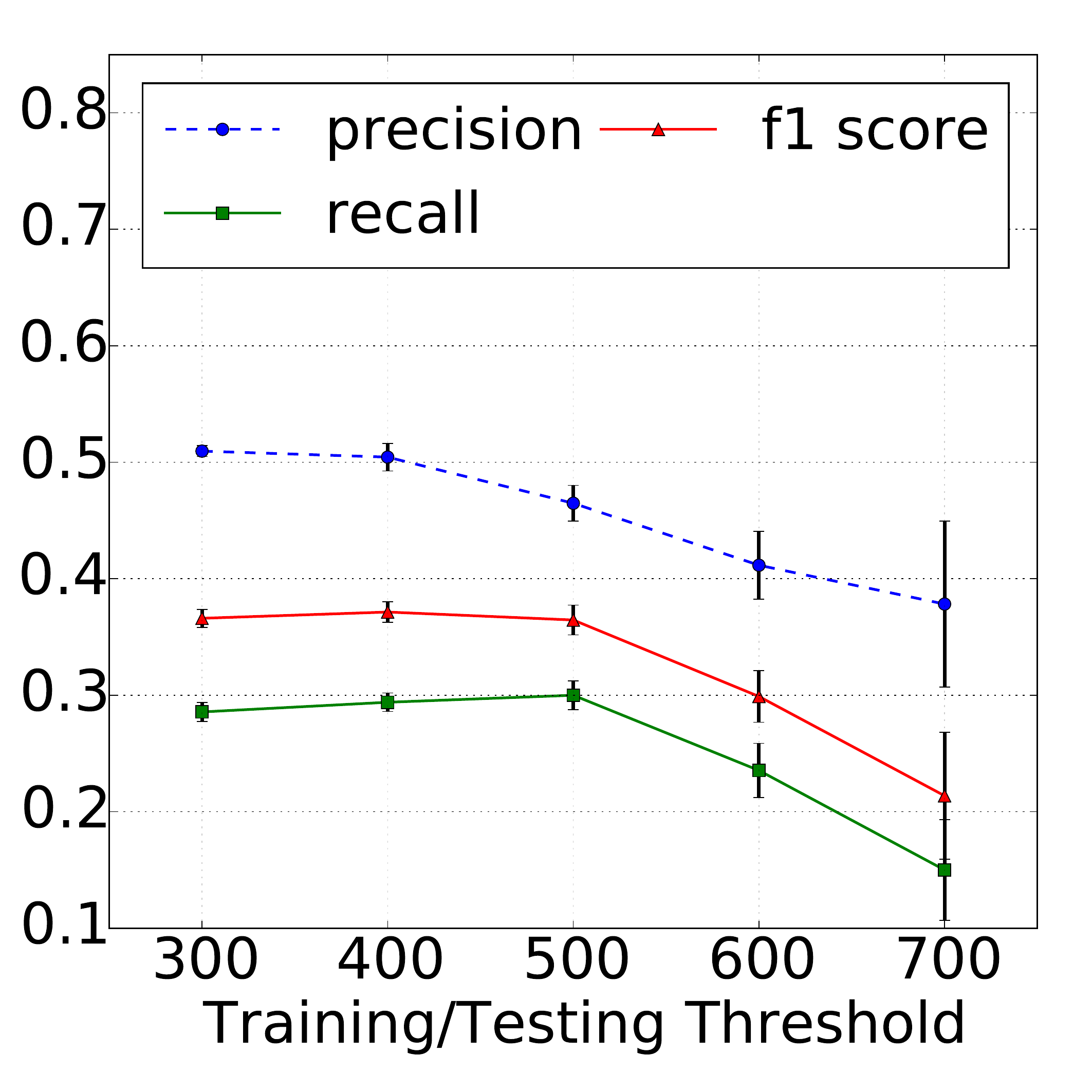}
		\caption{Classification results for features in group $B_m $(Louvain)}
		\label{fig:SIZE_B}
	\end{subfigure}
	
	\begin{subfigure}[b]{0.49\textwidth}
		\centering
		\includegraphics[width=39mm]{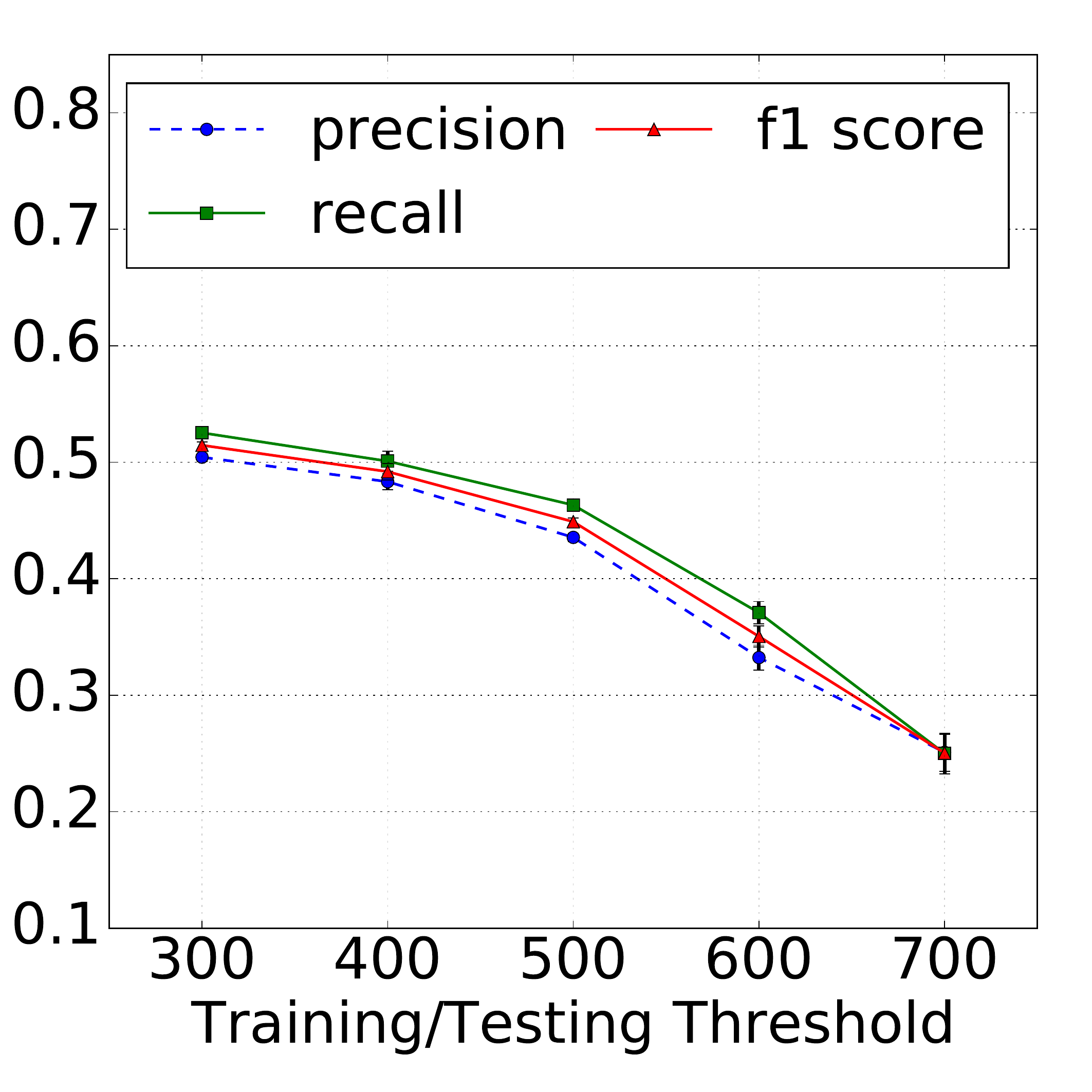}
		\caption{Classification results for features in group $ A_m  $(Infomap)}
		\label{fig:SIZE_A_infomap}
	\end{subfigure}
	\hfill
	\begin{subfigure}[b]{0.49\textwidth}
		\centering
		\includegraphics[width=39mm]{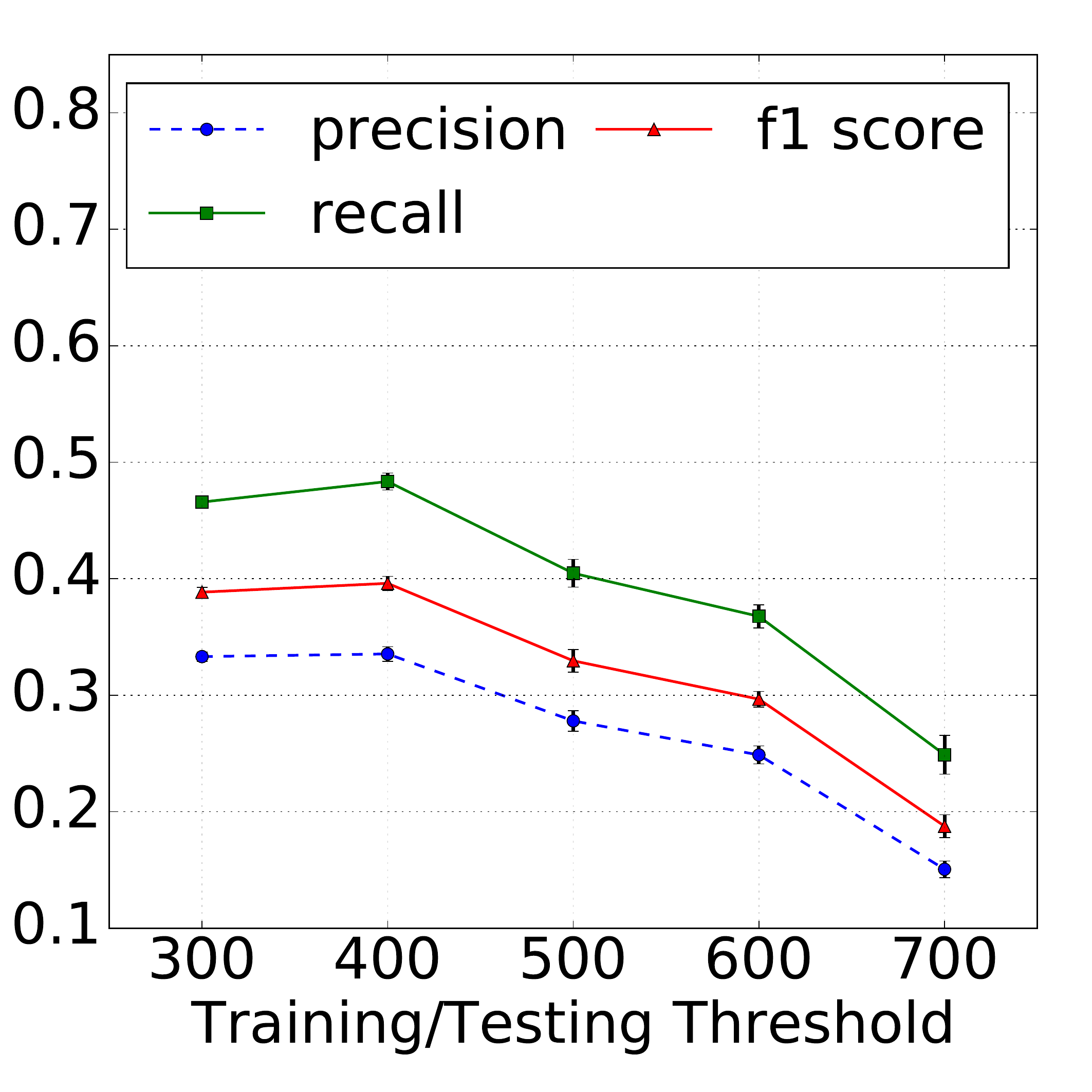}
		\caption{Classification results for features in group $B_m $(Infomap)}
		\label{fig:SIZE_B_infomap}
	\end{subfigure}
	
	\begin{subfigure}[b]{0.49\textwidth}
		\centering
		\includegraphics[width=39mm]{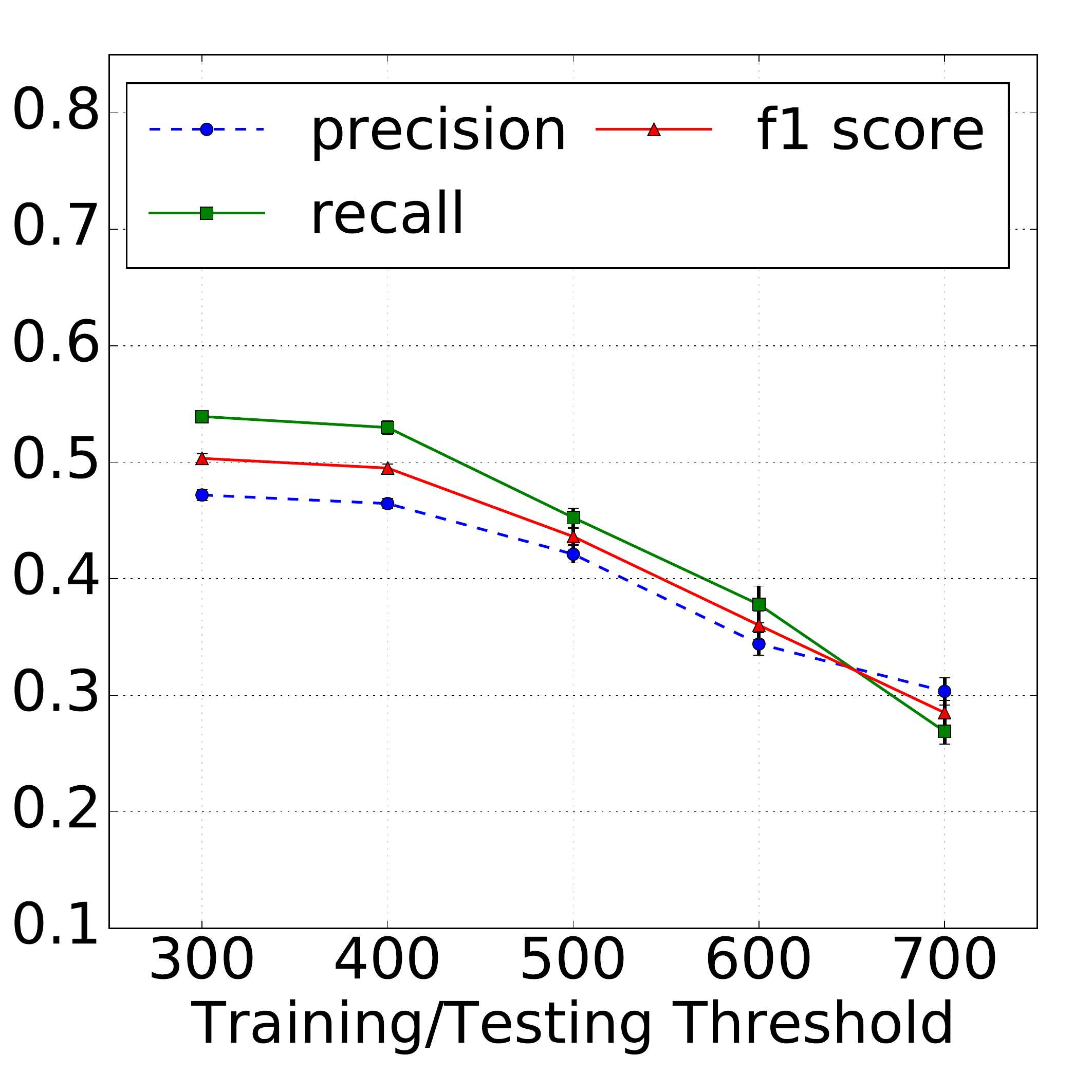}
		\caption{Classification results for features in group $ A_m  $(SLM)}
		\label{fig:SIZE_A_slm}
	\end{subfigure}
	\hfill
	\begin{subfigure}[b]{0.49\textwidth}
		\centering
		\includegraphics[width=39mm]{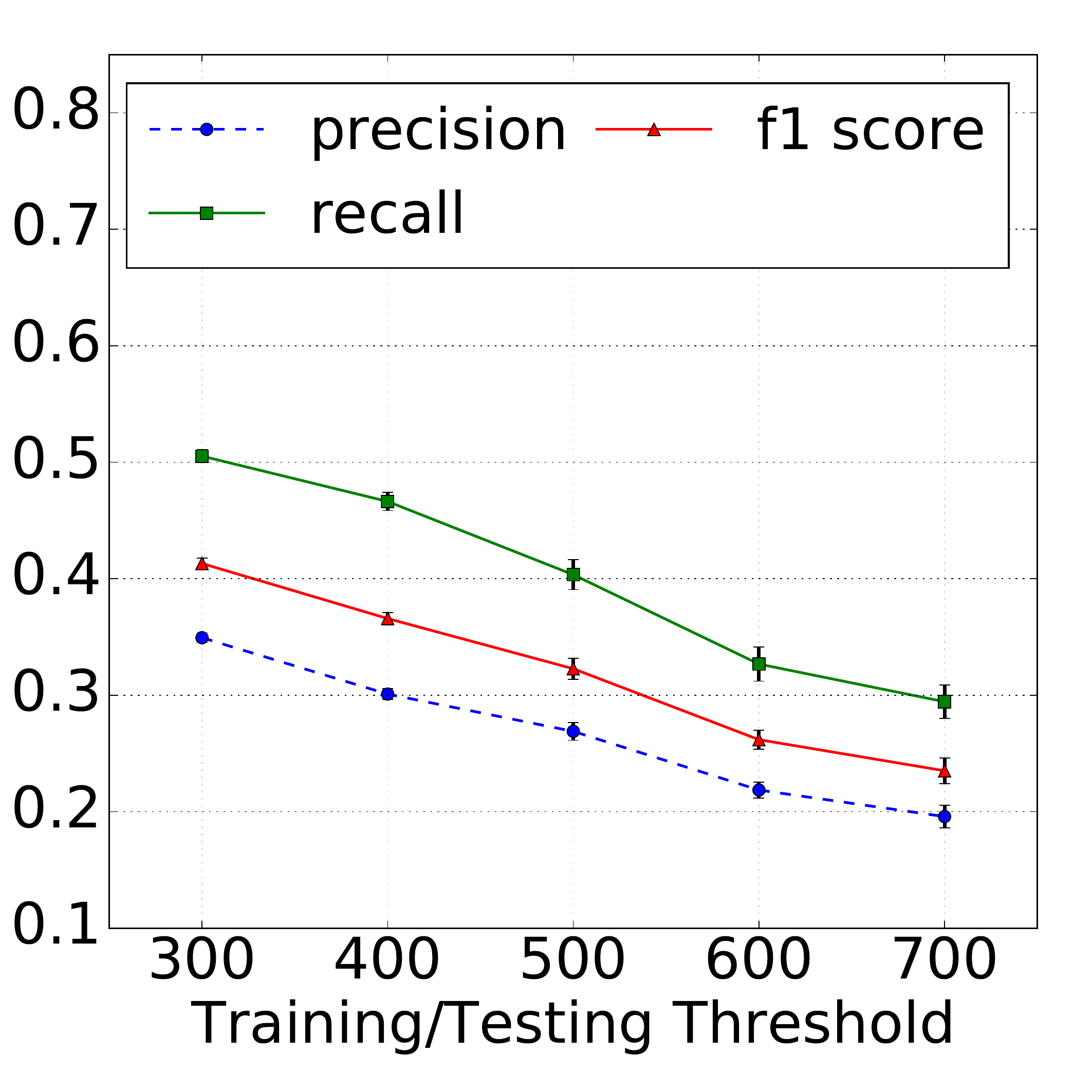}
		\caption{Classification results for features in group $B_m $(SLM)}
		\label{fig:SIZE_B_slm}
	\end{subfigure}
	
	\caption{Prediction results based on groups of features extracted for $ m=50 $ when $ TH = \left\{300,400,500,600,700\right\} $. Error bars represent one standard deviation.}
	\label{thChgSz}
\end{figure}

\paragraph{Size-based prediction.}  We studied cascades of size 50 that reached 500 for this task.  There are 13,285 cascades that can reach the size $ m = 50 $ while only 200 out of them reached the size of 500.  
Maintaining the threshold $ TH = 500 $, Figure~\ref{fig:pred_Size} shows random forest classifier trained with features in group $ A_m $ can outperform the other groups with any of the three community detection algorithms.  
The tradeoff between precision and recall can be achieved by changing the training threshold $TH_{tr}$ while maintaining the testing threshold $TH_{ts}=500$ (see Figure~\ref{tngSizeRes}). We also note that the average final size of viral cascades correctly classified by the classifier increases with the training threshold.  With threshold $ TH \in \left\{300,400,500,600,700\right\} $ on both training and testing samples, the features introduced in this paper ($A_m$) consistently outperform those previously introduced ($B_m$) -- see Figure~\ref{thChgSz}.
The fact that features in $B_m$ are not able to maintain their predictability over different $TH$ can be explained as that they only count the number of users on recently exposed users instead of taking the community structure of them or the decay of probability to repost over time into consideration.
As shown in Figure~\ref{fig:to_Size},~\ref{fig:to_Size_infomap} and~\ref{fig:to_Size_slm}, while the trends relating to this tradeoff are similar among the various community detection algorithms, the Louvain algorithm led to superior performance for precision and F1.  Infomap and SLM generally outperformed Louvain in terms of recall for both feature sets.  We also note that our features outperform those of Weng et al. regardless of the testing/training thresholds and the selected community finding algorithm.

\begin{figure}[!t]
	\centering
	\includegraphics[width=84mm]{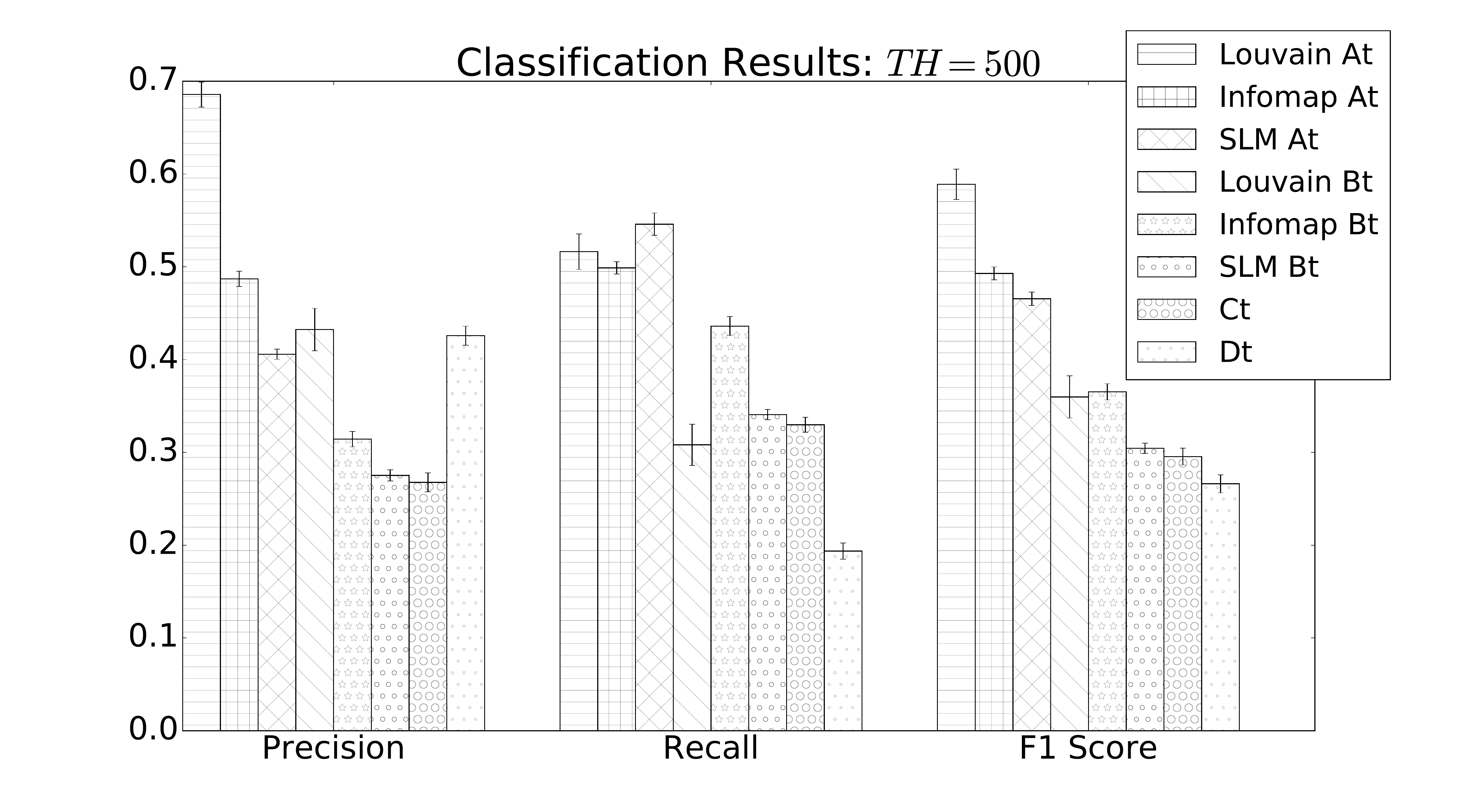}
	
	\caption{Classification results based on groups of features ($A_t$,$B_t$,$C_t$,$D_t$) extracted with three community detection algorithms (Louvain, Infomap and SLM) when $ t = 60 $ for fixed $ TH_{tr} = 500$, $TH_{ts} = 500 $. Error bars represent one standard deviation.}
	\label{fig:pred_TIME}
\end{figure}

\begin{figure}[!tbp]
	\begin{subfigure}[H]{0.49\textwidth}
		\centering
		\includegraphics[width=39mm]{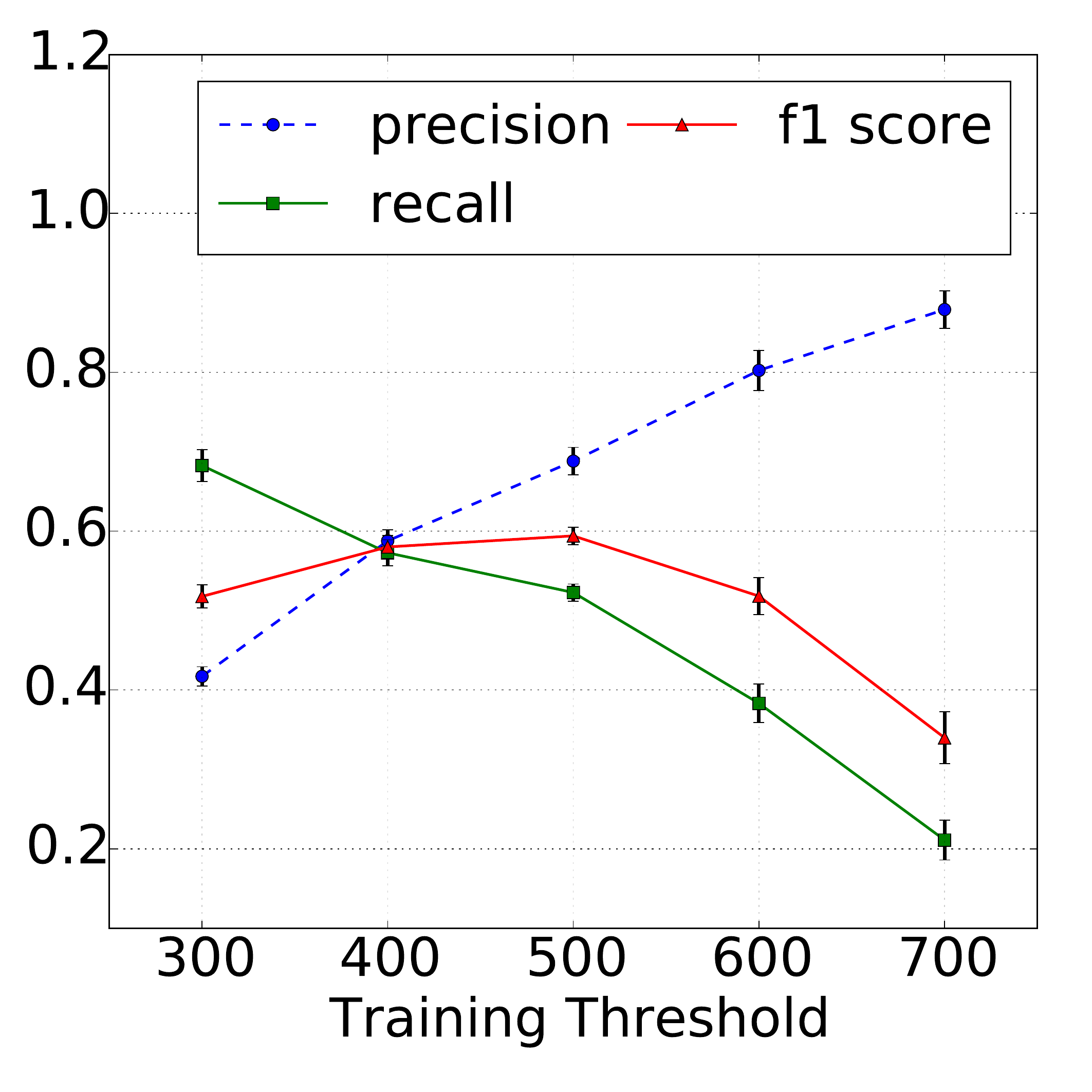}
		\caption{Precision, recall, and F1 score for different training thresholds, using Louvain algorithm.}
		\label{fig:to_Time}
	\end{subfigure}
	\hfill
	\begin{subfigure}[H]{0.49\textwidth}
		\centering
		\includegraphics[width=39mm]{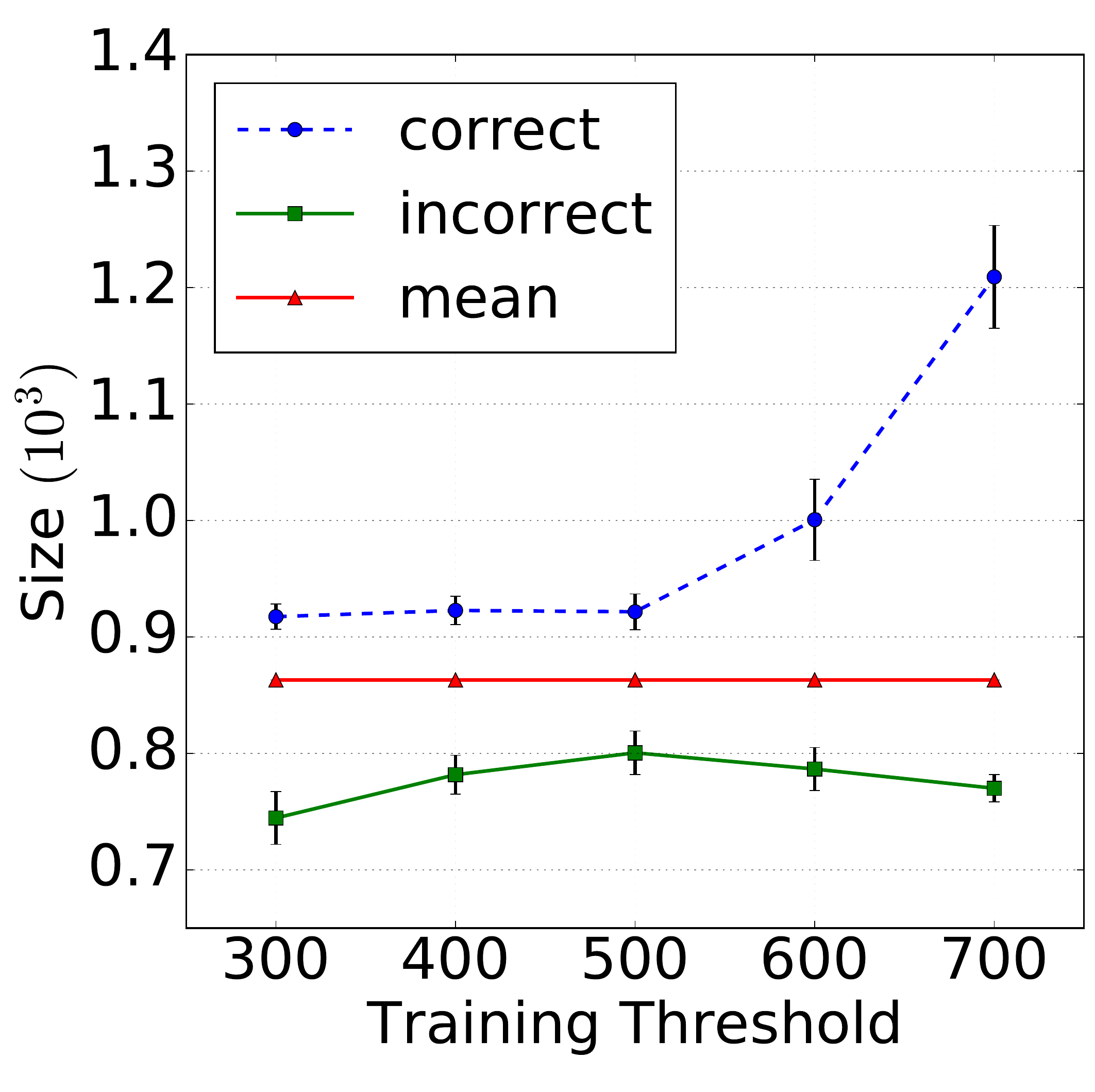}
		\caption{Average final size of viral cascades (correctly classified, mean and incorrectly classified),  using Louvain algorithm.}
		\label{fig:tr_th_Time}
	\end{subfigure}
	
	\begin{subfigure}[H]{0.49\textwidth}
		\centering
		\includegraphics[width=39mm]{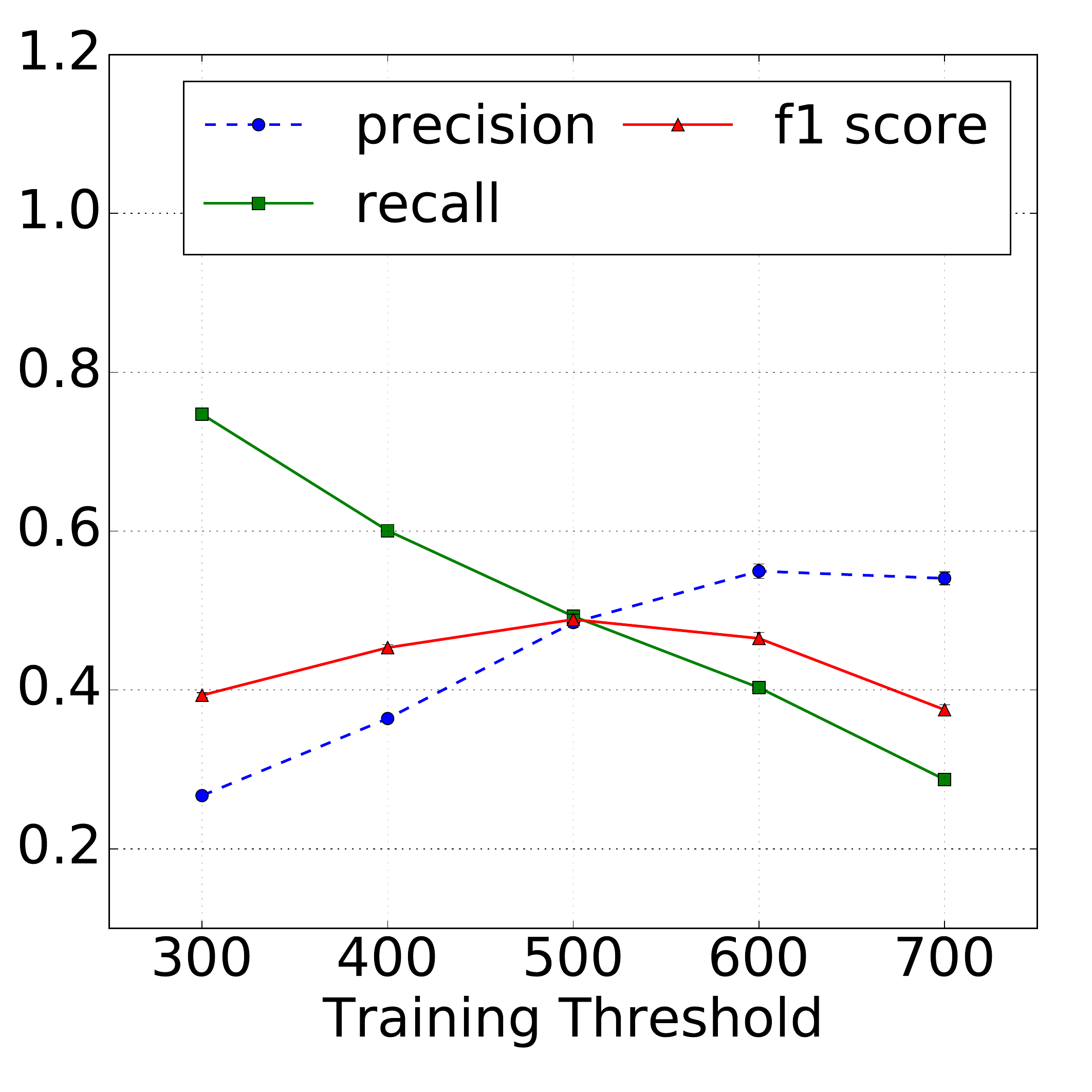}
		\caption{Precision, recall, and F1 score for different training thresholds, using Infomap algorithm.}
		\label{fig:to_Time_infomap}
	\end{subfigure}
	\hfill
	\begin{subfigure}[H]{0.49\textwidth}
		\centering
		\includegraphics[width=39mm]{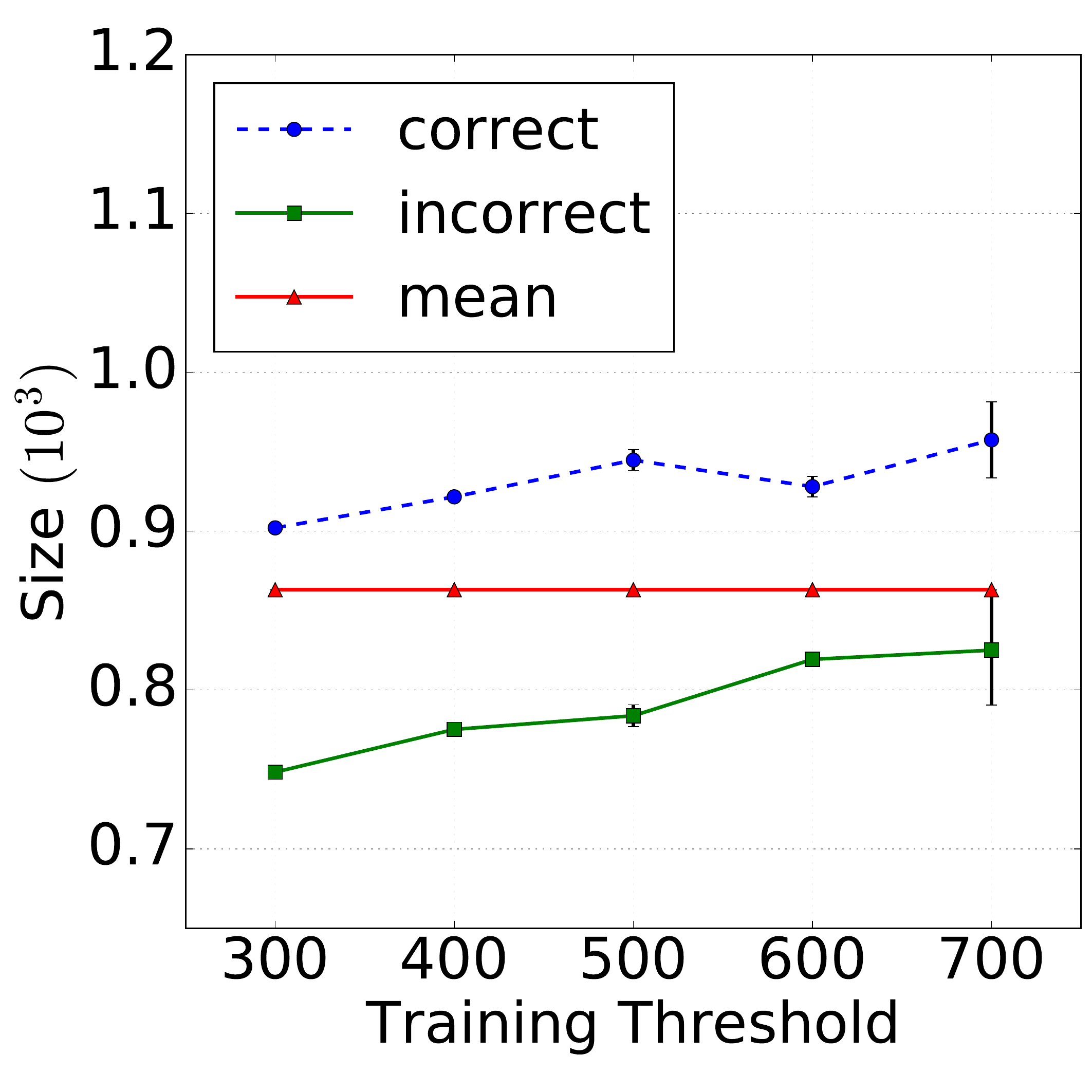}
		\caption{Average final size of viral cascades (correctly classified, mean and incorrectly classified),  using Infomap algorithm.}
		\label{fig:tr_th_Time_infomap}
	\end{subfigure}
	
	\begin{subfigure}[H]{0.49\textwidth}
		\centering
		\includegraphics[width=39mm]{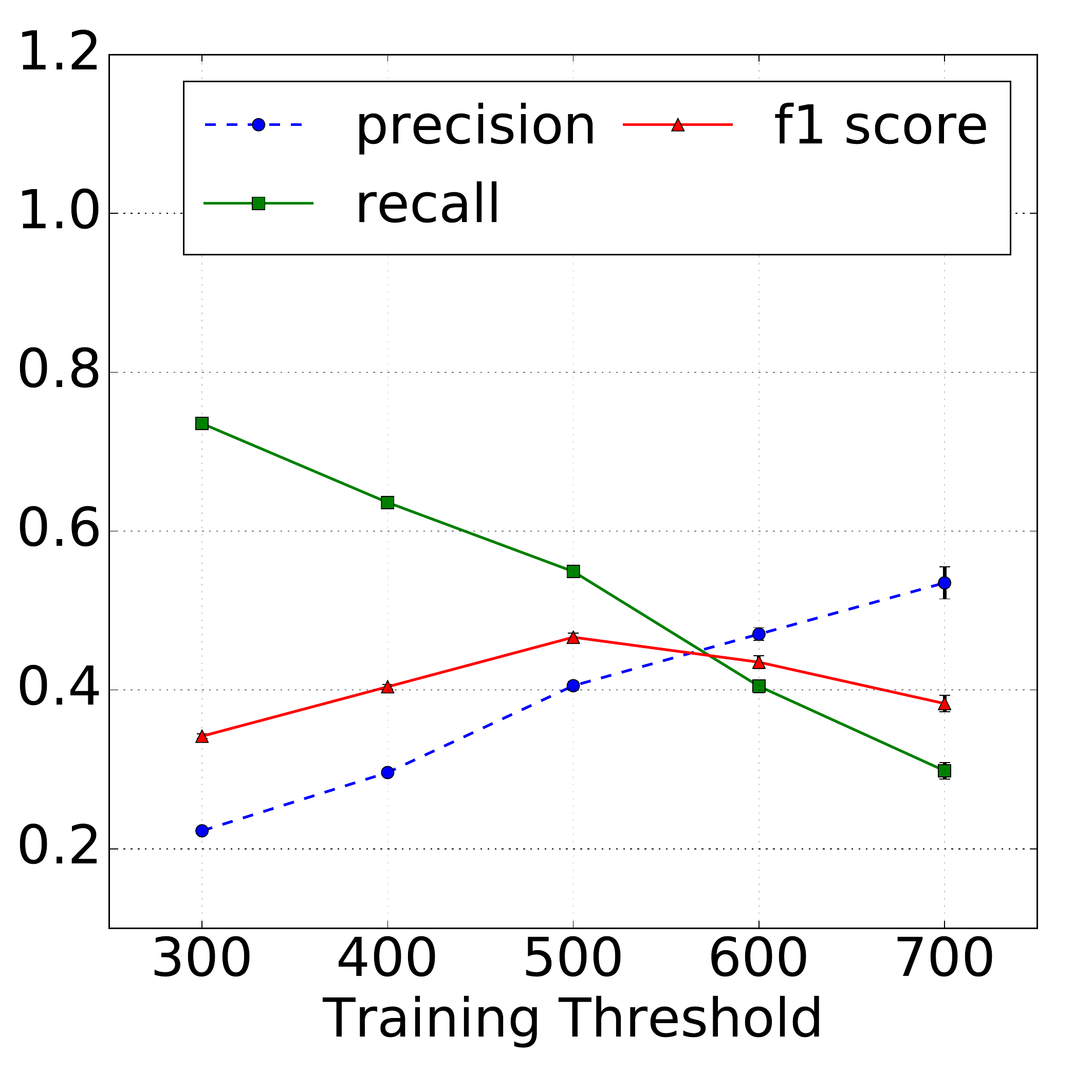}
		\caption{Precision, recall, and F1 score for different training thresholds, using SLM algorithm.}
		\label{fig:to_Time_slm}
	\end{subfigure}
	\hfill
	\begin{subfigure}[H]{0.49\textwidth}
		\centering
		\includegraphics[width=39mm]{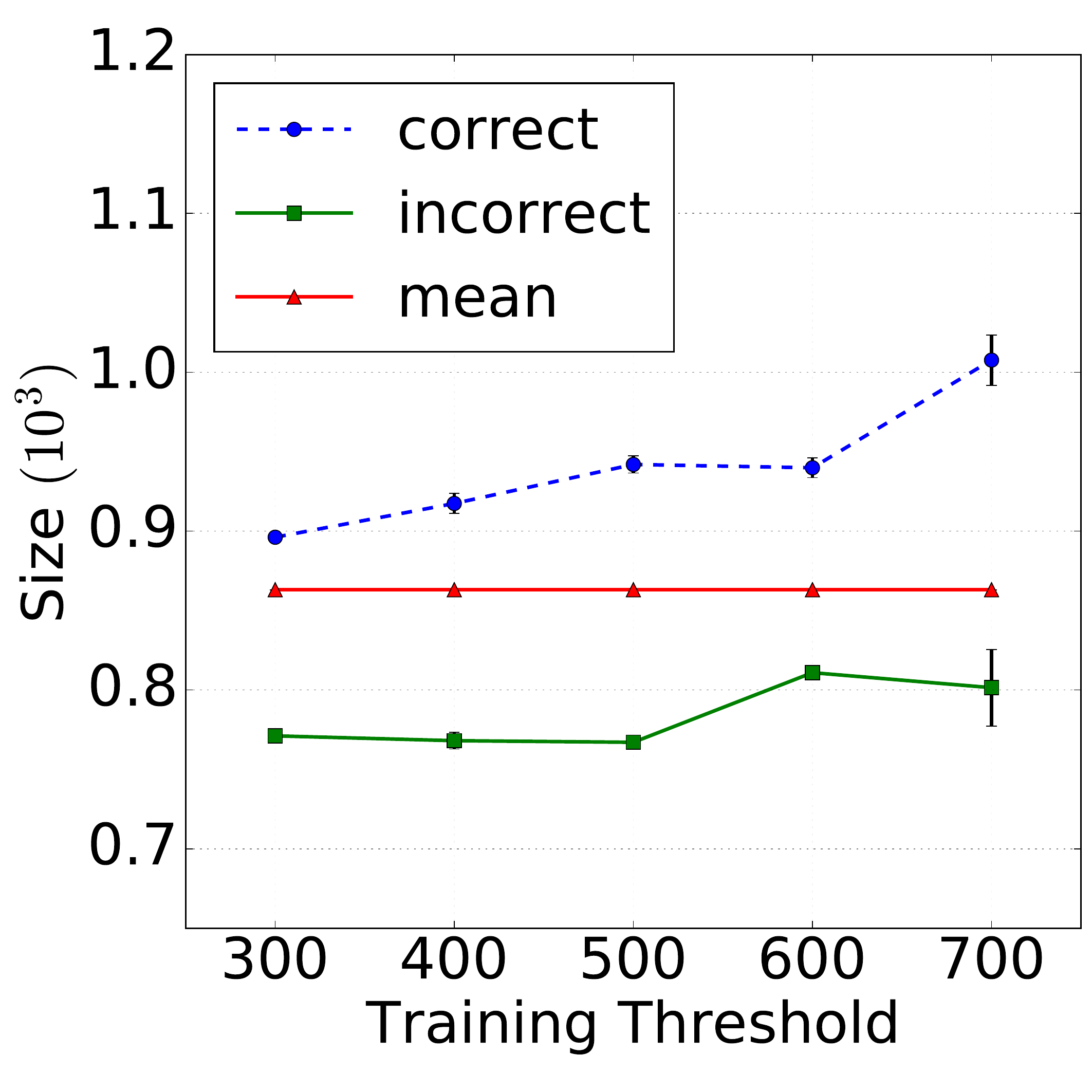}
		\caption{Average final size of viral cascades (correctly classified, mean and incorrectly classified),  using SLM algorithm.}
		\label{fig:tr_th_Time_slm}
	\end{subfigure}
	\caption{Prediction results when $TH_{tr} \in \left\{300,400,500,600,700\right\}$ for $ A_t $ (Louvain, Infomap and SLM). Error bars represent one standard deviation.}
\end{figure}

\begin{figure}[!tbp]
	\begin{subfigure}[H]{0.49\textwidth}
		\centering
		\includegraphics[width=39mm]{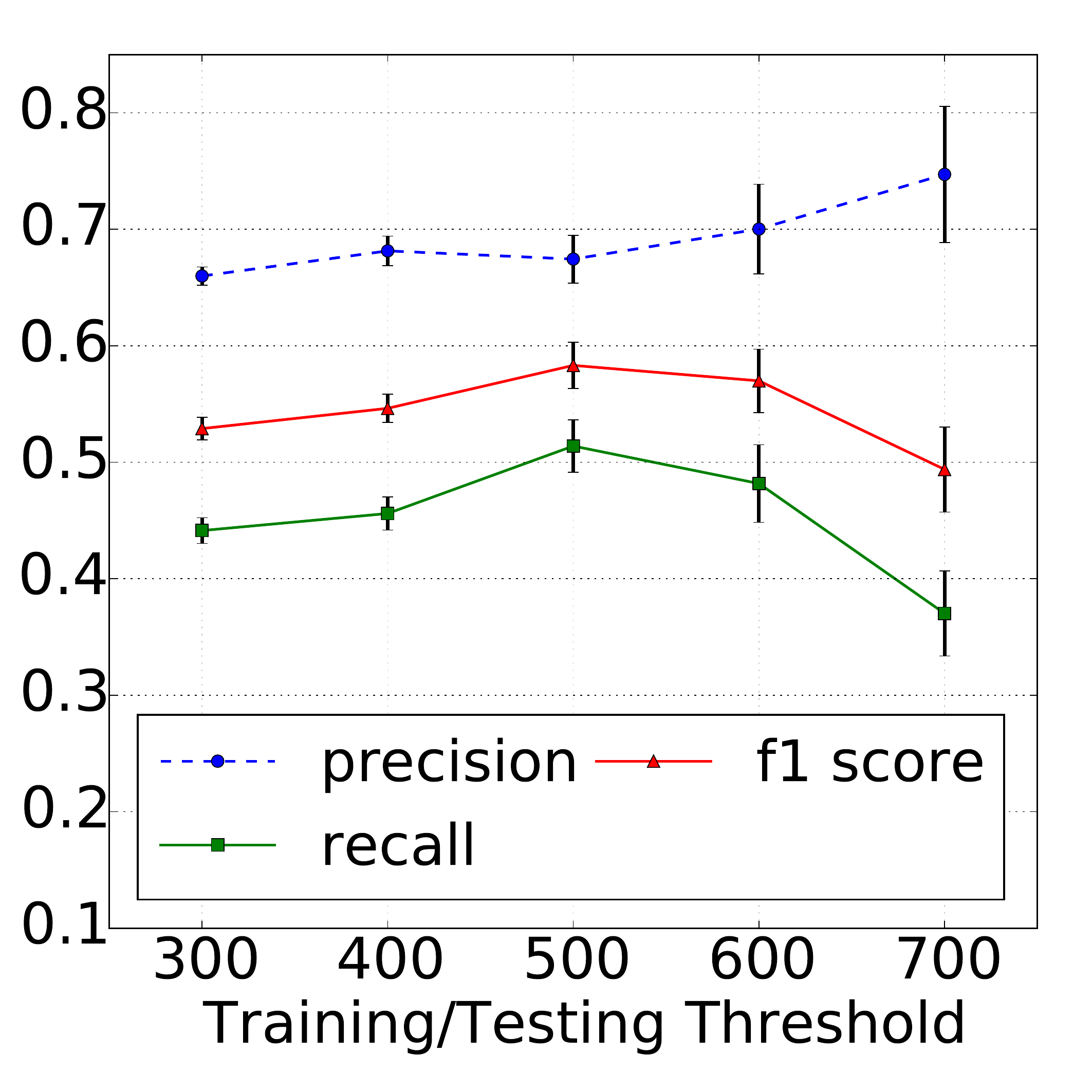}
		\caption{Classification results for features in group $ A_t $ (Louvain)}
		\label{fig:TIME_A}	
	\end{subfigure}%
	\hfill
	\begin{subfigure}[H]{0.49\textwidth}
		\centering
		\includegraphics[width=39mm]{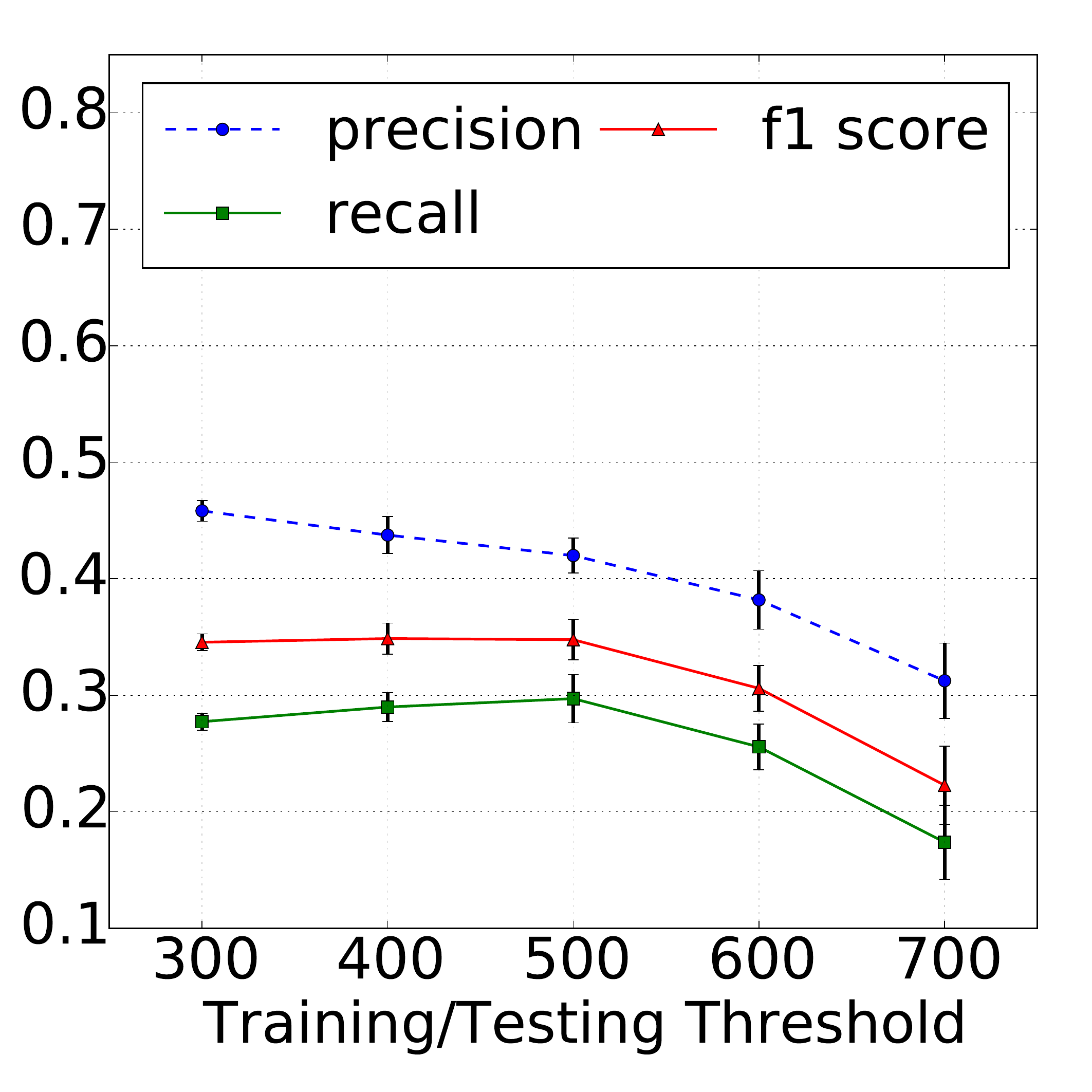}
		\caption{Classification results for features in group $ B_t $ (Louvain)}
		\label{fig:TIME_B}
	\end{subfigure}
	
	\begin{subfigure}[H]{0.49\textwidth}
		\centering
		\includegraphics[width=39mm]{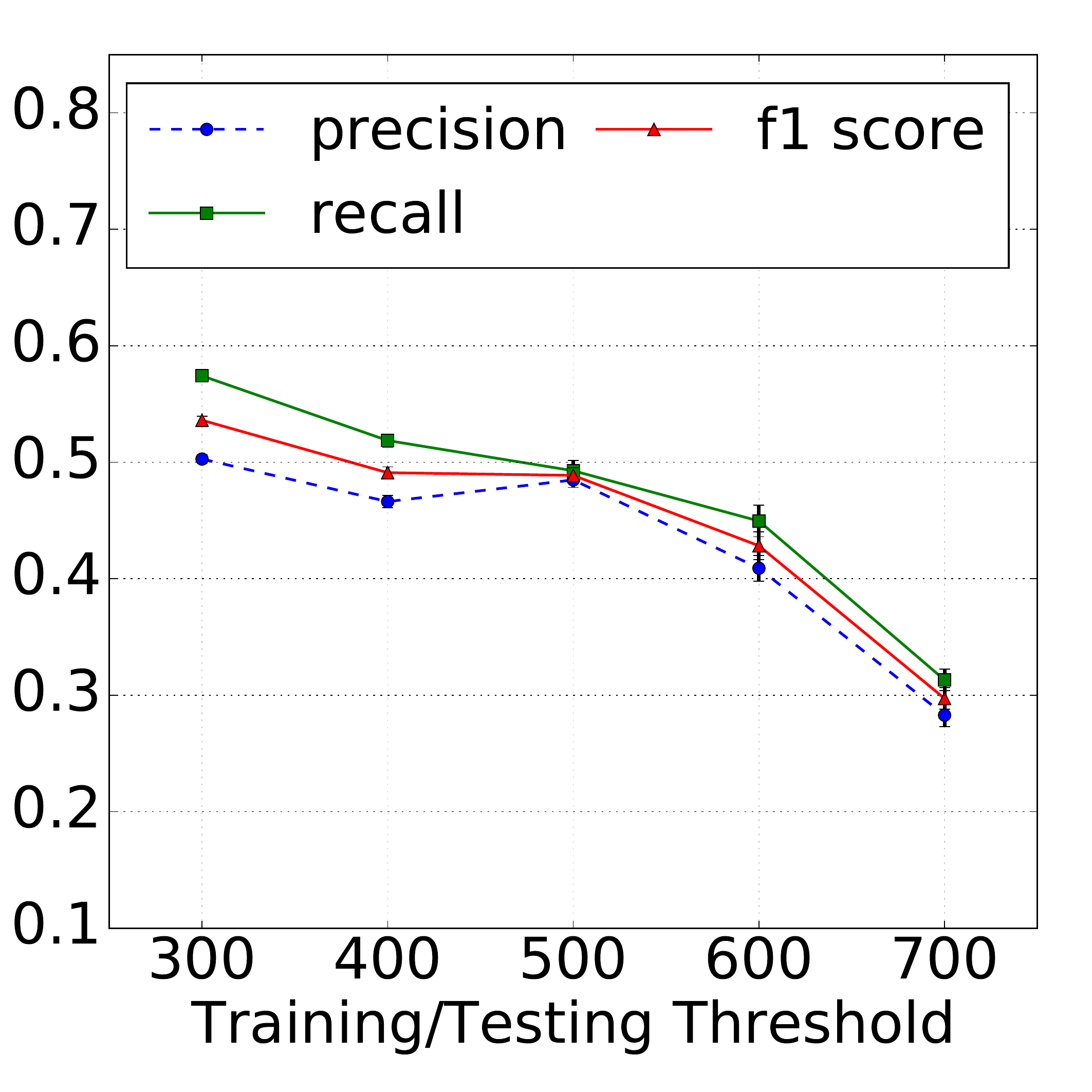}
		\caption{Classification results for features in group $ A_t $ (Infomap)}
		\label{fig:TIME_A_infomap}	
	\end{subfigure}%
	\hfill
	\begin{subfigure}[H]{0.49\textwidth}
		\centering
		\includegraphics[width=39mm]{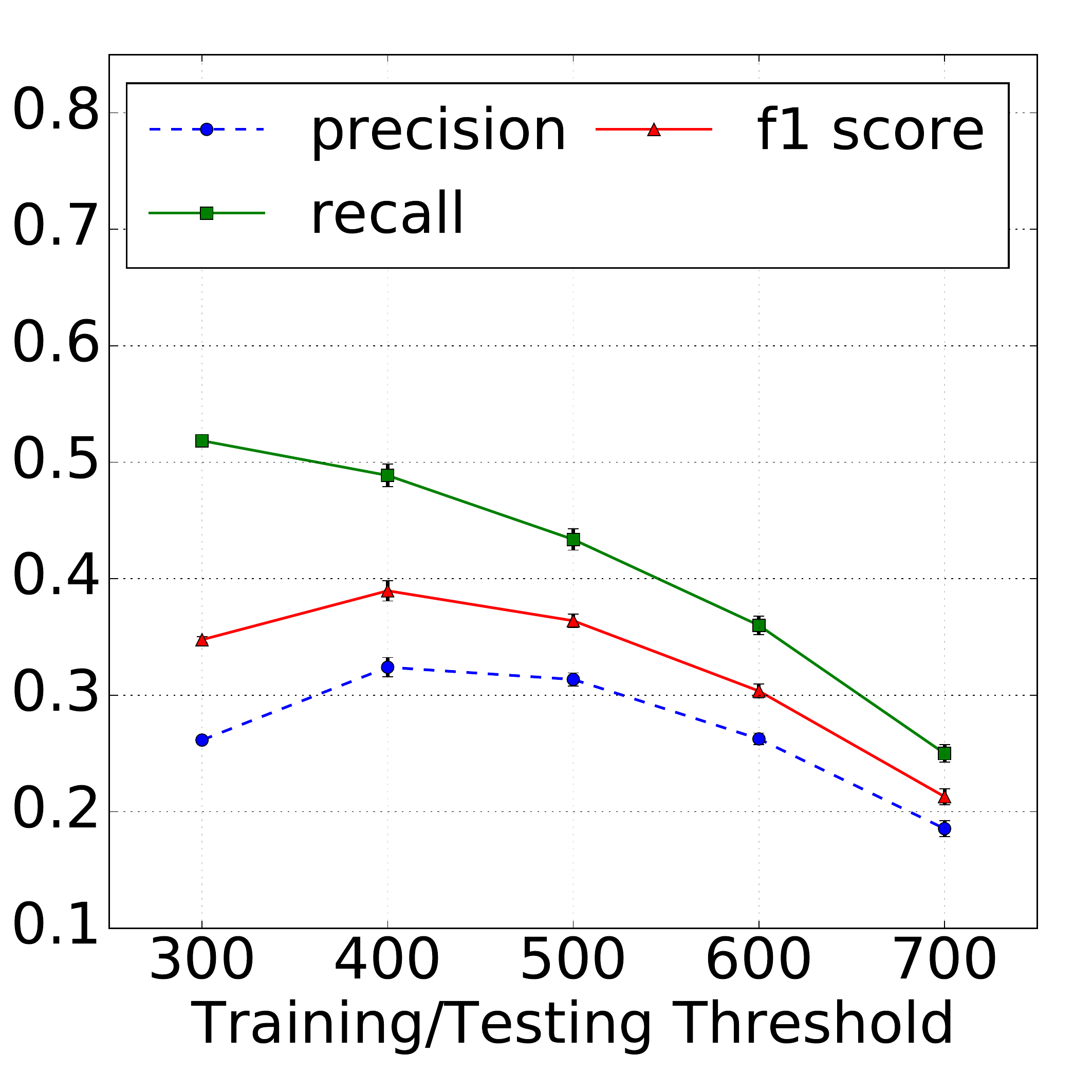}
		\caption{Classification results for features in group $ B_t $ (Infomap)}
		\label{fig:TIME_B_infomap}
	\end{subfigure}
	
	\begin{subfigure}[H]{0.49\textwidth}
		\centering
		\includegraphics[width=39mm]{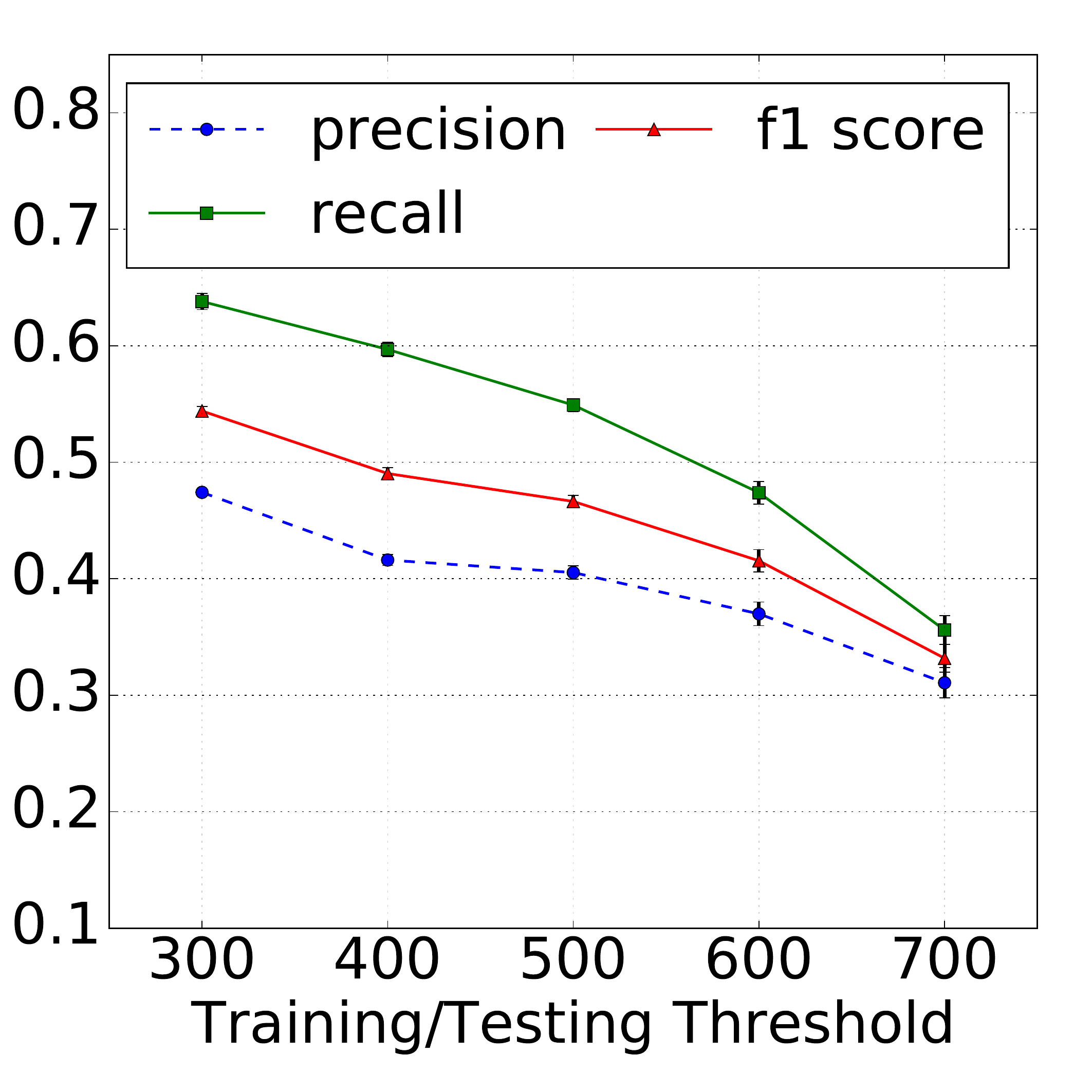}
		\caption{Classification results for features in group $ A_t $ (SLM)}
		\label{fig:TIME_A_slm}	
	\end{subfigure}%
	\hfill
	\begin{subfigure}[H]{0.49\textwidth}
		\centering
		\includegraphics[width=39mm]{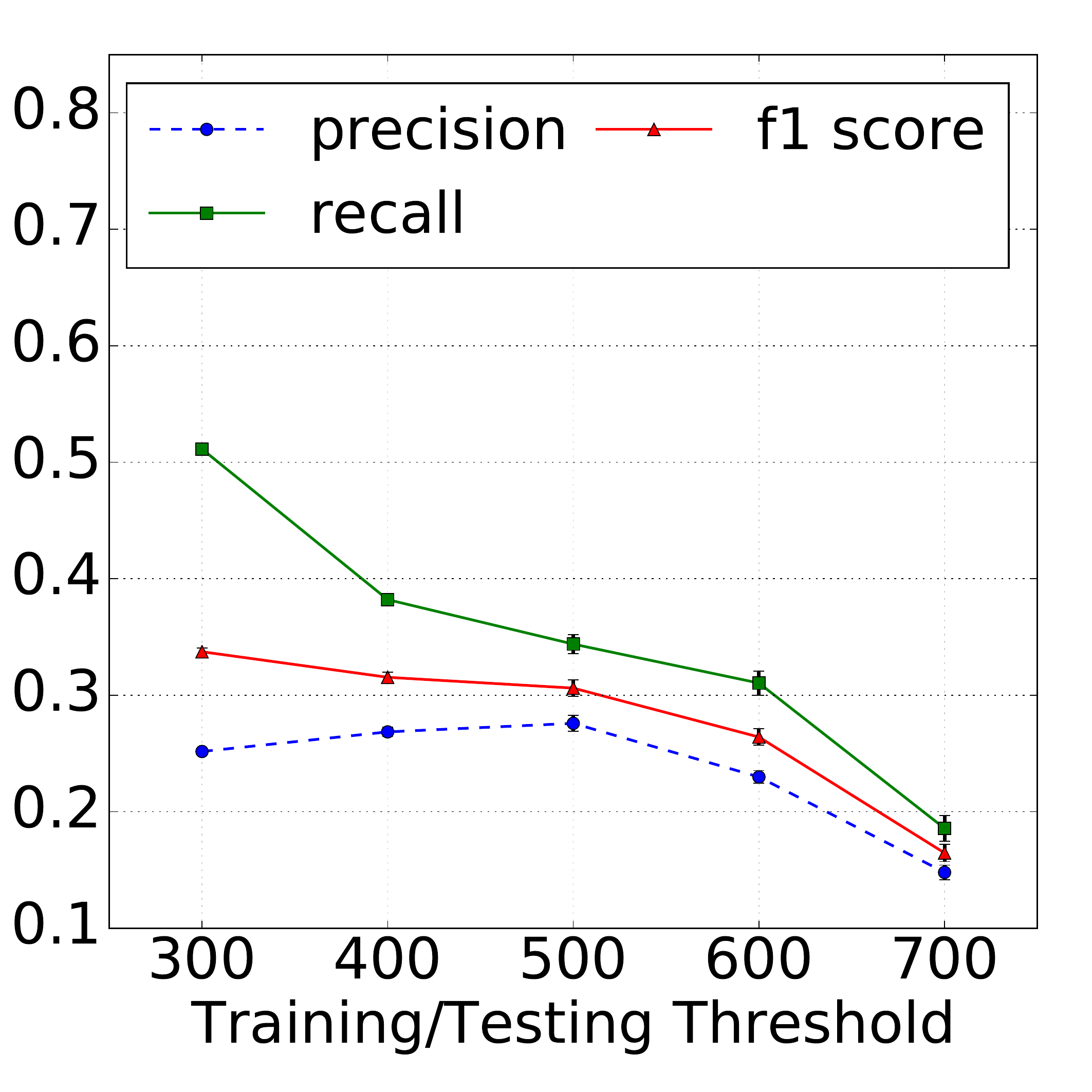}
		\caption{Classification results for features in group $ B_t $ (SLM)}
		\label{fig:TIME_B_slm}
	\end{subfigure}
	
	\caption{Prediction results based on groups of features extracted for $ t=60 $(min) for $ TH \in \left\{300,400,500,600,700\right\} $. Error bars represent one standard deviation.}
	\label{fig:Time_AB}
\end{figure}

\paragraph{Time-based prediction.}  As shown in Table~\ref{tab:samples}, there are 3,444 cascades in our dataset reached the size of $ m = 5 $ within $ 60 $ (min) with only $ 5\% $ from the minority class.  When the threshold is kept as $ TH = 500 $ for both training set and testing set, we obtain the results shown in Figure~\ref{fig:pred_TIME} again showing that the features introduced in this paper ($A_t$) outperform the other feature sets in terms of recall, precision and \textit{F1} score, no matter which community detection algorithm is used.  
By modifying threshold for training samples only, two phenomenon are discovered. 
First, a tradeoff between precision and recall can be manipulated by controlling the training threshold ($TH_{tr}$). This is shown in Figure~\ref{fig:to_Time}, ~\ref{fig:to_Time_infomap}, ~\ref{fig:to_Time_slm}.  Second, as shown in Figure~\ref{fig:tr_th_Time},~\ref{fig:tr_th_Time_infomap},~\ref{fig:tr_th_Time_slm}, with $TH_{tr}$ increasing, the average final size of correctly classified viral cascades also grows. 
Furthermore, we modify the threshold for training and testing sets together to show the reliability of features in group $ A_t $ is better than ones in $ B_t $ (See Figure~\ref{fig:Time_AB}). 
Here, we noted similar trends with regard to both feature sets and community finding algorithms as found in the size-based tests.

\subsection{Feature Investigation}
\begin{figure}[H]
	\centering
	\subcaptionbox{Classification results for subsets of $ A_m $(Louvain): \textit{ol} means overlap, \textit{gini} means gini impurity, \textit{at} represents average time to adoption ($ \frac{1}{50}\sum_{i=1}^{50} t(i) $).}%
	[.48\textwidth]{\includegraphics[width=39mm]{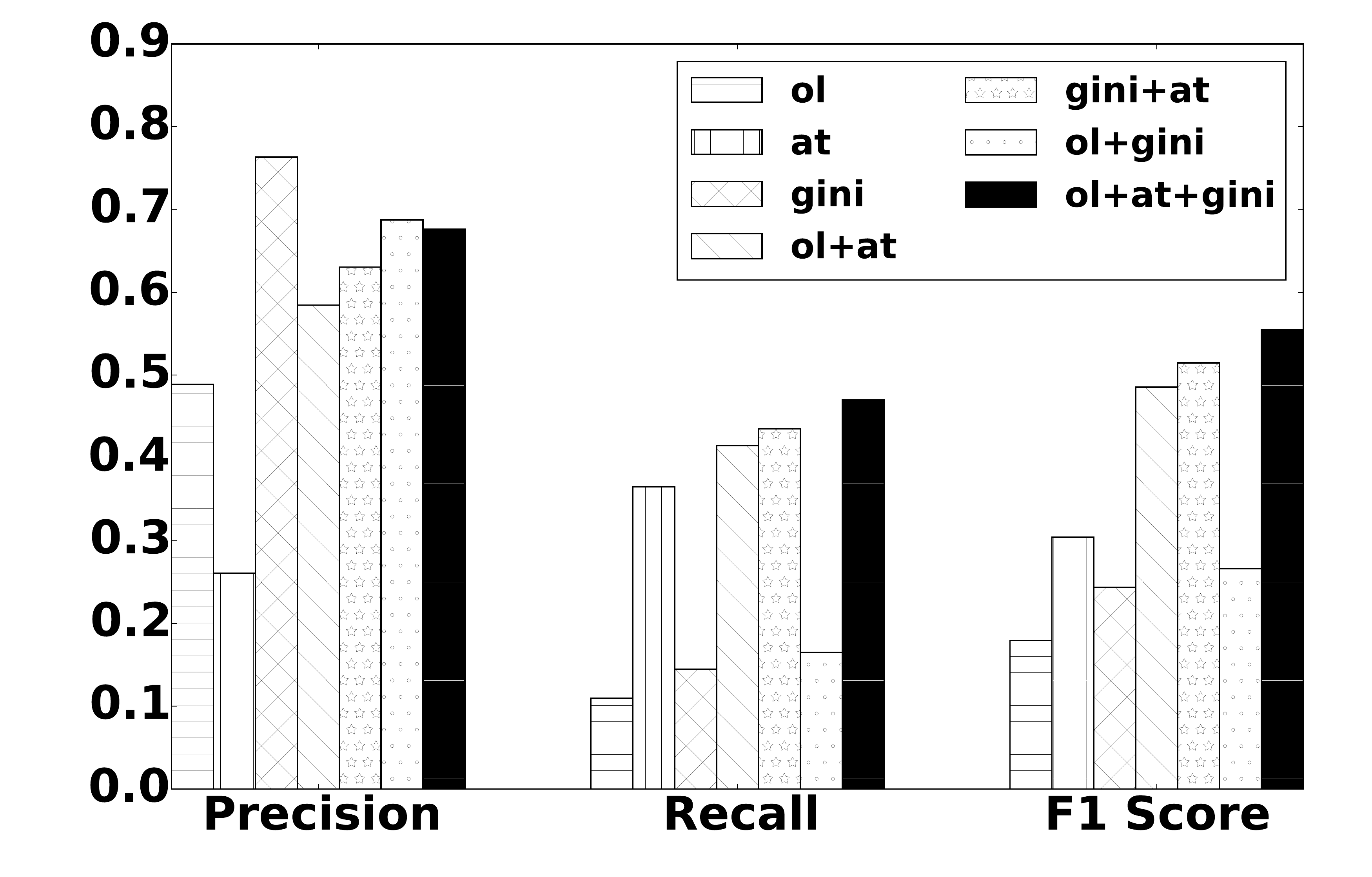}}
	\label{fig:Fsele_Size}
	\hfill
	\subcaptionbox{Classification results for subsets of $ A_t $(Louvain): \textit{ol} means overlap, \textit{gini} means gini impurity, \textit{cs} represents number of adopters ($ |U_{60}| $).}
	[.48\textwidth]{\includegraphics[width=39mm]{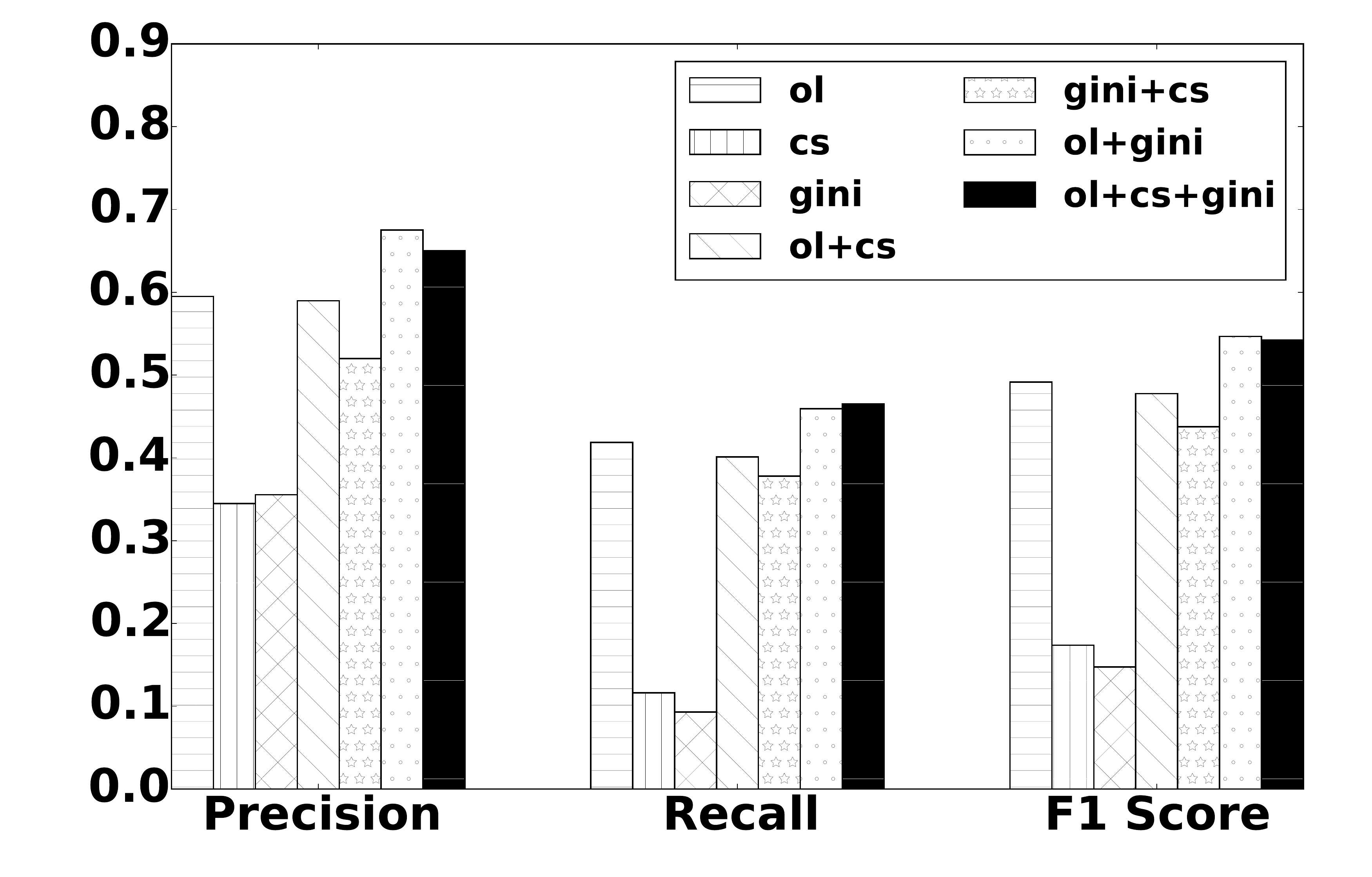}}
	\label{fig:Fsele_Time}
	\caption{Classification results (random forest with SMOTE) based on subsets of features from $A_m$ and $A_t$ (by Louvain algorithm)}
	\label{fig:Fsele}
\end{figure}

Here we investigate the importance of each feature in $ A_m $ (Louvain) and $ A_m $ (Louvain) as communities detected by Louvain algorithm achieves the best classification results out of the three.  
With $ TH_{tr} = 500 $ and $ TH_{ts} = 500 $, we trained 200 randomized logistic regressions models (100 for $A_m$ and 100 for $A_t$) - with each assigning weights to the features in those sets.  
We then categorized the features with weight larger than $ 0.01 $ (on average) into groups such as overlap, gini impurity, etc. Then, we performed classification on the basis of single feature group or combination of such groups.  The average weights assigned are shown in Table~\ref{tab:fsele} while classification results (by random forest with SMOTE) are depicted in Figure~\ref{fig:Fsele} for groups and combinations of them.  As shown, overlaps can make significant contribution to the prediction tasks. Intuitively, communication between two sets of nodes is more likely to happen in their shared communities - which is consistent with the results of \citep{ugander2012structural}. This implies that the larger overlap value, the more likely one set would repost from the other
For example, we can infer that viral cascades tend to have larger $ O(U_t,\mathcal{F}_t) $ value therefore adopters in them have larger chance to motivate the recently exposed users to repost than non-viral cascades. Figure~\ref{fig:S_ol} and Figure~\ref{fig:T_ol} provide evidence of this phenomenon. 

\begin{table}[!tbp]
	\setlength\extrarowheight{7pt}
	\centering
	\begin{tabular}{|c|c|c|c|c|}
		\hline
		Group Name & Features($A_m$(Louvain)) & Weights & Features($A_t$(Louvain)) & Weights \\
		\hline
		\multirow{3}{*}{Gini Impurity} & $I_G(F_{t(50)})$ &  0.020 & $I_G(U_{60})$ & 0.039
		\\
		& $I_G(N_{t(50)})$ & 0.021 & $I_G(U_{40})$ & 0.049 
		\\
		
		& $I_G(N_{t(30)})$ & \textbf{0.521} & $I_G(F_{40})$ & 0.331
		\\ 
		\hline
		\multirow{6}{*}{Overlap} &
		
		$O(U_{t(30)},F_{t(30)})$ & \textbf{0.503} & $O(U_{60},F_{60})$ & \textbf{0.500}
		\\ 
		
		& $O(U_{t(30)},N_{t(30)})$ & 0.037 & $O(U_{60},N_{60})$ & \textbf{0.538}
		\\  
		
		& $O(F_{t(30)},N_{t(30)})$ & 0.227 & $O(F_{60},N_{60})$ & 0.409\\
		& $O(U_{t(50)},F_{t(50)})$ & \textbf{0.500} & $O(U_{40},F_{40})$ & \textbf{0.628}
		\\
		
		& \multirow{2}{*}{$O(F_{t(50)},N_{t(50)}) $}  & \multirow{2}{*}{0.257} &  $O(U_{40},N_{40})$ & \textbf{0.509}
		\\ 
		&  &  & $O(F_{40},N_{40})$ & 0.288  \\
		\hline
		Baseline  & $ \frac{1}{50}\sum_{i=1}^{50} t(i) $ & \textbf{1.0} & $|U_{60}|$ & 0.072 \\
		\hline
	\end{tabular}
	\caption{Weights of features assigned by randomized logistic regression models}
	\label{tab:fsele}
\end{table}

\section{Related Work}
\label{sec:rwSec}
Early works about popularity prediction with data driven approach simplified the problem of cascade prediction as modeling one step information propagation \cite{galuba2010outtweeting,bian2014predicting,zhang2013social} or as predicting the near term popularity \cite{gupta2012predicting}.
As the real pioneer of cascade prediction, the work \citep{watts} devised a regression model for this task and was one of the first attempts to explore this problem.
They noted that the severe imbalance of the data due to a power-law relationship between cascade size and frequency (which we also observed) hindered the creation of useful model - they obtained an $R^2$ value of only $0.3$ for their regression model.  The later work of \citep{jenders13} also studies the problem, again taking a machine learning approach and identify several useful features to obtain relatively high precision and recall.  However, in their evaluation, they artificially balance the dataset - they ensure that each fold had equal amounts of viral and non-viral tweets.
The work of \citep{cheng2014can} predicts ``viral'' cascades with high precision and recall, but defines ``viral'' as cascades that can double in size (which also has the effect of balancing the classes in the dataset).  The very recent work of \citep{weng2014predicting} also looks at predicting viral cascades and does leverage some community-based features, some of which are also inspired by structural diversity - though their structural diversity features are more limited than in this study - we perform a comparison with their structural diversity method (see previous section).
In a nutshell, there are two main points differing our work from the ones mentioned in this section: (1). the method proposed by this paper does not need the content of microblogs or the underlying topology based on friendship relationships (2). this method is able to provide a reliable performance in prediction of order-of-magnitude increase of cascade size.
In a conference version of this paper \citep{guo2015toward} we described the basics of tis approach.  However that work did not include time-based results, examination of various underlying community finding algorithms and how each sub-group of features performs in independent classification experiments. 

In addition to the work on cascades, there is much related work on structural diversity.  This concept was first studied in \citep{ugander2012structural} and later explored in the work of \citep{zhang2013social,shakWsdm14,li2015influential,bao2013cumulative,bao2013incorporating,huang2013top}. However, these papers leverage structural diversity for a variety of other social network applications including the creation of new diffusion models, the study of peer influence, identifying influential nodes, and ranking communities.  Finally, we note that the popular work on diffusion in the areas of computer science~\citep{kkt03}, physics~\citep{sprd10}, and biology~\citep{nowak05} have led to a ground swell of research on this topic over the past decade, please see \citep{shakBk} for a review of major results.

\section{Conclusion}
In this paper, we explored the effect of structural diversity on a diffusion process which allowed us to predict viral cascades.  Moving forward, we look to integrate our structural-diversity approach with content information (which we believe will further increase performance) as well as study how to best operationalize this method in a system to detect viral cascades in near-real time.

\section{Acknowledgment}
Some of the authors of this paper are supported by by AFOSR Young Investigator Program (YIP) grant FA9550-15-1-0159, ARO grant W911NF-15-1-0282, and the DoD Minerva program.
Portions this work were also disclosed in U.S. provisional patent 62/201,517.
A non-provisional patent is currently being filed.

\bibliographystyle{spbasic}
\bibliography{ref.bib}

\end{document}